\documentclass[11pt]{article}

\usepackage{LatexCustom}

\title{}
\author{}

\begin{document}
%

\vspace{1truecm}
\renewcommand{\thefootnote}{\fnsymbol{footnote}}
\begin{center}

{\huge \bf{Quantum Corrections to Generic Branes: DBI, NLSM, and More}}

\end{center} 

\vspace{1truecm}
\thispagestyle{empty}
\centerline{\Large Garrett Goon$,{}^{\rm a}$\footnote{\href{mailto:ggoon@andrew.cmu.edu}{\tt ggoon@andrew.cmu.edu}} Scott Melville$,{}^{\rm b, c}$\footnote{\href{mailto:scott.melville@damtp.cam.ac.uk}{\tt scott.melville@damtp.cam.ac.uk}} and Johannes Noller${}^{\rm b,d}$
\footnote{\href{mailto:johannes.noller@damtp.cam.ac.uk}{\tt 	johannes.noller@damtp.cam.ac.uk}} }
\vspace{.7cm}

\centerline{\it ${}^{\rm a}$Department of Physics, Carnegie Mellon University}
\centerline{\it 5000 Forbes Ave, Pittsburgh, PA 15217, USA}

\vspace{.3cm}

\centerline{\it ${}^{\rm b}$Department of Applied Mathematics and Theoretical Physics}
\centerline{\it Cambridge University, Cambridge, CB3 0WA, UK}

\vspace{.3cm}

\centerline{ \it ${}^{\rm c}$Emmanuel College, University of Cambridge }
\centerline{ \it St Andrews Street, Cambridge CB2 3AP, UK }

\vspace{.3cm}

\centerline{\it ${}^{\rm d}$Institute for Theoretical Studies}
\centerline{\it ETH Z\"urich, Clausiusstrasse 47, 8092 Z\"urich, Switzerland}

\vspace{1cm}

\begin{abstract}

We study quantum corrections to hypersurfaces of dimension $d+1>2$ embedded in generic higher-dimensional spacetimes.  Manifest covariance is maintained throughout the analysis and our methods are valid for arbitrary co-dimension and arbitrary bulk metric. A variety of theories which are prominent in the modern amplitude literature arise as special limits: the scalar sector of Dirac-Born-Infeld theories and their multi-field variants, as well as generic non-linear sigma models and extensions thereof. Our explicit one-loop results unite the leading corrections of all such models under a single umbrella.  In contrast to naive computations which generate effective actions  that appear to violate the non-linear symmetries of their classical counterparts, our efficient methods maintain manifest covariance at all stages and make the symmetry properties of the quantum action clear.     We provide an explicit comparison between our compact construction and other approaches and demonstrate the ultimate physical equivalence between the superficially different results.

\end{abstract}

\newpage

\tableofcontents

\renewcommand*{\thefootnote}{\arabic{footnote}}
\setcounter{footnote}{0}

\newpage

\section{Introduction}

Brane actions describe the dynamics of hypersurfaces embedded into larger-dimensional spacetimes. These objects appear in many different contexts, such as in the study of thin films and soap bubbles \cite{Morgan:2007zza}, inflationary model-building \cite{Silverstein:2003hf,Alishahiha:2004eh},  as fundamental elements of string theory \cite{Polchinski:1996na}, and the celebrated AdS/CFT correspondence \cite{Maldacena:1997re}.  In a quantum mechanical setting, the $S$-matrix amplitudes associated to various brane models of the types we will consider are distinguished in the space of all quantum field theories. For instance, the scalar Dirac-Born-Infeld  (DBI) theory and the non-linear sigma model (NLSM) both arise as special limits of the classical action we consider in the following.  As is well known, DBI and NLSM amplitudes display many special properties, as they:
\begin{itemize}
\item Are constructible via soft-bootstrap methods, due to their ``exceptional" soft-scaling behavior  \cite{Cheung:2014dqa,Cheung:2016drk,Elvang:2018dco,Low:2019ynd}
\item Exhibit non-trivial single- and double-soft limits inherited from non-linear symmetries \cite{ArkaniHamed:2008gz,Cachazo:2015ksa,Cachazo:2016njl,Padilla:2016mno,Guerrieri:2017ujb,Li:2017fsb,Bogers:2018kuw,Bogers:2018zeg,Rodina:2018pcb,Yin:2018hht,Roest:2019oiw,Bonifacio:2019rpv}
\item Belong to the handful of theories which appear in double-copy relations \cite{Bern:2019prr}
\item Admit CHY representations \cite{Cachazo:2014xea,Cheung:2017ems}
\end{itemize}
The preceding references are only a partial list and many straddle the different categories above.

In this paper, we study quantum corrections to generic brane models, starting from the universal action which describes the brane bending modes of generic hypersurfaces.   Our methods apply to any system of spacetime dimension $d+1>2$. In particular, the co-dimension of the system and the bulk metric with which the higher-dimensional spacetime is endowed are both left entirely arbitrary in our analysis.   By taking various limits, our general results smoothly interpolate between a variety of models which appear in the modern amplitudes literature, such as DBI and the NLSM, and one of our central results is the compact and manifestly covariant functional determinant \eqref{Polyakov1PIFunctionalDeterminant} which encodes all one-loop corrections for the systems of interest. Explicit formulas for the corresponding logarithmic divergences in $d+1=4,6$ are given for various cases.

A technical aspect of the analysis is that naive one-loop computations of the quantum effective action will give results that do not respect the symmetries of the universal brane action.  It is well-known that this can occur when the symmetries of the original sytem are non-linear \cite{Gerstein:1971fm,Weinberg1,Weinberg2}, as is the case for generic brane systems.  In order to yield manifestly invariant results, we borrow techniques from non-Abelian gauge theory \cite{Abbott:1981ke} and NLSM analyses \cite{Honerkamp:1971sh,AlvarezGaume:1981hn} which were specifically developed to address this issue and we develop a covariant perturbation theory which utilizes the natural geometry of hypersurfaces.  We then use the  covariant heat kernel techniques reviewed in \cite{Barvinsky:1985an} to compute explicit expressions for the corresponding one-loop, logarithmic divergences in various cases.  Working in the limited context of a single DBI scalar, we explicitly compare and contrast the results of the covariant and non-covariant computations, emphasize the efficiency and elegance of the covariant method, and demonstrate their ultimate physical equivalence

\textbf{An Example:}
The universal action for a DBI scalar field is commonly written as
\begin{align}
S_{\rm DBI} &= - \int d^{d+1} x \, \sqrt{1 + ( \partial \phi )^2 } \approx \int d^{d+1} x \, \left( -1 - \frac{1}{2} ( \partial \phi )^2 + \frac{1}{8} ( \partial \phi )^4 + ...      \right) \; . 
\label{eqn:SDBI}
\end{align}
The structure of the action is protected by the following non-linear symmetry transformation:
\begin{align}
 \delta_{\rm DBI} \; \phi = b^\mu \left( x_\mu + \phi \partial_\mu \phi \right) \; . 
 \label{eqn:DBI_symm}
\end{align}
where $b^\mu$ is a constant, infinitesimal parameter.  When one-loop corrections to the corresponding quantum effective action, $\Gamma[\phi]$, are computed starting from the action as written in \eqref{eqn:SDBI} (by using, e.g., Feynman diagrams or heat kernel methods), it is found that the divergent stuctures do \textit{not} respect the symmetry \eqref{eqn:DBI_symm}.  For instance, in $d+1=4$ the leading, off-shell divergences are $\Ocal(\phi^{4})$: 
\begin{align}
 \Gamma[\phi] &\supset \tfrac{1}{30 (4 \pi)^2 \varepsilon} \,\int \mathrm{d}^4x \, \Big[-  \phi^{\mu} \phi_{\alpha}{}^{\beta} \phi^{\alpha\nu} \phi_{\mu\nu\beta} + \tfrac{1}{2} \phi^{\mu} \phi^{\nu\beta} \phi^{\alpha}{}_{\alpha} \phi_{\mu\nu\beta} -  \tfrac{47}{4} \phi^{\mu} \phi_{\alpha\nu} \phi^{\alpha\nu} \phi_{\mu}{}^{\beta}{}_{\beta}\nn 
 & -  \tfrac{11}{4} \phi^{\mu} \phi^{\nu}{}_{\nu} \phi^{\alpha}{}_{\alpha} \phi_{\mu}{}^{\beta}{}_{\beta} -  \tfrac{1}{4} \phi^{\mu} \phi^{\alpha} \phi_{\mu}{}^{\nu\beta} \phi_{\alpha\nu\beta} -  \tfrac{29}{4} \phi^{\mu} \phi^{\alpha} \phi_{\mu}{}^{\nu}{}_{\nu} \phi_{\alpha}{}^{\beta}{}_{\beta}+\ldots
 \Big]\ ,
 \label{eqn:DBI_G4_Intro}
\end{align} where $\ldots$ contains terms with two derivatives on each $\phi$ (see \eqref{eqn:DBI_G4} for the full expression) and $\phi_{\mu\ldots\nu}\equiv \partial_{\mu}\ldots\partial_{\nu}\phi$. DBI invariance demands that \eqref{eqn:DBI_G4_Intro} be symmetric under the field-independent part of \eqref{eqn:DBI_symm}, $ \phi \longrightarrow \phi+ b^\mu  x_\mu $, and it is straightforward to check that this test fails.  An extensive discussion of this system is continued in Sec.~\ref{sec:NaiveDBI}.

The naive computation sketched above is clearly unsatisfactory. For one, the loss of manifest DBI invariance leads to an unwanted (and unnecessary, as we will show) proliferation of divergent structures.  For instance, the one-loop computation generates divergences $\propto \partial^{8} \phi^4$, schematically, and there exist a plethora of independent operators of this general form, only a small subset of which could have arisen from operators invariant under \eqref{eqn:DBI_symm}.  The gap in this counting grows as one goes higher in fields and/or loops.  Additionally, DBI is but one example in a family of closely-related, ``exceptional" scalar theories whose forms are dictated by non-linear symmetries.  One expects similar issues to arise for other models in this class and a conventional renormalization program would require treating each theory and its attendant, messy divergences on a case-by-case basis.

In this work, we have overcome these concerns by uniting a wide variety of theories under a single geometric framework and utilizing a scheme which manifestly preserves all relevant symmetries. For instance, we provide in equation~\eqref{TrLogdims3} an explicit, symmetry-preserving master formula for the one-loop divergences arising in \emph{any} $d+1=4$ dimensional scalar field theory with non-linear symmetries that can be realized via a conventional brane construction.

\textbf{Outline:} In Sec.~\ref{sec:BranesAndLoops} we briefly review the construction of brane actions and discuss their quantum corrections in general terms.  Sec.~\ref{sec:GenericBraneComputations} is devoted to the general analysis of loop-corrections to brane systems and contains our main results, while Sec.~\ref{sec:NaiveDBI} highlights the advantages of our method over naive approaches. In Sec.~\ref{sec:Applications} we discuss various limits of our general results, including DBI and NLSM, and perform multiple non-trivial checks on our formulas.  In Sec.~\ref{sec:Conclusions} we conclude.  Our conventions can be found in App.~\ref{app:Conventions}, while the remaining appendices contain reviews of relevant topics and details of various computations discussed in the main body of the paper.

\section{Branes and Loops\label{sec:BranesAndLoops}}

In this section we briefly review the construction of brane actions, the ingredients needed for computing their loop corrections, and the subtleties which can arise in such calculations.

\subsection{Brane Actions and the Universal Term\label{sec:Branes}}

The position of a $(d+1)$-dimensional brane in some larger $(D+1)$-dimensional spacetime can be described with the help of embedding functions $X^{A}(x^{\mu})$, $A\in\{0,\ldots, D\}$, $\mu \in \{0,\ldots,d\}$ which associate each point on the brane, $x^{\mu}$, to a point in the larger spacetime, $X^{A}$.    We will denote the brane by $M_{d+1}$ and the bulk by $\Mcal_{D+1}$, so that the $X^{A}$ are maps $X^{A}: M_{d+1}\longrightarrow \Mcal_{D+1}$.  In a string theory context, such objects are referred to as D$p$-branes, where $p=d$.

If $\Mcal_{D+1}$ is endowed with a metric $\Gcal_{AB}$, the embedding functions induce a natural metric $g_{\mu\nu}$ on $M_{d+1}$ via the pullback:
\begin{align}
g_{\mu\nu}(x)&=\frac{\partial X^{A}}{\partial x^{\mu }}\frac{\partial X^{B}}{\partial x^{\nu}}\Gcal_{AB}(X)\ .\label{InducedMetric}
\end{align}
Associated to $g_{\mu\nu}$ and $\Gcal_{AB}$ are the Riemann curvature tensors $R_{\mu\nu\rho\sigma}$ and $\Rcal_{ABCD}$, respectively.
The factors $\frac{\partial X^{A}}{\partial x^{\mu }}$ also define $(d+1)$ bulk vectors $e_{\mu}=e_{\mu}{}^{A}\frac{\partial}{\partial X^{A}}\equiv  \frac{\partial X^{A}}{\partial x^{\mu }}\frac{\partial}{\partial X^{A}}$ which are tangent to $M_{d+1}$. Orthogonal to the $e_{\mu}$ are $(D-d)$ normal vectors whose derivative along the brane determine the $(D-d)$ independent extrinsic curvature tensors $K_{\mu\nu}{}^{A}$.  The preceding ingredients transform covariantly under both brane and bulk diffeomorphisms and there exists a covariant derivative on the brane $\Dcal_{\mu}$ which respects both such transformations.    For more on the geometry of generic hypersurfaces, see App.~\ref{app:Hypersurfaces}.

Invariant actions describing brane dynamics are built from diffeomorphism invariant combinations of the natural geometric building blocks:
\begin{align}
S&=\int\rd^{d+1}x\, \sqrt{-g}\, \mathcal{L}\left (g_{\mu\nu},R_{\mu\nu\rho\sigma},\Dcal_{\mu}, e_{\mu}{}^{A},K_{\mu\nu}{}^{A},\Gcal_{AB}, \Rcal_{ABCD},\nabla_{A}\right )\ ,\label{GenericBraneAction}
\end{align}
with all indices contracted in the natural manner and all bulk quantities are pulled back to $M_{d+1}$. 
The dynamical degrees of freedom in the preceding actions are the $D+1$ functions $X^{A}(x^{\mu})$.  Bulk diffeomorphisms simply correspond to field-redefinitions of the $X^{A}$'s, from this point of view, while brane diffeomorphisms are true gauge symmetries.   It is common in the literature to use this gauge freedom to fix ``unitary gauge" in which the first $d+1$ $X^{A}$'s are locked to the brane coordinates, $X^{\mu}(x)=x^{\mu}$, while the remaining fields are dynamical, $X^{a}=\phi^{a}$, $a\in \{d+1,\ldots,D\}$.  Any isometries of the bulk metric $\Gcal_{AB}$ turn into global symmetries of the action \eqref{GenericBraneAction}.  The $\phi^{a}$ degrees of freedom are sometimes referred to as ``branons" in the literature, as in \cite{Creminelli:2000gh,Contino:2001nj,Cembranos:2003mr}.  While more general high-energy studies of branes often include couplings to additional fields, such as the dilaton or various gauge bosons, in the present work we exclusively focus on actions which describe the brane's translational degrees of freedom.

The universal term in the brane action is simply the lowest dimension operator contained in \eqref{GenericBraneAction}, which is the volume element contribution\footnote{Dimensionally, a $\sim\Lambda^{d+1}$ prefactor should be included in $S_{\rm universal}$. We set this scale to unity here and in analogous actions below for clarity of presentation. It is trivial to restore such factors in later expressions.}
\begin{align}
S_{\rm universal}\equiv-\int\rd^{d+1}x\, \sqrt{-g}=-\int\rd^{d+1}x\, \sqrt{-\det \left (\partial_{\mu }X^{A}\partial_{\nu}X^{B}\Gcal_{AB}(X)\right )}\ .\label{UniversalBraneAction}
\end{align} 
The remainder of this paper is devoted to studying one-loop corrections to the action \eqref{UniversalBraneAction}.

\subsection{Divergences, Field Variables, and the Quantum Effective Action\label{sec:QuantumActionGeneral}}

Our goal is to compute the quantum corrections to the universal brane action \eqref{UniversalBraneAction}, focusing on one-loop results, at present. Ideally, the output of any such computation would be valid for arbitrary choices of bulk-metric $\Gcal_{AB}$ and bulk dimension $D+1$ and maintain manifest covariance with respect both brane and bulk diffeomorphisms.   While the standard lore is that the divergences respect all symmetries of the underlying action $S$, this statement is not without its subtleties, as we address in following sections.

The framework we find most useful for computing the divergences is the quantum effective action. Given an action $S[\phi]$ depending on some set of fields $\phi$, one probe of the quantum properties of the system is the quantum effective action\footnote{For notational simplicity, we use $\phi$ to denote the fields appearing in both the classical and quantum effective actions, despite the fact that these are logically distinct quantities.} $\Gamma[\phi]$ which can be defined through the path integral as 
\begin{align}
\exp\left (i \Gamma[\phi]\right )&=\int_{\rm 1PI} \Dcal \varphi\, \exp\left (iS[\phi+\varphi]\right )\ ,\label{1PIActionDef}
\end{align}
schematically, where the subscript on the integral indicates that when computing $\Gamma[\phi]$ diagrammatically, only diagrams which are 1-particle-irreducible (1PI) in $\varphi$-lines are included in the sum \cite{Abbott:1981ke}.  When $S[\phi]$ enjoys gauge symmetries, additional gauge-fixing and ghost-terms are required for a proper definition of \eqref{1PIActionDef}, as usual.

 In the one-loop approximation, \eqref{1PIActionDef} simplifies to 
\begin{align}
\exp\left (i \Gamma[\phi]\right )&\approx \exp\left (iS[\phi]-\frac{c}{2}\Tr\ln\left (\frac{\delta^{2}S\left [\phi+\varphi\right ]}{\delta\varphi(x)\delta\varphi(y)}\Big|_{\varphi=0}\right )+\ldots\right )\nn
&=\exp \left (iS[\phi]+i\Gamma_{1}[\phi]+\ldots\right )\label{1PIActionOneLoop}
\end{align}
where $c$ is a number which depends on the nature of the fields $\phi$.  While the preceding expression for $\Gamma[\phi]$ is valid regardless of whether $\phi$ extremizes $\Gamma[\phi]$, we will restrict the following calculations to the case where $\phi$ is a saddle point of the action $S[\phi]$.  In addition to being technically advantageous to impose this condition, it is only the on-shell value of $\Gamma[\phi]$ which is physical. For instance, $S$-matrix amplitudes and tunneling rates follow from on-shell computations using $\Gamma[\phi]$ and in gauge-theory contexts, $\Gamma[\phi]$ is only gauge-independent when evaluated on-shell; see, e.g., \cite{Nielsen:1975fs,Fukuda:1975di,Aitchison:1983ns,Hart:1984jy,Georgi:1991ch,Isidori:2001bm,Andreassen:2014eha,DiLuzio:2015iua,Andreassen:2016cvx}.  At one-loop order it is sufficient to use the \textit{tree-level} on-shell conditions in \textit{both} terms in \eqref{1PIActionOneLoop}, since the tree-level solution already extremizes $S[\phi]$.

The form of $\Gamma_{1}[\phi]$ is in general sensitive to the details of the calculation.  Gauge choices, field parameterizations, and regularization procedures can each affect its functional form.  In particular, divergent terms in $\Gamma_{1}[\phi]$ are not guaranteed to respect any non-linear symmetries that $S[\phi]$ may enjoy \cite{Weinberg1,Weinberg2}.   This fact is familiar from the study of NLSMs, for instance, where wisely chosen\footnote{The essential point is that $\Gamma[\phi]$ is only guaranteed to share the symmetries of the original problem when they act on $\phi$ in a linear manner and so a wise scheme will use field variables which have this property. More precisely, if one can choose the fluctuation field $\varphi$ appearing in \eqref{1PIActionDef} such that its transformation is linear, then the corresponding $\Gamma[\phi]$ will share the symmetries of $S[\phi]$.  This is one of the central properties of the background-field-method \cite{Abbott:1981ke}. \label{foot:1PISymmetries}} computational schemes are required to keep all symmetries manifest \cite{Gerstein:1971fm,AlvarezGaume:1981hn,Howe:1986vm,Mukhi:1985vy}.  However, since the underlying physics cannot depend on the preceding choices, all possible answers must somehow be physically equivalent.  Concentrating on the divergent pieces of $\Gamma_{1}[\phi]$, one expects that any required counterterms which appear to break the original symmetries of the problem must be related to a manifestly symmetric divergence after the use of field-redefinitions (equivalent to the use of on-shell conditions in $\Gamma_{1}[\phi]$) and integrations-by-parts, since the effect of such counterterms on $S$-matrix elements is insensitive to these operations.  We demonstrate this phenomenon explicitly in a concrete example in Sec.~\ref{sec:NaiveDBI}.  In the following section, we realize the goal stated in the beginning of this section by setting up a covariant form of brane perturbation theory and writing the corresponding functional determinant whose form manifestly respects all symmetries of the universal brane action and encodes all one-loop corrections to the system.

\section{Covariant Computations: Generic Branes\label{sec:GenericBraneComputations}}

In this section, we develop the covariant perturbation theory appropriate for computing quantum corrections to brane actions via the background-field-method \cite{Abbott:1981ke}, including appropriate gauge-fixing terms. The culmination of these efforts is the compact functional determinant \eqref{Polyakov1PIFunctionalDeterminant}. The methods we use are familiar from the string theory literature \cite{AlvarezGaume:1981hn,Howe:1986vm,Mukhi:1985vy}.  However, because we only consider the cases where $d+1>2$, Weyl invariance is lost and the present analysis differs from classic string theory scenarios.

 \subsection{Nambu-Goto and Polyakov Actions\label{sec:PolyakovAction}}
 
 The square-root in the universal, Nambu-Goto action \eqref{UniversalBraneAction} can be avoided by integrating in an auxiliary metric  $g_{\mu\nu}$, which results in the Polyakov action
\begin{align}
S_{\rm Poly}[g,X]&=\int\rd^{d+1}x\, \sqrt{-g}\left (-\frac{1}{2}g^{\mu\nu}\partial_{\mu}X^{A}\partial_{\nu}X^{B}\Gcal_{AB}(X)+\frac{(d-1)}{2}\right )\ .\label{PolyakovAction}
\end{align}
In the following, we present in detail the one-loop calculation arising from \eqref{PolyakovAction}. We have also carried out the computation starting from the original action \eqref{UniversalBraneAction}, which we comment on in Sec.~\ref{sec:NGVsPolyakov}. It turns out that Polyakov form of the action does not prove particularly advantageous for our computation, as we comment on in Sec.~\ref{sec:NGVsPolyakov}.

In \eqref{PolyakovAction}, the $g$ and $X$ equations of motion set
\begin{align}
g_{\mu\nu}&=\partial_{\mu }X^{A}\partial_{\nu}X^{B}\Gcal_{AB}(X)\nn
0&=\nabla^{\mu}\nabla_{\mu} X^{A}+\nabla_{\mu}X^{B}\nabla^{\mu}X^{C} \Gamma^{A}_{BC}=\Dcal^{\mu}e_{\mu}{}^{A}=-K_{\mu}{}^{\mu A}\label{PolyakovTreeLevelEOM}\ ,
\end{align}
where $\nabla_{\mu}$ is the brane-covariant derivative with respect to $g_{\mu\nu}$, $\Dcal_{\mu}$ is covariant with respect to diffeomorphisms of both the bulk and brane \eqref{app:eq:BraneCovariantDerivative}, and all other ingredients are addressed in Sec.~\ref{sec:Branes} and/or in App.~\ref{app:Hypersurfaces}.  The on-shell conditions \eqref{PolyakovTreeLevelEOM} will be used to simplify the computation of the one-loop functional determinant.

 \subsection{The 1PI Computation: Polyakov Action}
 
We now carry out the one-loop computation described in Sec.~\ref{sec:QuantumActionGeneral} for the Polyakov action \eqref{PolyakovAction}.

\subsubsection{Field Variables}

Though physical observables are insensitive to the choice of field variables and regularization procedure, some options are nevertheless wiser than others.  For the case of brane actions depending on fields $X^{A}$ which represent bulk coordinates, the naive way to introduce fluctuations $\chi^{A}$ for the background-field-method computation of Sec.~\ref{sec:QuantumActionGeneral} is to simply shift $X^{A}\longrightarrow X^{A}+\chi^{A}$ in \eqref{PolyakovAction}.  However, because the $\chi^{A}$ introduced in this manner are not proper tensors, this choice breaks manifest covariance.    A more elegant choice is to introduce $\chi^{A}$ by using normal coordinates around the point $X^{A}$ in which case the $\chi^{A}$'s can be taken as tangent vectors to geodesics emanating from $X^{A}$ \cite{Honerkamp:1971sh,AlvarezGaume:1981hn,Mukhi:1985vy} and are thus properly tensorial.  From the point of view of the quantum computation, this field choice realizes the goal explained in Foot.~\ref{foot:1PISymmetries} and ensures the covariance of the quantum effective action. The explicit construction is reviewed in App.~\ref{app:NormalCoordinates} and the resulting terms which are quadratic in fluctuations are
\begin{align}
S_{\rm Poly}^{(2)}&=\int\rd^{d+1}x\, \sqrt{-g}\,\Big(-\frac{1}{2}\Dcal^{\mu}\chi^{A}\Dcal_{\mu}\chi_{A}+\frac{1}{2}R_{ABCD}\chi^{A}\chi^{C}e_{\mu}{}^{B}e^{\mu D}\nn
&\quad-\frac{1}{2}h e^{\mu A}\Dcal_{\mu}\chi_{A}-\frac{1}{4}h_{\mu\nu}h^{\mu\nu}+\frac{1}{8}h^{2}+e^{\mu A}h_{\mu\alpha}\Dcal^{\alpha}\chi_{A}\Big)\label{PolyakovActionQuadratic}\ ,
\end{align}
where fluctuations of $g_{\mu\nu}$, denoted by $h_{\mu\nu}$, were introduced by sending $g_{\mu\nu}\longrightarrow g_{\mu\nu}+h_{\mu\nu}$.  The on-shell conditions on background fields \eqref{PolyakovTreeLevelEOM} were also imposed above.  The action \eqref{PolyakovActionQuadratic} manifestly respects all expected symmetries which act on the fields in the natural manner.
 
 \subsubsection{The Functional Determinant}

 The one-loop correction $\Gamma_{1}[g,X]$ then arises from the functional determinant
 \begin{align}
 \Gamma_{1}[g,X]&=\frac{i}{2}\ln  \Det \left ( \frac{\delta^{2} S'_{\rm Poly}}{\delta \phi(x)\delta\phi(y)}\right )+{\rm ghost-determinants}
 \end{align} 
 where
 $S'_{\rm Poly}=S_{\rm Poly}^{(2)}+S_{\rm GF}$ with $S_{\rm Poly}^{(2)}$ as in \eqref{PolyakovActionQuadratic} and $S_{\rm GF}$ a gauge-fixing term (discussed below), where the final piece above comes from the usual FP ghosts, and where we have defined
 \begin{align}
 \frac{\delta^{2} S'_{\rm Poly}}{\delta \phi(x)\delta\phi(y)}&\equiv \begin{pmatrix}
 \frac{\delta^{2} S'_{\rm Poly}}{\delta h^{\alpha\beta}( \chi ) \delta h^{\mu\nu}(y)} & \frac{\delta^{2} S'_{\rm Poly}}{\delta h^{\alpha\beta}( \chi ) \delta  \chi ^{B}(y)} \\
 \frac{\delta^{2} S'_{\rm Poly}}{ \delta  \chi ^{A}( \chi ) \delta h^{\mu\nu}(y)} & \frac{\delta^{2} S'_{\rm Poly}}{\delta   \chi ^{A}( \chi )\delta   \chi ^{B}(y)}
 \end{pmatrix}\ .\label{SPrimeDerivativeMatrix}
 \end{align}
Variational derivatives such as $ \frac{\delta^{2} S'_{\rm Poly}}{\delta \phi(x)\delta\phi(y)}$ are defined so that $ x$-variation is taken first so that $\frac{\delta^{2} S'_{\rm Poly}}{\delta \phi(x)\delta\phi(y)}$ naturally acts to the right on a $y$-valued object. The $S_{\rm Poly}^{(2)}$-dependent parts of \eqref{SPrimeDerivativeMatrix} are
 \renewcommand{\arraystretch}{1.25} 
  \begin{align}
 \frac{\delta S_{\rm Poly}^{(2)}}{\delta \phi(x)\delta\phi(y)} &=\left (\begin{array}{c|c}
  \frac{1}{4}g_{\alpha\beta}g_{\mu\nu}-\frac{1}{2}g_{\alpha(\mu}g_{\nu)\beta} &  -\frac{1}{2}e^{\mu }{}_{B}g_{\alpha\beta}\Dcal_{\mu}+e_{(\alpha| B|}\Dcal_{\beta)}\nn \hline
  \frac{1}{2}e_{\alpha A}g_{\mu\nu}\Dcal^{\alpha}+K_{\mu\nu A}-2e_{\alpha A}g^{\alpha}{}_{(\mu}\Dcal_{\nu)} & \Gcal_{AB}\Dcal^{2}+e_{\alpha}{}^{C}e^{\alpha D}\Rcal_{ACBD}
 \end{array}\right ) \  .
 \end{align}
  where the tree-level on-shell conditions $K_{\mu}{}^{\mu A}=0$ were used to simplify and the $\sqrt{-g}\delta^{d+1}(x-y)$ factor was left implicit.
  
Let us define $G_{\mu\nu\alpha\beta}( x ,y)$ to be the inverse of $\frac{\delta S'_{\rm Poly}}{\delta h^{\alpha\beta}( \chi ) \delta h^{\mu\nu}(y)}$:
\begin{align}
\int\rd^{d+1}z\,\frac{\delta^{2} S'_{\rm Poly}}{\delta h^{\alpha\beta}( x ) \delta h^{\mu\nu}(z)}G^{\mu\nu\rho\sigma}(z,y)=\delta^{(\rho}_{(\alpha}\delta^{\sigma)}_{\beta)}\delta^{d+1}(x-y)\ ,\label{GPropagatorDef}
\end{align}
it is then useful to insert ${\bf 1}$ into the determinant in the (schematic) form
\begin{align}
{\bf 1}&=\begin{pmatrix}
\frac{\delta^{2}S'}{\delta h^{2}} & 0\\
0 & {\bf 1}
\end{pmatrix}
\cdot 
\begin{pmatrix}
G & 0\\
0 & {\bf 1}
\end{pmatrix}\ ,
\end{align}
after which the determinant usefully factorizes as
\begin{align}
 \Det  \left ( \frac{\delta^{2} S'_{\rm Poly}}{\delta \phi(x)\delta\phi(y)}\right )&=  \Det \left (\frac{\delta^{2}S'_{\rm Poly}}{\delta h^{2}}\right )\times  \Det \left (\frac{\delta^{2}S'_{\rm Poly}}{\delta  \chi ^{2}}-\frac{\delta^{2}S'_{\rm Poly}}{\delta  \chi  \delta h}G \frac{\delta^{2}S'_{\rm Poly}}{\delta h\delta  \chi }\right )\ ,\label{PolyakovDetFactored}
\end{align}
with proper index  placements and arguments left implicit.  

Note that $\frac{\delta^{2}S_{\rm Poly}^{(2)}}{\delta h^{2}}$ is local in position space.   If we choose our gauge fixing term such that $S_{\rm GF}$ does not involve derivatives of $h_{\mu\nu}$, then $\frac{\delta^{2}S'_{\rm Poly}}{\delta h^{2}}$ will also be local in position space and the contribution of this term in the action will be
\begin{align}
 \Gamma_{1}\supset \frac{i}{2}\Tr \ln \left (\frac{\delta^{2}S'_{\rm Poly}}{\delta h^{2}}\right )\propto \delta^{d+1}(0)\ ,
 \end{align}
 which is vanishing in any scale-free regularization scheme, which we assume throughout.  The vanishing of $\delta^{d+1}(0)$ is used repeatedly below.  If we further assume that $S_{\rm GF}$ is entirely independent of $h_{\mu\nu}$, then the propagator in \eqref{GPropagatorDef} can be explicitly computed from $S_{\rm Poly}^{(2)}$ alone, 
 \begin{align}
 G_{\alpha\beta\rho\sigma}(x,y)&=\frac{1}{\sqrt{-g}}\left (-2g_{\alpha(\rho}g_{\sigma)\beta}+\frac{2}{d-1}g_{\alpha\beta}g_{\rho\sigma}\right )\delta^{d+1}(x-y)\ ,\label{hPropagator}
 \end{align}
  and the final factor in \eqref{PolyakovDetFactored} becomes
 \begin{align}
 &\quad \Det \left (\frac{\delta^{2}S'_{\rm Poly}}{\delta  \chi ^{2}}-\frac{\delta^{2}S'_{\rm Poly}}{\delta  \chi  \delta h}G \frac{\delta^{2}S'_{\rm Poly}}{\delta h\delta  \chi }\right )\nn
 &=\Det\left (\Pcal_{\perp}^{AB}\Dcal^{2}+2K_{\alpha\beta}{}^{A}e^{\alpha B}\Dcal^{\beta}-\Pcal_{\perp}^{AC}\Pcal_{\parallel}^{DE}\Rcal_{CDE}{}^{B}+\frac{\delta^{2}S_{\rm GF}}{\delta\chi_{A}\delta \chi_{B}}\right)\ ,\label{PolyakovFunctionalDeterminant}
 \end{align}
 where $\Pcal_{\perp}$ and $\Pcal_{\parallel}$ are projectors onto the spaces normal and tangent to the brane reviewed in App.~\ref{app:Hypersurfaces}:
 \begin{align}
 \Pcal_{\perp}^{AB}=\Gcal^{AB}-e^{\mu A}e_{\mu}{}^{B}\ , \quad \Pcal_{\parallel}^{AB}=e^{\mu A}e_{\mu}{}^{B}\ .
 \end{align}

 \subsubsection{Projections and Gauge-Fixing}
 
   It is then convenient to project \eqref{PolyakovFunctionalDeterminant} by inserting $\mathbf{1}=\Pcal_{\perp}+\Pcal_{\parallel}$ and separating the argument of \eqref{PolyakovFunctionalDeterminant} into its various distinct components. Using the shorthand $\mathcal{O}+\frac{\delta^{2}S_{\rm GF}}{\delta\chi^{2}}$ for the operator in \eqref{PolyakovFunctionalDeterminant} and noting that $\Pcal_{\parallel}\cdot\mathcal{O}=0$, due to the explicit $\Pcal_{\perp}$ projectors and the fact that $K_{\mu\nu}{}^{A}$ is normal to the brane, \eqref{PolyakovFunctionalDeterminant} is equivalently written as
   \begin{align}
&\quad  \Det  \left (\mathcal{O}+\frac{\delta^{2}S_{\rm GF}}{\delta\chi^{2}}\right )\nn
&= \Det\begin{pmatrix}
\Pcal_{\parallel}\cdot\frac{\delta^{2} S_{\rm GF}}{\delta \chi^{2}}\cdot \Pcal_{\parallel} & \Pcal_{\parallel}\cdot\frac{\delta^{2} S_{\rm GF}}{\delta \chi^{2}}\cdot \Pcal_{\perp}\\
\Pcal_{\perp}\cdot \left( \mathcal{O}+\frac{\delta^{2}S_{\rm GF}}{\delta\chi^{2}} \right)\cdot \Pcal_{\parallel} & \Pcal_{\perp}\cdot \left( \mathcal{O}+\frac{\delta^{2}S_{\rm GF}}{\delta\chi^{2}} \right)\cdot \Pcal_{\perp}
\end{pmatrix}\ .\label{PolyakovFunctionalDeterminantProjected}
\end{align}

The above suggests that it is wise to choose a gauge-fixing term which obeys $\frac{\delta^{2} S_{\rm GF}}{\delta \chi^{2}}\cdot \Pcal_{\perp}=0$, since such a choice factorizes the preceding determinant:
\begin{align}
\frac{\delta^{2} S_{\rm GF}}{\delta \chi^{2}}\cdot \Pcal_{\perp}=0\implies  \Det  \left (\mathcal{O}+\frac{\delta^{2}S_{\rm GF}}{\delta\chi^{2}}\right )=\Det \left (\Pcal_{\parallel}\cdot\frac{\delta^{2} S_{\rm GF}}{\delta \chi^{2}}\cdot \Pcal_{\parallel} \right )\times	\Det\left ( \Pcal_{\perp}\cdot \mathcal{O}\cdot \Pcal_{\perp}\right )\ .\label{PolyakovFunctionalDeterminantProjectedFactored}
\end{align}
The following gauge-fixing function $G_{\mu}(\chi,X)$ realizes this goal while simultaneously preserving manifest covariance:
\begin{align}
G_{\mu}=\chi_{A}\nabla_{\mu}X^{A}&\implies S_{\rm GF}=\int\rd^{d+1}x\, \sqrt{-g}\frac{1}{2\xi}\left (\chi\cdot \nabla_{\mu}X\right )\left (\chi\cdot \nabla^{\mu}X\right )\nn
&\implies \frac{1}{\sqrt{-g}}\frac{\delta^{2}S_{\rm GF}}{\delta \chi_{A}\delta \chi_{B}}=\frac{1}{\xi}\Pcal_{\parallel}^{AB}\ .\label{PolyakovGFTerm}
\end{align}

The gauge-fixing term in \eqref{PolyakovGFTerm} is additionally convenient as its one-loop contributions to the 1PI action are completely trivial when a scale-free regularization scheme is used.  The first factor in \eqref{PolyakovFunctionalDeterminantProjectedFactored} produces a term
\begin{align}
\Gamma\supset \frac{i}{2}\Tr\ln \frac{1}{\xi}\Pcal_{\parallel}\ ,
\end{align}
and since $\Pcal_{\parallel}$ is a diagonal operator in position space, the above is $\propto \delta^{d+1}(0)$  and hence trivial for aforementioned reasons.  The FP determinant associated to \eqref{PolyakovGFTerm} contributes trivially along the same lines
\begin{align}
\Gamma\supset -i\Tr\ln \partial_{\mu}\left (\chi+X\right )\cdot\partial_{\nu}X\propto\delta^{d+1}(0)\longrightarrow 0\ .
\end{align}

\subsubsection{Final Form}

Therefore, the only surviving contribution to the one-loop effective action is
\begin{align}
\Gamma_{1}=\frac{i}{2}\Tr\ln \left [\Pcal_{\perp}^{A}{}_{Y} \left (\Pcal_{\perp}^{YZ}\Dcal^{2}+2K_{\alpha\beta}{}^{Y}e^{\alpha Z}\Dcal^{\beta}-\Pcal_{\perp}^{YU}\Pcal_{\parallel}^{VW}\Rcal_{UVW}{}^{Z}\right )\Pcal_{\perp Z}{}^{B}\right ]\ .
\end{align}
The trace is taken over the space of tensors normal to the brane and the above can be more naturally written in terms of the covariant derivative which maps normal tensors to normal tensors: $\Dcal_{\perp}\equiv \Pcal_{\perp}\cdot \Dcal\cdot\Pcal_{\perp}$, schematically; see \eqref{app:eq:DPerpParallel}.  After translation, we find
\begin{align}
\Gamma_{1}&=\frac{i}{2}\Tr \ln\left ( \Pcal_{\perp}^{AB}\Dcal_{\perp}^{2}+K_{\alpha\beta}{}^{A}K^{\alpha\beta B}-\Pcal_{\perp}^{AY}\Rcal_{YVWX}\Pcal_{\parallel}^{WV}\Pcal_{\perp}^{XB}\right )\ .\label{Polyakov1PIFunctionalDeterminant}
\end{align}
Since the number of physical degrees of freedom is given by the co-dimension of the brane, $D-d$, it is pleasing that the ultimate functional trace is over a subspace of the same dimensionality: $\tr \Pcal_{\perp}^{AB}=D-d$. Functional traces of precisely the above form are well-studied and the logarithmically divergent terms are known in various dimensions; see, e.g.,  \cite{Gilkey:1975iq,Barvinsky:1985an,Avramidi:1986mj} and App.~\ref{app:CovariantHeatKernels} for a review.

\subsection{Nambu-Goto vs.~Polyakov\label{sec:NGVsPolyakov}}

Starting with the Polyakov form of the action was not necessary or even necessarily helpful for this calculation.  If we had started with the square-root form of the action \eqref{UniversalBraneAction} and introduced $\chi$ by shifting $g_{\mu\nu}=\partial_{\mu}X^{A}\partial_{\nu}X^{B}\Gcal_{AB}$ by $g_{\mu\nu}\longrightarrow g_{\mu\nu}+h_{\mu\nu}$ with
\begin{align}
h_{\mu\nu}=2\Dcal_{(\mu}\chi^{A}e_{\nu)A}+\Dcal_{\mu}\chi^{A}\Dcal_{\nu}\chi_{A}-\Rcal_{ABCD}e_{\mu}{}^{A}e_{\nu}{}^{C}\chi^{B}\chi^{D}\ ,
\end{align}
as follows from \eqref{app:eq:NormalCoordinateMetricExpansion}, and added the gauge-fixing term \eqref{PolyakovGFTerm} to the action, a straightforward calculation shows that we would have arrived at precisely the same ultimate result \eqref{Polyakov1PIFunctionalDeterminant} without the need for introducing an independent $g_{\mu\nu}$ field.  The calculation started with the Polyakov action \eqref{PolyakovAction} was presented in order to make better contact with standard string theory methods. It is possible that the Polyakov-like form of the action would prove more advantageous when studying actions beyond the universal form \eqref{UniversalBraneAction} or when computing to higher-orders in loops.

\subsection{Explicit Results}

We can compute the logarithmic divergences arising from \eqref{Polyakov1PIFunctionalDeterminant} using the well-known heat kernel results reviewed in App.~\ref{app:CovariantHeatKernels}.   Using dimensional regularization in $d+1-2\varepsilon$ dimensions, one-loop divergences occur when $d+1=2n$, $n\in\mathbb{Z}$ and are given by \eqref{app:eq:LogDivergencesFromHeatKernel} 
\begin{align}
\Tr\ln\Ocal\supset -\frac{1}{\varepsilon}\frac{i}{(4\pi)^{n}}\int\rd^{2n}x\, \sqrt{-g(x)}\,\tr\left [\mathsf{a}_{n}(x)\right ]\ ,
\end{align}
where the $a_{n}$'s are the Seeley-DeWitt coefficients associated to the operator appearing in \eqref{Polyakov1PIFunctionalDeterminant} and which are reviewed in App.~\ref{app:CovariantHeatKernels}.

\subsubsection{General Formula $d+1=4$}

When $d+1=4$, we find that after using the Gauss-Codazzi relations \eqref{app:eq:GaussCodazzi} to remove all instances of the brane Riemann curvature $R_{\mu\nu\rho\sigma}$, the coefficient which controls the divergence is 
\begin{align}
\tr\left [\mathsf{a}_{2}\right ]&=\left (\frac{3-D}{60}\right )\langle\Kcal^{A}{}_{B}{}^{A}{}_{B}\rangle +\left (\frac{18-D}{45}\right )\langle\Kcal^{[AB]}{}_{AB}\rangle\nn
&\quad +\left (\frac{31+3D}{120}\right ) \langle \Kcal^{A}{}_{A}\rangle\langle \Kcal^{B}{}_{B}\rangle- \left (\frac{42+D}{45}\right )\langle \Kcal^{[A}{}_{A}\rangle\langle \Kcal^{B]}{}_{B}\rangle\nn
&\quad -  \frac{1}{3} e^{\alpha A} e^{\beta B} K_{\alpha}{}^{\gamma C} K_{\beta \gamma}{}^{D} \Rcal_{ABCD} + \frac{1}{45} \left(-3 + D\right) e^{\alpha A} e^{\beta B} e^{\gamma C} e^{\delta D} K_{\alpha \beta}{}^{E} K_{\gamma \delta E} \Rcal_{ACBD} \nn
&\quad+ \langle \Kcal^{AB}\rangle\Pcal_{\parallel}^{CD} \Rcal_{ACBD} + \frac{1}{90} \left(-3 + D\right) e^{\alpha A} e^{\beta B} K_{\alpha}{}^{\gamma C} K_{\beta \gamma C} \Pcal_{\parallel}^{DE} \Rcal_{ADBE} \nn
&\quad-  \frac{1}{6}\langle \Kcal^{A}{}_{A}\rangle \Pcal_{\perp}^{BC} \Pcal_{\parallel}^{DE} \Rcal_{BDCE} + \frac{1}{36} \left(9 -  D\right)\langle \Kcal^{A}{}_{A}\rangle \Pcal_{\parallel}^{BC} \Pcal_{\parallel}^{DE} \Rcal_{BDCE} \nn
&\quad-  \frac{1}{12} \Pcal_{\perp}^{AB} \Pcal_{\perp}^{CD} \Pcal_{\parallel}^{EF} \Pcal_{\parallel}^{GH} \Rcal_{ACEG} \Rcal_{BDFH} + \frac{1}{180} \left(-3 + D\right) \Pcal_{\parallel}^{AB} \Pcal_{\parallel}^{CD} \Pcal_{\parallel}^{EF} \Pcal_{\parallel}^{GH} \Rcal_{ACEG} \Rcal_{BDFH}\nn
&\quad + \frac{1}{2} \Pcal_{\perp}^{AB} \Pcal_{\perp}^{CD} \Pcal_{\parallel}^{EF} \Pcal_{\parallel}^{GH} \Rcal_{AECF} \Rcal_{BGDH} + \frac{1}{180} \left(3 -  D\right) \Pcal_{\parallel}^{AB} \Pcal_{\parallel}^{CD} \Pcal_{\parallel}^{EF} \Pcal_{\parallel}^{GH} \Rcal_{ACBE} \Rcal_{DGFH}\nn
&\quad + \frac{1}{6} \Pcal_{\perp}^{AB} \Pcal_{\parallel}^{CD} \Pcal_{\parallel}^{EF} \Pcal_{\parallel}^{GH} \Rcal_{ACBD} \Rcal_{EGFH} + \frac{1}{72} \left(-3 + D\right) \Pcal_{\parallel}^{AB} \Pcal_{\parallel}^{CD} \Pcal_{\parallel}^{EF} \Pcal_{\parallel}^{GH} \Rcal_{ACBD} \Rcal_{EGFH}\nn
&\quad+\Dcal_{\perp}^{2}\left (\frac{1}{30} (8 -  D) \langle \Kcal^{A}{}_{A}\rangle + \frac{1}{6} \Pcal_{\perp}^{AB} \Pcal_{\parallel}^{CD} \Rcal_{ACBD} + \frac{1}{30} (-3 + D) \Pcal_{\parallel}^{AB} \Pcal_{\parallel}^{CD} \Rcal_{ACBD}\right )\ ,\label{TrLogdims3}
\end{align}
where we used the condensed notation $\langle \Kcal^{ABC}\rangle\equiv K_{\mu }{}^{\nu A}K_{\nu}{}^{\rho B}K_{\rho}{}^{\mu C}$ and similar for traces over spacetime indices of extrinsic curvatures where possible.

\subsubsection{Flat Bulk Formula $d+1=6$}

When $d+1=6$, an expression similar to \eqref{TrLogdims3} may be derived for the general case, but due to its length we will not reproduce it here.  In the simplified case where the bulk is flat, $\Rcal_{ABCD}=0$, relevant to the DBI and multi-field DBI \cite{Hinterbichler:2010xn} scenarios, the expressions are more manageable and the result can be written in terms of the following basis:
\begin{align}
\tr\left [\mathsf{a}_{3}\right ]&=a_{1}\langle\Kcal^{A}{}_{A}{}^{B}{}_{B}{}^{C}{}_{C}\rangle+a_{2}\langle\Kcal^{[A}{}_{A}{}^{B}{}_{B}{}^{C]}{}_{C}\rangle+a_{3}\langle \Kcal^{[ABC]}{}_{ABC}\rangle+a_{4}\langle \Kcal^{[AB]C}{}_{ABC}\rangle+a_{5}\langle \Kcal^{[A}{}_{B}{}^{C]}{}_{A}{}^{B}{}_{C}\rangle\nn
&\quad+b_{1}\langle \Kcal^{ABC}\rangle\langle \Kcal_{ABC}\rangle+b_{2}\langle \Kcal^{[A}{}_{B}{}_{C}\rangle \langle \Kcal_{A}{}^{B]C}\rangle+b_{3}\langle \Kcal^{[A}{}_{A}\rangle\langle \Kcal^{B]}{}_{B}{}^{C}{}_{C}\rangle\nn
&\quad+b_{4}\langle \Kcal^{[A}{}_{A}\rangle\langle \Kcal^{B]C}{}_{BC}\rangle+b_{5}\langle \Kcal^{A}{}_{A}\rangle\langle \Kcal^{[BC]}{}_{BC}\rangle\nn
&\quad+c_{1}\langle \Kcal^{A}{}_{A}\rangle^{3}+c_{2}\langle \Kcal^{[A}{}_{A}\rangle\langle \Kcal^{B}{}_{B}\rangle\langle \Kcal^{C]}{}_{C}\rangle+c_{3}\langle \Kcal^{[A}{}_{A}\rangle\langle \Kcal^{B]}{}_{B}\rangle\langle \Kcal^{C}{}_{C}\rangle\nn
&\quad+d_{1}\langle \Kcal^{A}{}_{A}\rangle \nabla_{\alpha}K_{\mu\nu}{}^{ B}\nabla^{\alpha}K^{\mu\nu}{}_{B}+d_{2}K^{\alpha\gamma [A}K_{\gamma \beta A}\nabla_{\alpha}K_{\mu\nu}{}^{B]}\nabla^{\beta}K^{\mu\nu}{}_{B}\nn
&\quad+d_{3}K^{\beta\gamma [A}K_{\gamma \alpha A}\nabla_{\alpha}K_{\mu\nu}{}^{B]}\nabla^{\beta}K^{\mu\nu}{}_{B}+\left ({\rm total \ derivatives}\right )\label{TrLogdims6Flat}
\end{align}
where the set of total derivatives includes the dimension $d+1=6$ topological term \eqref{app:eq:GaussBonnet6D} and we employed the same condensed notation representing traces as was used in \eqref{TrLogdims3}.  The explicit results of the computation give:
\begin{align}
\begin{pmatrix}[1.5]
a_{1} \\
a_{2} \\
a_{3} \\
a_{4} \\
b_{1} \\
b_{2} \\
b_{3} \\
b_{4} \\
b_{5} \\
c_{1}\\
c_{2}\\
c_{3}\\
c_{4}\\
d_{1}\\
d_{2}\\
d_{3}
\end{pmatrix}
&=\begin{pmatrix}[1.5]
\frac{23
   D-325}{9450} \\ \frac{105725-523
   D}{94500} \\ \frac{89
   D-57775}{31500} \\ -\frac{2 (24
   D-8975)}{7875} \\ \frac{-2
   D-2125}{7875} \\ \frac{8
   D+65}{14175} \\ \frac{31
   D+30925}{47250} \\ \frac{41
   D-5275}{6750} \\ \frac{34675-89
   D}{47250} \\ \frac{113
   D-2077}{3780} \\ \frac{377-67
   D}{7560} \\ \frac{47275-89
   D}{94500} \\ \frac{261
   D-3650}{15750} \\ \frac{13
   D+145}{2100} \\ -\frac{26}{15} \\ \frac{D+205}{525}
\end{pmatrix}\ .
\end{align}
A perturbative check of this result is discussed in Sec.~\ref{sec:DBIExplicitResults}.

\section{Non-Covariant Calculations: DBI Example\label{sec:NaiveDBI}}

In this section we perform a naive, one-loop computation of \eqref{1PIActionOneLoop} using heat-kernel methods for the concrete case of co-dimension-1 Dirac-Born-Infeld (DBI) in $d+1=4$ in order to demonstrate the disadvantages of non-covariant approaches to the problem in comparison to the covariant analysis of Sec.~\ref{sec:GenericBraneComputations}. The analogous $d+1=6$ computation is discussed in App.~\ref{app:Naive6DDBICalculation}.

\subsection{DBI Review\label{sec:DBIReview}}

DBI describes\footnote{The usage of ``DBI" varies throughout the literature and elsewhere it can refer to generalized versions of \eqref{UniversalBraneAction} where additional fields appear in the determinant, for instance. Our usage of DBI strictly refers to the model \eqref{DBIActionNoPrefactor}.} the co-dimension-1 limit of \eqref{UniversalBraneAction} in which the bulk spacetime is flat: $D=d+1$ and $\Gcal_{AB}=\eta_{AB}$. 
The DBI effective field theory (EFT) possesses a host of remarkable properties.   Its amplitudes have exceptional soft-limit behavior \cite{Cheung:2014dqa,Cachazo:2015ksa,Cheung:2016drk,Padilla:2016mno,Guerrieri:2017ujb,Roest:2019oiw} and are one of the distinguished theories which arise in double-copy constructions (see \cite{Bern:2019prr} for a review).  Interesting perspectives on their symmetry properties can be found in \cite{Pajer:2018egx,Grall:2019qof,Cheung:2020qxc} and some phenomenological features are discussed in \cite{Goon:2010xh,Bonifacio:2019rpv}. The DBI literature is vast and the preceding works represent only a select fraction of the whole.

After fixing unitary gauge, $X^{A}(x)=(x^{\mu},\phi)$, the various geometric ingredients discussed in Sec.~\ref{sec:Branes} become
\begin{gather}
g_{\mu\nu}=\eta_{\mu\nu}+\partial_{\mu}\phi\partial_{\nu}\phi\ , \quad
g^{\mu\nu}=\eta^{\mu\nu}-\gamma^{2}\partial^{\mu}\phi\partial^{\nu}\phi\ , \quad
\sqrt{-g}=\gamma^{-1}\nn
 K_{\mu\nu}=-\gamma\partial_{\mu}\partial_{\nu}\phi\  ,\quad
\gamma\equiv\left (1+(\partial\phi)^{2}\right )^{-1/2}\ , \label{DBIQuantities}
\end{gather}
where here and below $(\partial\phi)^{2}\equiv \eta^{\mu\nu}\partial_{\mu }\phi\partial_{\nu}\phi$.  The Gauss-Codazzi relations \eqref{app:eq:GaussCodazzi} further determine that
\begin{align}
  R[g]_{\mu\nu\rho\sigma}=K_{\mu\rho}K_{\nu\sigma}-K_{\mu\sigma}K_{\nu\rho} \ , \quad \nabla_{\mu}K_{\nu\rho}=\nabla_{\nu}K_{\mu\rho}\ ,\label{DBIGaussCodazzi}
  \end{align}
  where $\nabla_{\mu}$ is the covariant derivative with respect to the DBI metric in \eqref{DBIQuantities}. See \cite{deRham:2010eu,Goon:2010xh,Goon:2011uw}, for instance, for expanded discussions of the geometry of DBI.

The universal part of the action \eqref{UniversalBraneAction} is
\begin{align}
S_{\rm DBI}&\equiv-\int\rd^{d+1}x\,  \sqrt{1+(\partial\phi)^{2}}\nn
&\approx -\int\rd^{d+1}x\, \left (1+\frac{1}{2}(\partial\phi)^{2}-\frac{(\partial\phi)^{4}}{ 8 }+\frac{(\partial\phi)^{6}}{ 16 }+\ldots\right )\ .\label{DBIActionNoPrefactor}
\end{align}
The above action inherits the $ISO(d+1,1)$ symmetries of the bulk $\eta_{AB}$ metric which act on the $X^{A}$'s in the usual, linear manner: $\delta X^{A}\longrightarrow \omega^{A}{}_{B}X^{B}+\epsilon^{A}$.  After fixing unitary gauge, however, the realization becomes non-linear on $\phi$ as the symmetries of the gauge-fixed action arise as those combinations of $X^{A}$ transformations and brane diffeomorphism which preserve the gauge condition $X^{\mu}=x^{\mu}$; see, e.g., \cite{Goon:2010xh} for a longer discussion.  While a $ISO(d,1)$ subgroup acts on the field as $\phi(x^{\mu})\longrightarrow \phi(x^{\mu}+\omega^{\mu}{}_{\nu}x^{\nu}+\epsilon^{\mu})$, the remaining symmetries of \eqref{DBIActionNoPrefactor} are
  \begin{align}
\delta_{\rm DBI}\phi(x)&= c+b^{\mu}\left (x_{\mu}+\phi\partial_{\mu}\phi\right )\ ,\label{DeltaDBI}
\end{align}
where $c$ and $b^{\mu}$ are constants. Under \eqref{DeltaDBI}, $g_{\mu\nu}$ transforms under a diffeomorphism:
\begin{align}
\delta g_{\mu\nu}=\pounds_{\xi}g_{\mu\nu}=2\nabla_{(\mu}\xi_{\nu)} \ ,\quad \xi_{\mu}=g_{\mu\nu}b^{\nu}\phi\ ,\label{DBIDiff}
\end{align}
and $K_{\mu\nu}$ transforms similarly.  The non-linear symmetries \eqref{DeltaDBI} fix the entire structure of \eqref{DBIActionNoPrefactor}, determining all relative coefficients in the expansion.  The equation of motion can be written as
\begin{align}
\frac{\delta S_{\rm DBI}}{\delta\phi(x)}&=-K=\frac{1}{\gamma}\square\phi\ ,\label{DBIEOM}
\end{align}
where $K=K^{\mu}{}_{\mu}$ and $\square$ the Laplacian for the DBI metric \eqref{DBIEOM}. The vacuum solution to \eqref{DBIActionNoPrefactor} is $\phi=$ constant, corresponding to a flat $g_{\mu\nu}=\eta_{\mu\nu}$ brane, and this configuration can be viewed as an instance of spontaneous symmetry breaking with pattern $ISO(d+1,1)\longrightarrow ISO(d,1)$.  Analyses of this symmetry breaking pattern in which \eqref{DBIActionNoPrefactor} arises from a coset construction can be found in \cite{Goon:2012dy,Creminelli:2014zxa}, for instance.

\subsection{One-Loop Corrections to DBI (Naive)\label{sec:4DNaiveDBICalculationExplicit}}

Now consider the one-loop corrections to the DBI action as computed via \eqref{1PIActionOneLoop}.  If the divergences respect the DBI symmetries, then $\Gamma_{1}$ can be written as a function of the extrinsic curvature of $K_{\mu\nu}$ and covariant derivatives thereof alone.  We will see that when $\Gamma_{1}$ is computed in the present field variables, this expectation is not manifestly realized.  As discussed previously, this stems directly from our choice of field variables for which the DBI symmetries act non-linearly \eqref{DeltaDBI}. However, we will also show that the divergence contains the same physical content as the manifestly invariant expressions in \eqref{TrLogdims3}, as anticipated by the discussion in Sec.~\ref{sec:QuantumActionGeneral}.  The following computation was considered in \cite{deRham:2014wfa} and below we present additional details of the calculation.  We first discuss the computation of the $\Ocal(\phi^{4})$ terms in $\Gamma_{1}[\phi]$ via traditional Feynman diagram methods and then move on to an all-orders-in-$\phi$ computation via a naive heat kernel application.

\noindent \textbf{Feynman diagrams:}  At low orders in $\phi$, it is feasible to compute $\Gamma_{1}[\phi]$ through standard Feynman methods and the result at $\Ocal(\phi^{4})$ is
\begin{align}
 \Gamma_1 &\supset  \tfrac{1}{30 (4 \pi)^2 \varepsilon} \,\int \mathrm{d}^4x \, \Big[
 \phi^{\mu}{}_{\mu} \phi_{\nu\beta} \phi_{\alpha}{}^{\beta} \phi^{\alpha\nu}- \tfrac{37}{8} \phi_{\mu\alpha} \phi^{\mu\alpha} \phi_{\nu\beta} \phi^{\nu\beta} -  \phi_{\mu}{}^{\nu} \phi^{\mu\alpha} \phi_{\nu\beta} \phi_{\alpha}{}^{\beta} -  \tfrac{11}{4} \phi^{\mu}{}_{\mu} \phi_{\nu\beta} \phi^{\nu\beta} \phi^{\alpha}{}_{\alpha} \nn
 &-  \tfrac{1}{8} \phi^{\mu}{}_{\mu} \phi^{\nu}{}_{\nu} \phi^{\alpha}{}_{\alpha} \phi^{\beta}{}_{\beta} -  \phi^{\mu} \phi_{\alpha}{}^{\beta} \phi^{\alpha\nu} \phi_{\mu\nu\beta} + \tfrac{1}{2} \phi^{\mu} \phi^{\nu\beta} \phi^{\alpha}{}_{\alpha} \phi_{\mu\nu\beta} -  \tfrac{47}{4} \phi^{\mu} \phi_{\alpha\nu} \phi^{\alpha\nu} \phi_{\mu}{}^{\beta}{}_{\beta}\nn 
 & -  \tfrac{11}{4} \phi^{\mu} \phi^{\nu}{}_{\nu} \phi^{\alpha}{}_{\alpha} \phi_{\mu}{}^{\beta}{}_{\beta} -  \tfrac{1}{4} \phi^{\mu} \phi^{\alpha} \phi_{\mu}{}^{\nu\beta} \phi_{\alpha\nu\beta} -  \tfrac{29}{4} \phi^{\mu} \phi^{\alpha} \phi_{\mu}{}^{\nu}{}_{\nu} \phi_{\alpha}{}^{\beta}{}_{\beta}
 \Big]\ .
 \label{eqn:DBI_G4}
\end{align} 
where $\phi_{\mu\ldots\nu}\equiv \partial_{\mu}\ldots\partial_{\nu}\phi$ as before and $\eta_{\mu\nu}$ was used in all contractions. It is straightforward to verify that the above divergence does not correspond to any DBI-invariant counterterm.    While one can in principle also compute at higher orders in $\phi$ with Feynman diagrams, such calculations quickly become burdensome, so we next turn to the heat kernel.

\noindent \textbf{Heat kernel:}
    In order to compute $\Gamma[\phi]$ via \eqref{1PIActionOneLoop}, we compute the $\mathcal{O}(\varphi^{2})$ terms in $S_{\rm DBI}[\phi+\varphi]$:
 \begin{align}
S_{\rm DBI}[\phi+\varphi]=-\frac{1}{2}\int\rd^{d+1}x\, \left (\sqrt{-\tilde{g}}\, \tilde{g}^{\mu\nu}\partial_{\mu}\varphi\partial_{\nu}\varphi\right )+\ldots\ ,
\end{align}
where the effective metric $\tilde{g}_{\mu\nu}$ is conformally related to the induced DBI metric \eqref{DBIQuantities}:
\begin{align}
\tilde{g}_{\mu\nu}&=\Omega^{2}g_{\mu\nu}\ , \quad \Omega\equiv\gamma^{\frac{2}{d-1}}\ .\label{DBIEffectiveMetricForFluctuations}
\end{align}
The one-loop correction to the effective action is then given by 
\begin{align}
 \Gamma_{1}[\phi]&= \frac{i}{2}\Tr\ln \tilde{\square}\ , 
\end{align}
with $\tilde{\square}$ the Laplacian associated to $\tilde{g}_{\mu\nu}(\phi)$ and the above can be computed through standard heat kernel methods.

Given two metrics $\tilde{g}_{\mu\nu}$ and $g_{\mu\nu}$ related as in \eqref{DBIEffectiveMetricForFluctuations}, the action of their respective Laplacians on a scalar quantity $S(x)$ are related through
\begin{align}
\tilde{\square}S=\Omega^{-2}\square S+(d-1)\Omega^{-3}\nabla_{\mu}\Omega\nabla^{\mu}S\ , \label{LaplacianConformalTransformation}
\end{align}
and hence
\begin{align}
\Tr\ln \tilde{\square}=\Tr\ln \Omega^{-2}+\Tr\ln \left (\square+(d-1)\Omega^{-1}\nabla^{\mu}\Omega\nabla_{\mu}\right )\ .\label{DBIFunctionalDet}
\end{align}
The $\Tr\ln \Omega^{-2}$ term is proportional to $\delta^{d+1}(0)$ which is vanishing in our scale-free regularization scheme. The final term in the remaining trace is not a DBI-covariant operator, as can by checked by computing
 \begin{align}
  \delta_{\rm DBI}  \Omega^{-1}\nabla_{\mu}\Omega-\pounds_{\xi}  \Omega^{-1}\nabla_{\mu}\Omega=-\frac{2}{(d-1)}b^{\alpha}\partial_{\alpha}\partial_{\mu}\phi\neq 0.
  \end{align}
  with $\delta_{\rm DBI}$ and $\xi^{\mu}$ as in \eqref{DeltaDBI} and \eqref{DBIDiff}, respectively. Therefore, \eqref{DBIFunctionalDet} will not generate DBI-invariant operators as it corresponds to the functional determinant of a non-DBI-covariant operator.

  Below, we verify this claim explicitly by computing the logarithmic divergences in \eqref{DBIFunctionalDet} in $d+1=4$ dimensions and confirm that the result is not DBI-invariant.  In App.~\ref{app:Naive6DDBICalculation}, we perform the analogous computation in $d+1=6$.  The logarithmically divergent terms arising from \eqref{DBIFunctionalDet}  in $d+1=4$ are given by\footnote{The Fadeev-Popov contributions which come from fixing unitary gauge can easily shown to be $\propto \delta^{d+1}(0)$ and hence trivial.} \eqref{app:eq:LogDivergencesFromHeatKernel}:
\begin{align}
\left (\Tr\ln \tilde{\square}\right )_{\rm log-div}&=-\frac{1}{\varepsilon}\frac{i}{(4\pi)^{2}}\int\rd^{4}x\,\sqrt{-\tilde{g}}\,[\mathsf{a}_{2}(x)]\nn
&=-\frac{1}{\varepsilon}\frac{i}{(4\pi)^{2}}\int\rd^{4}x\,\sqrt{-\tilde{g}}\, \left (\frac{1}{30}\tilde{\square} \tilde{R}+\frac{1}{180}\tilde{R}_{\mu\nu\rho\sigma}\tilde{R}^{\mu\nu\rho\sigma}-\frac{1}{180}\tilde{R}_{\mu\nu}\tilde{R}^{\mu\nu}+\frac{1}{72}\tilde{R}^{2}\right )\ ,\label{a2MetricOnly}
\end{align}
where $[\mathsf{a}_{2}(x)]$ is the second Seeley-DeWitt coefficient \eqref{app:eq:anHeatKernelResults}. Evaluating \eqref{a2MetricOnly} for the metric \eqref{DBIEffectiveMetricForFluctuations} and rephrasing the result in terms $g_{\mu\nu}$, its associated curvature, and covariant derivative, we ultimately find
\begin{align}
\sqrt{-\tilde{g}}\, [\mathsf{a}_{2}(x)]&= \sqrt{-g}\,\Big(- \frac{1}{180} R_{\alpha \beta} R^{\alpha \beta} + \frac{1}{72} R^2 + \frac{1}{180} R_{\alpha \beta \gamma \delta} R^{\alpha \beta \gamma \delta} + \frac{1}{30} \nabla_{\alpha}\nabla^{\alpha}R\nn
&\quad -  \frac{2 R \nabla_{\alpha}\nabla^{\alpha}\gamma}{9 \gamma} -  \frac{\nabla_{\alpha}\gamma \nabla^{\alpha}R}{15 \gamma} + \frac{R \nabla_{\alpha}\gamma \nabla^{\alpha}\gamma}{18 \gamma^2} + \frac{97 \nabla_{\alpha}\nabla^{\alpha}\gamma \nabla_{\beta}\nabla^{\beta}\gamma}{90 \gamma^2}\nn
&\quad -  \frac{53 \nabla_{\alpha}\gamma \nabla^{\alpha}\gamma \nabla_{\beta}\nabla^{\beta}\gamma}{45 \gamma^3} + \frac{4 \nabla^{\alpha}\gamma \nabla_{\beta}\nabla^{\beta}\nabla_{\alpha}\gamma}{5 \gamma^2} -  \frac{\nabla_{\beta}\nabla^{\beta}\nabla_{\alpha}\nabla^{\alpha}\gamma}{5 \gamma}\nn
&\quad -  \frac{34 R_{\alpha \beta} \nabla^{\alpha}\gamma \nabla^{\beta}\gamma}{45 \gamma^2} + \frac{\nabla_{\alpha}\gamma \nabla^{\alpha}\gamma \nabla_{\beta}\gamma \nabla^{\beta}\gamma}{15 \gamma^4}-  \frac{4 \nabla^{\alpha}\gamma \nabla_{\beta}\nabla_{\alpha}\gamma \nabla^{\beta}\gamma}{45 \gamma^3} \nn
&\quad -  \frac{R_{\alpha \beta} \nabla^{\beta}\nabla^{\alpha}\gamma}{45 \gamma} + \frac{\nabla_{\beta}\nabla_{\alpha}\gamma \nabla^{\beta}\nabla^{\alpha}\gamma}{45 \gamma^2}\Big) \label{DBINaiveLogDivergences4D}\ ,
\end{align}
where the $\nabla_{\mu}$'s are again the covariant derivatives with respect to the metric in \eqref{DBIQuantities} and above they act on $\gamma$ as though it were a scalar. No on-shell conditions or integrations by parts were used in evaluating \eqref{DBINaiveLogDivergences4D}; it is a fully off-shell expression.  The result \eqref{DBINaiveLogDivergences4D} is not DBI-invariant since $\gamma=1/\sqrt{-g}$, in fact, does not transform as a scalar under \eqref{DeltaDBI}.

However, while the unadulterated form of \eqref{DBINaiveLogDivergences4D} does not obey the symmetries \eqref{DeltaDBI}, it is possible to massage the result into a DBI-symmetric form through the addition of non-DBI-symmetric total derivatives and the use of the tree-level equations of motion.   Specifically, after exhaustively adding all possible total-derivatives with the correct dimensions to the action with arbitrary coefficients, using the Gauss-Codazzi relations \eqref{app:eq:GaussCodazzi} to trade Riemann curvatures for extrinsic curvatures, the on-shell condition $ K=0$, and the identities
\begin{align}
 \nabla_{\mu}\nabla_{\nu}\phi=-\frac{1}{\gamma}K_{\mu\nu}\ , \quad\nabla_{\mu}\gamma=K^{\nu}{}_{\mu}\nabla_{\nu}\phi \ , \quad \nabla_{\mu}K^{\mu}{}_{\nu}=\nabla^{\nu}K\ ,
 \end{align}
 one finds that it is possible to dramatically simplify \eqref{DBINaiveLogDivergences4D} to the form
\begin{align}
\,\sqrt{-\tilde{g}}\,[\mathsf{a}_{2}(x)]\Big|_{\rm on-shell}&= \sqrt{-g}\,\frac{7}{10}\langle K^{4}\rangle +\left ({\rm total\ derivatives}\right ) \ ,\label{DBICovariantLogDivergences4D}
\end{align} 
in condensed trace notation: $\langle K^{3}\rangle \equiv K_{\mu }{}^{\nu}K_{\nu}{}^{\rho}K_{\rho}{}^{\mu}$ and similar.  The result \eqref{DBICovariantLogDivergences4D} precisely agrees with the DBI limit of the general formula \eqref{TrLogdims3}; see Sec.~\ref{sec:DBIExplicitResults}.  
The total derivatives added to the action in order to simplify were $\mathcal{L}_{\rm TD}=\nabla_{\alpha}J^{\alpha}-\frac{43}{360}\mathcal{L}^{(4)}_{\rm GB}[g]$ where
\begin{align}
J^{\alpha}&=- \frac{1}{3} K^{\beta \lambda} \nabla^{\alpha}K_{\beta \lambda} + \frac{29 K^{\alpha}{}_{\beta} K_{\lambda \delta} K^{\lambda \delta} \nabla^{\beta}\phi}{90 \gamma} -  \frac{K_{\beta}{}^{\delta} \nabla^{\alpha}K_{\lambda \delta} \nabla^{\beta}\phi \nabla^{\lambda}\phi}{45 \gamma^2} \nn
&\quad+ \frac{K^{\alpha}{}_{\beta} K_{\lambda}{}^{\varepsilon} K_{\delta \varepsilon} \nabla^{\beta}\phi \nabla^{\lambda}\phi \nabla^{\delta}\phi}{45 \gamma^3}\nn
&=\frac{13}{90} \nabla^{\alpha}\square\ln\gamma -  \frac{29}{90} \nabla^{\alpha}\ln\gamma\square\ln\gamma + \frac{1}{45} \square\nabla^{\alpha}\ln\gamma-  \frac{29}{90} \nabla^{\alpha}\ln\gamma\left (\nabla\ln\gamma\right )^{2} \nn
&\quad  + \frac{14}{45} \nabla_{\beta}\nabla^{\alpha}\ln\gamma \nabla^{\beta}\ln\gamma \ ,\label{DBI4DNaiveTotalDerivativeTerm}
\end{align}
with equality holding on-shell, and where $\mathcal{L}^{(4)}_{\rm GB}$ is the dimension $d+1=4$ topological Gauss-Bonnet term, explicitly given by
\begin{align}
\mathcal{L}_{\rm GB}[g]&\equiv -\frac{1}{4}\epsilon^{\mu\nu\rho\sigma}\epsilon^{\alpha\beta\kappa\lambda}R_{\mu\nu\alpha\beta}R_{\rho\sigma\kappa\lambda}\nn
&= R_{\mu\nu\rho\sigma}^{2}-4R_{\mu\nu}^{2}+R^{2}\label{GaussBonnet4D}\ ,
\end{align}
for a metric $g_{\mu\nu}$.  Equivalently, the contact amplitudes computed from \eqref{DBINaiveLogDivergences4D} and \eqref{DBICovariantLogDivergences4D} agree and it therefore follows that in an $S$-matrix element, the counterterm needed to subtract the divergence expressed in \eqref{DBINaiveLogDivergences4D} is physically equivalent to the much simpler \eqref{DBICovariantLogDivergences4D}.  This can be explicitly seen from the $\Ocal(\phi^{4})$ terms computed with Feynman diagrams in \eqref{eqn:DBI_G4} which contribute to the four-point, on-shell amplitude\footnote{One might raise the point that a traditional, on-shell, one-loop, 4-pt amplitude computed with Feynman diagrams can also be used to determine the logarithmic divergence for the theory \eqref{DBIActionNoPrefactor} by matching to the quartic terms arising from the unique on-shell counterterm $\sim \langle K^{4}\rangle$. In this method, no symmetry-breaking expressions analogous to \eqref{DBINaiveLogDivergences4D} are ever encountered. While this is true, there also exist disadvantages to such an amplitude-based calculation relative to the heat kernel based approach.  In the present example, one might wish to additionally compute the one-loop 6-pt, 8-point, etc.~amplitudes in order to confirm that the divergences in these cases are consistent with the full non-linear structure of the $\sim \langle K^{4}\rangle$ counterterm, for instance, and the complexity of these computations grows with valence; see \cite{Guerrieri:2017ujb} for the 4- and 6-point computations.    More generally, DBI is the simplest model of the many possible brane theories and increasing the brane dimension, co-dimension, and including non-trivial bulk curvature are all features which complicate the calculations and increase the number of amplitudes needed to determine the counterterms.  In contrast, our computations cover all of these extensions simultaneously and automatically ensure the proper non-linear structure.  These are simply the same reasons why analogous methods \cite{Appelquist:1980ae,AlvarezGaume:1981hn,Boulware:1981ns,Akhoury:1982hv,Gasser:1983yg,Gaillard:1985uh,Mukhi:1985vy,Howe:1986vm,Costa:1988ef,Alonso:2016oah} are celebrated in a NLSM context.} as
\begin{align}
\mathcal{A}_{4} \supset \frac{7}{10240 \pi^2 \varepsilon} \;  \left(  s^4 + t^4 + u^4  \right)\ ,
\end{align}
in agreement with \cite{Guerrieri:2017ujb}.  This contribution is also produced by the off-shell-inequivalent term
\begin{align}
 \Gamma^{\rm on-shell}_1 [ \phi ] = \frac{1}{2} \frac{1}{(4\pi)^2 \varepsilon} \int d^4 x \,  \frac{7}{10} \phi^\mu_\nu \phi^\nu_\alpha \phi^\alpha_\beta \phi^\beta_\mu \; ,  \label{eqn:DBI_G4_OnShell}
\end{align}
which are precisely the $\Ocal(\phi^{4})$ divergences corresponding to \eqref{DBICovariantLogDivergences4D}. An analogue of this procedure for the case of a four-dimensional non-linear sigma model can be found in \cite{Appelquist:1980ae}.

The manipulations leading from \eqref{DBINaiveLogDivergences4D} to \eqref{DBICovariantLogDivergences4D} were essentially an extensive exercise in guess-and-check.  While the $d+1=4,6$ DBI computations were manageable with extensive use of \texttt{Mathematica} and \texttt{xAct/xTensor} \cite{xAct,Nutma:2013zea}, we note that DBI is only the simplest of all possible brane models: the hypersurface is co-dimension-1 and the bulk metric is flat.  We expect that extending the preceding method to higher-co-dimension cases with non-trivial bulk metrics would quickly be found to be infeasible and generically inferior to the covariant methods of Sec.~\ref{sec:GenericBraneComputations}, as the $d+1=6$ case in App.~\ref{app:Naive6DDBICalculation} already makes abundantly clear.

We close this section by noting that the second form of the total-derivative current in \eqref{DBI4DNaiveTotalDerivativeTerm} is intriguing.  We are capturing only the logarithmic divergences above via the naive application of the heat kernel formulas of App.~\ref{app:CovariantHeatKernels} and perhaps there exists a more refined method by which the $\sim\nabla^{\#}\ln \gamma$ factors would naturally arise, even when computing using the naive variables chosen above.  Studying power-divergences in the DBI model provides further interesting findings along these lines.  The $d+1=4$ quadratic divergences would necessarily be of the form
\begin{align}
\left (\Tr\ln \tilde{\square}\right )_{\Lambda^{2}-{\rm div}}\sim \tilde{\Lambda}^{2}\int d^{4}x\, \sqrt{-\tilde{g}}\, \tilde{R} \ ,
\end{align}
for some energy scale $\tilde{\Lambda}$ and the above is not DBI-covariant.  However, if it were possible to manipulate the calculation by pulling $\tilde{\Lambda}$ inside the integral and scaling $\tilde{\Lambda}\longrightarrow \Lambda \Omega(x)^{-2}$ with $\Omega(x)$ the conformal factor in \eqref{DBIEffectiveMetricForFluctuations}, then DBI-covariance would be restored, since
\begin{align}
\Omega(x)^{-2}\sqrt{-\tilde{g}}\tilde{R}=5\sqrt{-g} K_{\mu\nu}^{2}\ .
\end{align}
Similar results hold for power-divergences in $d+1=6$.  It seems plausible that similar manipulations would work for logarithmically divergent terms, but we leave further exploration of this question to future work.

\section{Applications\label{sec:Applications}}

In this section, we discuss a selection of models for which the results of Sec.~\ref{sec:GenericBraneComputations} are relevant.

\subsection{DBI\label{sec:DBIExplicitResults}}

The DBI limit \eqref{DBIActionNoPrefactor} is the simplest scenario, as discussed in the preceding section.  This system is co-dimension-1, meaning $D=d+1$, and hence any terms in \eqref{TrLogdims3} or \eqref{TrLogdims6Flat} involving anti-symmetrization over the bulk indices of $K_{\mu\nu}{}^{A}$ factors identically vanish.  Since the bulk curvature is also trivial, $\Rcal_{ABCD}=0$, very few terms remain.  We find
\begin{itemize}
\item When $d+1=4$:
\begin{align}
\tr\left [\mathsf{a}_{2}(x)\right ]&=\frac{7}{10}\langle K^{4}\rangle +\left ({\rm total \ derivatives}\right )\label{TrLogdims3DBILimit}\ ,
\end{align}
where the Gauss-Bonnet term \eqref{GaussBonnet4D} was used to remove the combination $\sim \left (K_{\mu\nu}K^{\mu\nu}\right )^{2}$ and where $\langle \ \cdot \ \rangle$ is the same shorthand for spacetime traces used \eqref{DBICovariantLogDivergences4D}.    When expressed in the typical, unitary gauge form reviewed in Sec.~\ref{sec:DBIReview}, the corresponding divergence in $\Gamma[\phi]$ is:
\begin{align}
\Gamma[\phi]&\supset\frac{7}{320\pi^{2}\varepsilon}\int\rd^{4}x\, \Big(\gamma^3 \langle \Phi^{4}\rangle-4\gamma^{5}\langle \partial\phi\!\cdot\! \Phi^{4}\!\cdot\!\partial\phi\rangle+2\gamma^{7}\langle \partial \phi\!\cdot\! \Phi^{2}\!\cdot\!\partial\phi\rangle ^{2}\nn
&\quad+4\gamma^{7}\langle \partial \phi\!\cdot\! \Phi\!\cdot\!\partial\phi\rangle\langle \partial \phi\!\cdot\! \Phi^{3}\!\cdot\!\partial\phi\rangle-4\gamma^{9}\langle \partial \phi\!\cdot\! \Phi\!\cdot\!\partial\phi\rangle^{2}\langle \partial \phi\!\cdot\! \Phi^{2}\!\cdot\!\partial\phi\rangle+\gamma^{11}\langle \partial \phi\!\cdot\! \Phi\!\cdot\!\partial\phi\rangle^{4}\Big)\ ,
\end{align}
where $\gamma$ is defined in \eqref{DBIQuantities}, $\Phi\equiv \partial_{\mu}\partial_{\nu}\phi$, and $\langle \ \cdot \ \rangle$ is shorthand for spacetime contractions via $\eta^{\mu\nu}$, e.g., $\langle \Phi^{2}\rangle \equiv \eta^{\mu\nu}\eta^{\alpha\beta}\partial_{\nu}\partial_{\alpha}\phi\partial_{\beta}\partial_{\mu}\phi$ and $\langle \partial\phi\!\cdot\!\Phi\!\cdot\!\partial\phi\rangle \equiv \eta^{\mu\nu}\eta^{\alpha\beta}\partial_{\mu}\phi\partial_{\nu}\partial_{\alpha}\phi\partial_{\beta}\phi$.
\item When $d+1=6$:
\begin{align}
\tr[\mathsf{a}_{3}(x)]\Big|_{D=6}&= - \frac{187}{9450} \langle K^{6}\rangle + \frac{113}{14175} \langle K^{3}\rangle^{2} -  \frac{5}{1512}\langle K^{2}\rangle^{3} + \frac{223}{2100} \langle K^{2}\rangle \nabla_{\varepsilon}K_{\gamma \delta} \nabla^{\varepsilon}K^{\gamma \delta}\nn
&\quad +\left ({\rm total \ derivatives}\right )\ ,
\end{align}
and we will not reproduce the unitary gauge form of the above for brevity.
\end{itemize}
As mentioned in Sec.~\ref{sec:NaiveDBI}, the logarithmic divergences which follow from \eqref{app:eq:LogDivergencesFromHeatKernel} are in perfect agreement with the $d+1=4$ result \eqref{DBICovariantLogDivergences4D} and the (extremely cumbersome) $d+1=6$ computation outlined in App.~\ref{app:Naive6DDBICalculation}.   The divergences for the multi-field DBI case studied, for instance, in \cite{Hinterbichler:2010xn}, can also easily be read off from our general formulas.  As a check of the $d+1=6$ formula \eqref{TrLogdims6Flat}, we have verified that the corresponding divergences generated in the four-point amplitude $\Acal_{4}$ precisely agree with those arising from a standard Feynman diagram calculation for a co-dimension-$N$ DBI system for arbitrary $N$.

\subsection{Product Manifolds, Non-Linear Sigma Models, and their Extensions\label{sec:ProductManifolds}}

The universal action \eqref{UniversalBraneAction} for a generic bulk metric $\Gcal_{AB}$ may not permit flat vacua where $g_{\mu\nu}=\eta_{\mu\nu}$.  A class of special bulk manifolds which do permit such solutions are product manifolds of the form $\Mcal_{D+1}=\mathsf{M}_{d+1}\times \Sigma_{D-d}$, where $\mathsf{M}_{d+1}$ is $(d+1)$-dimensional Minkowski space and $\Sigma$ is some $(D-d)$-dimensional manifold. 

 Let us consider a $(d+1)$-dimensional brane embedded in such a bulk and split the bulk $X^{A}$ coordinate as $X^{A}=\left (X^{\mu},\phi^{a}\right )$, $\mu\in\{0,\ldots,d\} $ and $a\in\{d+1,\ldots, D\}$.   By assumption, the bulk line element can be written in the form
\begin{align}
\rd s^{2}_{D+1}&=\eta_{\mu\nu}\rd X^{\mu}\rd X^{\nu}+\g_{ab}\left (\phi\right )\rd \phi^{a}\rd\phi^{b}	\ ,\label{ProductSpaceMetric}
\end{align}
and working in unitary gauge, $X^{\mu}=x^{\mu}$, the induced metric is
\begin{align}
g_{\mu\nu}(x)&=\eta_{\mu\nu}+\g_{ab}(\phi(x))\partial_{\mu }\phi^{a}(x)\partial_{\nu}\phi^{b}(x)\ .
\end{align}
The universal action is then
\begin{align}
S_{\rm universal}&=-\int\rd^{d+1}x\, \sqrt{-\det\left (\eta_{\mu\nu}+\g_{ab}(\phi)\partial_{\mu}\phi^{a}\partial_{\nu}\phi^{b}\right )}\nn
&\approx -\int\rd^{d+1}x\,\left (\frac{1}{2}\partial_{\mu}\phi\cdot\partial^{\mu}\phi-\frac{1}{4}\left (\partial_{\mu}\phi\cdot\partial_{\nu}\phi\right )\left (\partial^{\mu}\phi\cdot\partial^{\nu}\phi\right )+\frac{1}{8}\left (\partial_{\mu}\phi\cdot\partial^{\mu}\phi\right )^{2}+\ldots\right )\ ,\label{UniversalActionProductSpace}
\end{align}
where Greek indices were raised and lowered with $\eta^{\mu\nu}$ and we used the shorthand $\partial_{\mu}\phi\cdot\partial_{\nu}\phi\equiv \g_{ab}\partial_{\mu}\phi^{a}\partial_{\nu}\phi^{b}$ in the second line.  Further comments on the functional form of the action \eqref{UniversalActionProductSpace} can be found in the conclusions, Sec.~\ref{sec:Conclusions}.

The action \eqref{UniversalActionProductSpace} is intimately related to the non-linear sigma model (NLSM) and extensions thereof.  As is well known, the universal term for generic NLSMs which describe the Goldstone fields arising from spontaneous symmetry breaking takes on the form \cite{Coleman:1969sm,Callan:1969sn}
\begin{align}
S_{\rm NLSM}=-\frac{1}{2}\int\rd^{d+1}x\, \g_{ab}(\phi)\partial\phi^{a}\cdot\partial\phi^{b}\label{NLSM}
\end{align}
where the field-space metric $\g_{ab}$ arises from a coset analysis\footnote{Often, this action is expressed in the form $\Lcal_{\rm NLSM}\sim \tr \left [\partial_{\mu}U^{\dagger}\partial^{\mu}U\right ]$ where $U=e^{i\pi(x)\cdot Z}$ are coset elements in $G/H$ for a breaking pattern $G\longrightarrow H$, $\pi(x)$ are the Goldstone modes, and $Z$'s are group generators.}. For instance, for the symmetry breaking pattern $SO(N+1)\longrightarrow SO(N)$ the action takes on the above form with $\g_{ab}$ the metric on the $N$-sphere.  The square-root structure in \eqref{UniversalActionProductSpace} is also closely related\footnote{The action (5.17) in \cite{Cachazo:2014xea} reduces to \eqref{UniversalActionProductSpace} after rescaling their lagrangian by an overall factor of $\lambda^{2}$, sending $\ell\longrightarrow \lambda \ell$ and $\Phi\longrightarrow \Phi/\lambda$, and then taking $\lambda\longrightarrow 0$.} to the scalar sector of the ``extended Dirac-Born-Infeld" theory which was first discussed in \cite{Cachazo:2014xea} and has recently appeared in \cite{Low:2020ubn}.

As an initial check on our $d+1=4$ dimensional result \eqref{TrLogdims3},  we can evaluate the corresponding logarithmic divergence in the limit where \eqref{UniversalActionProductSpace} reduces to the NLSM and verify that it reproduces the well-known divergences of this latter model. The NLSM regime is isolated by introducing a formal counting parameter $\lambda$ and taking the following limit
\begin{align}
\lim_{\lambda\to 0}\frac{-1}{\lambda}\int\rd^{d+1}x\, \sqrt{1+\lambda\g_{ab}(\phi)\partial\phi^{a}\cdot\partial\phi^{b}}=S_{\rm NLSM}\ ,\label{NLSMActionFromLimit}
\end{align}
which holds up to a divergent, but field-independent term.    One-loop divergences to the action \eqref{NLSM} are well-studied \cite{Appelquist:1980ae,Boulware:1981ns,Akhoury:1982hv,Gasser:1983yg,Gaillard:1985uh,Costa:1988ef} and generate the following on-shell divergence \cite{Alonso:2016oah}:
\begin{align}
\Gamma_{1}^{\rm NLSM}&\supset  \frac{1}{\varepsilon}\frac{i}{2(4\pi)^{2}}\int\rd^{4}x\,\left (\frac{1}{2}\rcal_{aebf}\rcal_{c}{}^{e}{}_{d}{}^{f}-\frac{1}{12}\rcal_{acef}\rcal_{bd}{}^{ef}\right )\left (\partial\phi^{a}\cdot\partial\phi^{b}\right )\left (\partial\phi^{c}\cdot\partial\phi^{d}\right )\ ,\label{NLSMDivergencesdims3}
\end{align}
in our conventions.   Inserting the counting parameter $\lambda$ in front of $\g_{ab}$ as in \eqref{NLSMActionFromLimit},  it is straightforward to show that one can effectively replace the various geometric ingredients as
\begin{gather}
g_{\mu\nu}\longrightarrow \eta_{\mu\nu} \ , \quad g^{\mu\nu}\longrightarrow \eta^{\mu\nu}\nn
P_{\parallel}^{AB}\longrightarrow \partial X^{A}\cdot\partial X^{B}\ , \quad P_{\perp}^{AB}\longrightarrow \lambda^{-1}\g^{ab}\ , \quad e^{\alpha A}\longrightarrow \partial^{\alpha}\phi^{a}\nn
 \Rcal_{ABCD} \longrightarrow \lambda \rcal_{abcd}\ , \quad K_{\mu\nu}^{A}\longrightarrow-\Dcal_{\mu}\partial_{\nu}\phi^{a}
\end{gather}
at leading order in the limit. Making the above substitutions in \eqref{TrLogdims3}, the only terms which scale as $\Ocal(\lambda^{0})$ are
\begin{align}
\tr\left [\mathsf{a}_{2}\right ]&\supset-  \frac{1}{12} \Pcal_{\perp}^{AB} \Pcal_{\perp}^{CD} \Pcal_{\parallel}^{EF} \Pcal_{\parallel}^{GH} \Rcal_{ACEG} \Rcal_{BDFH} + \frac{1}{2} \Pcal_{\perp}^{AB} \Pcal_{\perp}^{CD} \Pcal_{\parallel}^{EF} \Pcal_{\parallel}^{GH} \Rcal_{AECF} \Rcal_{BGDH} \nn
&\quad+\Dcal_{\perp}^{2}\left (\frac{1}{6} \Pcal_{\perp}^{AB} \Pcal_{\parallel}^{CD} \Rcal_{ACBD}\right )+\Ocal(\lambda)\nn
&=\left (\frac{1}{2}\rcal_{aebf}\rcal_{c}{}^{e}{}_{d}{}^{f}-\frac{1}{12}\rcal_{acef}\rcal_{bd}{}^{ef}\right )\left (\partial\phi^{a}\cdot\partial\phi^{b}\right )\left (\partial\phi^{c}\cdot\partial\phi^{d}\right )+{\rm total \ derivatives}\ ,
\end{align}
and extracting the corresponding logarithmic divergence using \eqref{app:eq:LogDivergencesFromHeatKernel} and comparing to \eqref{NLSMDivergencesdims3}, we find perfect agreement.

Our full result \eqref{TrLogdims3} contains the generalization of the NLSM result \eqref{NLSMDivergencesdims3} to its brane-world extension \eqref{UniversalActionProductSpace}, when \eqref{TrLogdims3} is evaluated for the system \eqref{ProductSpaceMetric}.  The following results are useful for expressing \eqref{TrLogdims3} in terms of the natural geometry of \eqref{ProductSpaceMetric}:
\begin{itemize}
\item Fixing unitary gauge uses all of the brane diffeomorphism freedom, but the gauge-fixed action is still covariant under field-redefinitions of the $\phi^{a}$ amongst themselves.  For this reason, it is useful to define covariant derivatives of tensors $T^{\alpha a\ldots}$ whose $a$-type indices transform covariantly under such redefinitions (an example of which is $\partial_{\mu}\phi^{a}$) via
\begin{align}
 \dcal_{\mu}T^{a\alpha \ldots}=\partial_{\mu}T^{a\alpha\ldots}+ \Gamma^{a}_{bc}T^{c\alpha\ldots}\partial_{\mu}\phi^{b}+\ldots\ ,\label{dcalDef}
\end{align}
where $\Gamma^{a}_{bc}$ is the Christoffel symbol associated to $\g_{ab}$, in analogy to \eqref{app:eq:BraneCovariantDerivative}.
\item The Christoffel symbol $\Gamma^{\alpha}_{\mu\nu}$ associated to $g_{\mu\nu}$ is
\begin{align}
\Gamma^{\alpha}_{\mu\nu}&=g^{\alpha\rho}\g_{ab}\dcal_{\mu}\partial_{\nu}\phi^{b}\partial_{\rho}\phi^{a}\ ,
\end{align}
\item The extrinsic curvature is
\begin{align}
K_{\mu\nu}{}^{A}&=-\left (\dcal_{\mu}\partial_{\nu}\phi^{a}-g^{\alpha\rho}\g_{bc}\dcal_{\mu}\partial_{\nu}\phi^{b}\partial_{\rho}\phi^{c}\partial_{\alpha}\phi^{a}\right )\delta_{a}^{A}+g^{\alpha\rho}\g_{ab}\dcal_{\mu}\partial_{\nu}\phi^{b}\partial_{\rho}\phi^{a}\delta_{\alpha}^{A}
\end{align}
\item The inverse induced metric $g^{\mu\nu}$ is
\begin{align}
g^{\mu\nu}&=\eta^{\mu\nu}-\h_{ab}\partial^{\mu}\phi^{a}\partial^{\nu}\phi^{b}\ , \quad
\h_{ab}=\left (\g^{ab}+\partial^{\mu}\phi^{a}\partial_{\mu}\phi^{b}\right )^{-1}
\end{align}
\item Both $\Gamma^{A}_{BC}$ and $\Rcal_{ABCD}$ computed from $\Gcal_{AB}$ are vanishing unless all indices take on values corresponding to directions along the $\phi^{a}$'s.  That is, only components of the form $\Gamma^{a}_{bc}$ and $\Rcal_{abcd}$ are non-vanishing and are simply those calculated from $\g_{ab}$.  The projectors onto the space parallel to the brane takes on the form
\begin{align}
P_{\parallel}^{AB}=g^{\mu\nu}e_{\mu}{}^{A}e_{\nu}{}^{B}=\begin{pmatrix}
g^{\mu\nu} & g^{\mu\alpha}\partial_{\alpha}\phi^{b}\\
\partial_{\alpha}\phi^{a}g^{\alpha\nu} & g^{\alpha\beta}\partial_{\mu}\phi^{a}\partial_{\nu}\phi^{b}
\end{pmatrix}\ ,
\end{align}
meaning that contractions between $P_{\parallel}^{AB}$ and $\Rcal_{ABCD}$ factors can be expressed as contractions of $\partial_{\mu}\phi^{a}$ factors with $\rcal_{abcd}$ and $g_{\mu\nu}$.  Similar remarks hold for $P_{\perp}^{AB}=\Gcal^{AB}-P_{\parallel}^{AB}$ and $\Rcal_{ABCD}$. 
\end{itemize}

Explicit formulas for the scenario in which $\Sigma_{D-d}$ is an $N$-sphere, $S_{N}$ and $d+1=4$, are provided in App.~\ref{app:MTimesSNDivergences} and we close this section by comparing the logarithmic divergences arising from our general result \eqref{Polyakov1PIFunctionalDeterminant} to those from the corresponding lowest-order Feynman diagram calculation.  When $\Sigma_{D-4}=S_{N}$, the relevant geometric quantities are 
\begin{gather}
\g_{ab}(\phi)=\delta_{ab}+\frac{\phi_{a}\phi_{b}}{L^{2}-\phi^{c}\phi^{d}\delta_{cd}}\ ,\quad  \g^{ab}(\phi)=\delta^{ab}-\frac{\phi^{a}\phi^{b}}{L^{2}}\ , \quad  \rcal_{abcd}=\frac{1}{L^{2}}\left (\g_{ac}\g_{bd}-\g_{ad}\g_{bc}\right )\ ,\label{NSphereGeometry}
\end{gather}
where $L$ is the radius of $S_{N}$.  Up to $\Ocal(\phi^{4})$, the universal lagrangian is 
\begin{align}
\Lcal_{\rm universal}&\approx-  \tfrac{1}{2} \partial_{\mu}\phi_{a} \partial^{\mu}\phi^{a}-  \frac{\phi^{a} \phi^{b} \partial_{\mu}\phi_{b} \partial^{\mu}\phi_{a}}{2 L^2}  + \tfrac{1}{4} \partial_{\mu}\phi_{b} \partial^{\mu}\phi^{a} \partial_{\nu}\phi^{b} \partial^{\nu}\phi_{a} -  \tfrac{1}{8} \partial_{\mu}\phi_{a} \partial^{\mu}\phi^{a} \partial_{\nu}\phi^{b} \partial^{\nu}\phi_{b}\ ,
\end{align}
where Greek and Latin indices were raised and lowered with $\eta_{\mu\nu}$ and $\delta_{ab}$.  Computing the corresponding four-point amplitude is straightforward.  The tree-level result is
\begin{align}
\Acal_{4}^{\rm tree}&=  \left(\frac{t^2 + u^2-s^{2}}{4} -  \frac{ t + u}{L^2}\right)  \delta^{ab} \delta^{cd}+\left(\frac{u^2 + s^2-t^{2}}{4} -  \frac{ u + s}{L^2}\right)  \delta^{ac} \delta^{bd}\nn
&\quad +\left(\frac{s^2 + t^2-u^{2}}{4} -  \frac{ s + t}{L^2}\right) \delta^{ad} \delta^{bc} 
\end{align}
where particles $a,b,c,d$ were assigned momenta $p_{1},p_{2},p_{3},p_{4}$, respectively, and the $\Ocal(L^{-2})$ terms are the ordinary NLSM result.  A standard Feynman computation gives the following one-loop divergences
\begin{align}
\Acal_{4}^{{\rm 1-loop}}&\supset\frac{1}{96(4\pi)^{2}\varepsilon}\delta^{ab}\delta^{cd}\Big( \tfrac{2}{5} s^4 N + s^3 \left( \tfrac{2}{5} t (50 + N)+\frac{88 - 8 N}{L^2} \right) \nn
&\quad  + s^2 \left( \tfrac{2}{5} t^2 (90 + N)+\frac{80 t}{L^2}  + \frac{16 (-5 + 3 N)}{L^4}\right)+ \frac{16 s t \left(L^2 t ( 2 L^2 t+5)+4 \right)}{L^4}\nn
&\quad+16 t^2 \left (t^2+\frac{4}{L^4} \right ) \Big)   +{\rm permutations}\ ,\label{OneLoopMTimesSN4pt}
\end{align}
where only the pole terms were displayed.
From the results of App.~\ref{app:MTimesSNDivergences}, the $\Ocal(\phi^{4})$ parts of the predicted counterterm are 
\begin{align}
\Gamma_{1}&\supset\frac{1}{720(4\pi)^{2}\varepsilon}\int\rd^{4}x\,\Big( \frac{240 \partial_{\mu}\phi_{b} \partial^{\mu}\phi^{a} \partial_{\nu}\phi^{b} \partial^{\nu}\phi_{a}}{L^4} + \frac{60 (-7 + 3 N) \partial_{\mu}\phi_{a} \partial^{\mu}\phi^{a} \partial_{\nu}\phi^{b} \partial^{\nu}\phi_{b}}{L^4}\nn
&\quad + \frac{120 \partial^{\mu}\phi^{a} \partial_{\nu}\partial^{\rho}\phi_{a} \partial^{\nu}\phi^{b} \partial_{\rho}\partial_{\mu}\phi_{b}}{L^2} -  \frac{120 \partial_{\mu}\partial^{\rho}\phi_{a} \partial^{\mu}\phi^{a} \partial^{\nu}\phi^{b} \partial_{\rho}\partial_{\nu}\phi_{b}}{L^2} \nn
&\quad-  \frac{360 \partial_{\mu}\phi^{b} \partial^{\mu}\phi^{a} \partial_{\rho}\partial_{\nu}\phi_{b} \partial^{\rho}\partial^{\nu}\phi_{a}}{L^2} -  \frac{60 (-7 + N) \partial_{\mu}\phi_{a} \partial^{\mu}\phi^{a} \partial_{\rho}\partial_{\nu}\phi_{b} \partial^{\rho}\partial^{\nu}\phi^{b}}{L^2}\nn
&\quad - 2 (30 + N) \partial_{\mu}\partial^{\sigma}\phi^{b} \partial_{\nu}\partial^{\rho}\phi_{a} \partial^{\nu}\partial^{\mu}\phi^{a} \partial_{\sigma}\partial_{\rho}\phi_{b} + (60 - 4 N) \partial^{\nu}\partial^{\mu}\phi^{a} \partial_{\rho}\partial_{\mu}\phi^{b} \partial_{\sigma}\partial_{\nu}\phi_{b} \partial^{\sigma}\partial^{\rho}\phi_{a}\nn
&\quad + 4 (45 \!+\!N) \partial_{\nu}\partial_{\mu}\phi^{b} \partial^{\nu}\partial^{\mu}\phi^{a} \partial_{\sigma}\partial_{\rho}\phi_{b} \partial^{\sigma}\partial^{\rho}\phi_{a}+ 5 ( N\!-\!12) \partial_{\nu}\partial_{\mu}\phi_{a} \partial^{\nu}\partial^{\mu}\phi^{a} \partial_{\sigma}\partial_{\rho}\phi_{b} \partial^{\sigma}\partial^{\rho}\phi^{b}\Big)\label{OneLoopMTimesSN4ptHKPrediction}\ ,
\end{align}
where Greek and Latin indices were once again raised and lowered with $\eta_{\mu\nu}$ and $\delta_{ab}$.  The contact amplitude $\Acal_{4}$ computed from \eqref{OneLoopMTimesSN4ptHKPrediction} exactly reproduces \eqref{OneLoopMTimesSN4pt}, providing another non-trivial check of our general results.

\subsection{Conformal Galileons/DBI\label{sec:ConformalDBI}}

While the universal action \eqref{UniversalBraneAction} does not always admit flat vacua, $g_{\mu\nu}=\eta_{\mu\nu}$, in special cases it may be possible to add a \textit{lower-derivative} term to the action, such that the total system \textit{does} permit such flat solutions.

An important example is the case of a four-dimensional brane embedded in $AdS_{5}$, which provides a non-linear realization of the conformal group.  This model appears in a variety of contexts, e.g.,
\cite{Maldacena:1997re,Bellucci:2002ji,Silverstein:2003hf,Alishahiha:2004eh,Elvang:2012st,Hinterbichler:2012fr,DiVecchia:2017uqn,Guerrieri:2017ujb}.  The universal action for this theory, sometimes referred to as the conformal DBI or conformal galileon model, can be written in the form
\begin{align}
S_{\rm universal}&=-\int\rd^{4}x\, e^{-4\phi/L}\sqrt{1+e^{2\phi/L}(\partial\phi)^{2}}\ ,
\end{align}
where $L$ is the $AdS_{5}$ radius and we specialized to unitary gauge: $X^{\mu}=x^{\mu}$, $X^{5}=\phi(x)$.  A $\phi=$ constant configuration corresponds to a flat Minkowski configuration, but this is not a solution of the above. 

 A lower-derivative term which obeys the $AdS_{5}$ symmetries is
\begin{align}
S_{\rm WZ}&=\int\rd^{4}x\, e^{-4\phi/L}\label{WZTermConfGal}
\end{align}
and the combined $S_{\rm universal}+S_{\rm WZ}$ action allows for flat vacua.  The operator \eqref{WZTermConfGal} is Wess-Zumino (WZ) term which changes by a total derivative under the $AdS_{5}$ isometries.  In a string theory context, where \eqref{WZTermConfGal} is sometimes called a Chern-Simons term, it arises through the electric coupling of the brane to a four-form gauge field.  The fact that it allows for flat vacua corresponds to the so-called ``no-force" constraint; see \cite{Aharony:1999ti,Johnson:2005mqa,Baumann:2014nda}. The operator has a geometric interpretation as the bulk volume bounded by a flat $\phi=$ constant surface and a non-trivial surface defined by $\phi(x)$ \cite{Goon:2010xh}
 \begin{align}
S_{\rm WZ}&\sim \int^{\phi=\phi(x)}\rd^{5}X\, \sqrt{-\Gcal}\label{WZTermConfGalBulk}\ ,
\end{align}
meaning that \eqref{WZTermConfGal} is of the typical WZ form; compare to \cite{Altland:2006si}.

The heat-kernel analysis leading to \eqref{Polyakov1PIFunctionalDeterminant} does not directly apply to the full action $S_{\rm universal}+S_{\rm WZ}$, though it should still capture a subset of the divergences of this combined system.  Incorporating the effect of the WZ term into the functional determinant is left to future work.  While it is not possible to express \eqref{WZTermConfGal} itself in terms of natural geometric quantities defined on the embedded hypersurface, we expect that the normal-coordinate perturbations thereof will have such a representation, along the lines of what is found for the WZ terms for standard NLSMs \cite{Mukhi:1985vy,Akhoury:1990px} where the WZ operators can be interpreted as the existence of torsion \cite{Curtright:1984dz,Gates:1984nk} on the NLSM manifold.

 \section{Conclusions \label{sec:Conclusions}}

 In this work we have developed a formalism for deriving manifestly covariant quantum corrections for a generic class of brane systems in dimensions $d+1>2$. Included among this class are Dirac-Born-Infeld scalar theories, non-linear sigma models, and various generalizations thereof, which are associated to amplitudes which have special standing in various lines of modern $S$-matrix research \cite{Cachazo:2014xea,Cheung:2014dqa,Cachazo:2015ksa,Cachazo:2016njl,Cheung:2016drk,Padilla:2016mno,Guerrieri:2017ujb,Li:2017fsb,Bogers:2018kuw,Bogers:2018zeg,Elvang:2018dco,Rodina:2018pcb,Yin:2018hht,Roest:2019oiw,Low:2019ynd,Bern:2019prr,Bonifacio:2019rpv,Low:2020ubn}.  We developed a covariant form of perturbation theory using the natural geometric ingredients which describe hypersurfaces and their perturbations. One-loop results were our primary focus and one of our main results is the compact, covariant functional determinant \eqref{Polyakov1PIFunctionalDeterminant} which controls all one-loop corrections for systems of arbitrary co-dimension and arbitrary bulk metric.  The general form of the corresponding logarithmic divergences in $d+1=4$ are controlled by the explicit expression \eqref{TrLogdims3} and those for a limiting case in $d+1=6$ can be found in \eqref{TrLogdims6Flat}.  Our techniques carry significant advantages relative to naive approaches to the computation of the quantum effective action, as the latter can generate divergences which do not respect the symmetries of the original system: explicit examples of this phenomenon can be seen in Sec.~\ref{sec:NaiveDBI} and App.~\ref{app:Naive6DDBICalculation}.

Natural extensions of the present work include the following:
 \begin{itemize}
  \item We have only focused on the brane's own degrees of freedom which describe its motion in the higher-dimensional bulk spacetime, as described by \eqref{UniversalBraneAction}.  It is common to include couplings of gauge-bosons and other fields to the hypersurface in more general brane models and accounting for their effects\footnote{A study of one-loop corrections to specific brane models with additional fields can be found in \cite{Shmakova:1999ai,Wen:2020qrj,Elvang:2020kuj}.} would extend the applicability of our methods.
 \item It would be worthwhile to extend this work to higher-loop order.  In particular, when $d+1$ is odd, the first logarithmic divergences only occur at two-loops.  There is no in-principle obstruction to such computations and the covariant perturbation theory and gauge-fixing procedure we have developed are perfectly amenable to such higher-order computations. In the present work, we have used the on-shell equations of motion to simplify the computation, whereas higher-loop calculations would require working off-shell. The case of $d+1=3$ is particularly interesting, as shift-symmetric scalars can be dualized to 1-form gauge fields, resulting in Born-Infeld-Electrodynamics-like theories\footnote{See \cite{Elvang:2020kuj} for a recent amplitude-based study of related models.}.  
 \item The square-root form of the product-space, gauge-fixed action \eqref{UniversalActionProductSpace} is somewhat puzzling as it is not typically fixed by global symmetries\footnote{Apart from the special DBI $\g_{ab}(\phi)\longrightarrow\delta_{ab}$ limit in which case it is fixed by \eqref{DeltaDBI}.}.  Clearly, in the general case this structure must be understood as deriving from the fact that these theories also admit a geometric, diffeomorphism-invariant description.  This situation appears analogous to Born-Infeld Electrodynamics in which the action of a $U(1)$ gauge field is given by $\Lcal\sim \sqrt{-\det\left (\eta_{\mu\nu}+F_{\mu\nu}\right )}$, despite the lack of any non-linear symmetry for $A_{\mu}$ to enforce this structure \cite{Elvang:2018dco,Klein:2018ylk}.  Born-Infeld is exceptional in the space of vector EFTs due to its enhanced soft-limits \cite{Cheung:2018oki} and it would be interesting to check whether \eqref{UniversalActionProductSpace} is similarly exceptional in the space of multi-flavor scalar EFTs.
 \item As discussed in Sec.~\ref{sec:ConformalDBI},  Wess-Zumino or Chern-Simons terms can play an important role in particular brane models and amending our construction to accommodate such terms is a non-trivial and important goal. 
 \item There exist scalar theories which admit a more-involved geometric interpretation to which our methods do not immediately apply, namely the special galileons \cite{Novotny:2016jkh,Preucil:2019nxt}.  Some comments on these theories can be found in App.~\ref{app:SpecGal}. It would be interesting to more fully explore these systems along the lines emphasized in the present paper. 
\item Effective field theories with one or more light scalar degrees of freedom are used routinely in cosmology, for instance to describe dark energy in the late Universe, inflation in the early Universe, or modifications from General Relativity in strong gravity regimes. In order to manage the number of free Wilson coefficients, and to ensure that their tuning is radiatively stable, a large number of these scalar field models are endowed with a non-linearly realised symmetry. Since our general result for the one-loop effective action can be applied to any (multi-)scalar field theory which possesses any of a large class of braneworld non-linearly symmetries, we expect that our findings will find fruitful applications in such studies.
\end{itemize}
We leave such explorations for the future.

 \textbf{Acknowledgements}: We thank James Bonifacio, Kurt Hinterbichler, Riccardo Penco, David Stefanyszyn, and  Mark Trodden for helpful discussions.  The Mathematica packages \texttt{xAct} \cite{xAct} and \texttt{xTras} \cite{Nutma:2013zea} were used extensively in the course of this work. The work of GG is partially supported by the National Science Foundation under Grant No. PHY-1915611.  SM is supported by an Emmanuel College Research Fellowship. This work is supported in part by STFC under grants ST/L000385, ST/L000636 and ST/P000681/1.  JN is supported by an STFC Ernest Rutherford Fellowship, grant reference ST/S004572/1, and also acknowledges support from Dr.~Max R\"ossler, the Walter Haefner Foundation, the ETH Z\"urich Foundation, and King's College Cambridge.

\appendix

\section{Conventions \label{app:Conventions}}

We use mostly plus signature and our curvature conventions are $R^{\rho}{}_{\sigma\mu\nu}\equiv \partial_{\mu}\Gamma^{\rho}_{\nu\sigma}+\ldots$, $R_{\mu\nu}\equiv R^{\rho}{}_{\mu\rho\nu}$, and $R=R^{\mu}{}_{\mu}$. We work in $d+1$ spacetime dimensions when possible.  Our conventions for describing hypersurface embeddings are described in detail in  App.~\ref{app:Hypersurfaces}. Fourier conventions: $f(k)=\int\rd^{d+1}x\, e^{-ik\cdot x}f(x)$, $f(x)=\int\rd^{d+1}\tilde{k}\, e^{ik\cdot x}f(k)$, where $\tilde{k}\equiv k/(2\pi)$ and $\tilde{\delta}^{d+1}(\bfk)=(2\pi)^{d+1}\delta^{d+1}(\bfk)$ were defined to minimize explicit $(2\pi)$-factors.  Symmetrization and anti-symmetrization is defined with a $1/n!$, e.g.~$T_{(\mu\nu)}=\frac{1}{2!}(T_{\mu\nu}+T_{\nu\mu})$. Scattering amplitudes for $n$ particles $\Acal_{n}$ are defined through
\begin{align}
 \langle \bfk_{1}\ldots|\Tcal|\bfp_{1}\ldots\rangle\equiv \tilde{\delta}^{d+1}\left (\sum k-\sum p\right )\Acal_{n}(p_{1},\ldots;k_{1},\ldots)
 \end{align}
  where $S\equiv\mathbf{1}+i\Tcal$ is the usual $S$-matrix operator.   Mandelstam conventions are $s=-(p_{1}+p_{2})^{2}$, $t=-(p_{1}+p_{3})^{2}$ and $u=-(p_{1}+p_{4})^{2}$.  When computing in dimensional regularization, we will use $-2\varepsilon$ to denote the deviation from integer dimensions, e.g., we shift $d+1\to d+1-2\varepsilon$ with $d\in \mathbb{Z}$. Heat kernel calculations generally depend on both matrix and functional determinants or traces.  In such contexts, we use $ \Det$ to denote the combination of a matrix and a functional determinant, while $\det$ is reserved for purely functional determinants.  Similar conventions are used for traces: $\Tr$ vs $\tr$ and delta-function factors are left implicit in all such functional determinants and traces. The notation $[A(x)]=\lim_{x'\to x}A(x,x')$ to denote coincidence limits.

\section{Hypersurfaces \label{app:Hypersurfaces}}

 A review of the relevant geometry needed to describe higher-co-dimension hypersurfaces. Appendix A of \cite{Hinterbichler:2010xn} has a more extensive discussion, which we partially follow here.

 \subsection{Embedding Functions and Terminology}

A $(d+1)$-dimensional submanifold $M_{d+1}$ embedded within a $(D+1)$-dimensional manifold $\Mcal _{D+1}$ can be specified via embedding functions $X^{A}(x^{\mu})$, $A\in \{0,\ldots,D\}$ and $\mu\in\{0,\ldots, d\}$, where $X^{A}$ and $x^{\mu}$ are coordinates on $\Mcal_{D+1}$ and $M_{d+1}$, respectively.  The manifold $M_{d+1}$ is said to be of co-dimension-$(D-d)$.  Throughout this paper, we will refer to $M_{d+1}$ and the brane and $\Mcal _{D+1}$ as the bulk.

\subsection{Tangent Vectors, Normal Vectors, Projectors, and Extrinsic Curvatures}

The embedding functions directly define the $(d+1)$ independent tangent vectors to $M_{d+1}$ whose bulk components are given by
\begin{align}
 e_{\mu}=\partial_{\mu} X^{A}(x)\partial_{A}\ .
 \end{align}
 Orthogonal to these are $D-d$ normal vectors $n^{A}{}_{i}$, where $i\in \{d+1, \ldots, D\}$ labels the various normal vectors. 
  If $\Mcal_{D+1}$ is equipped with a metric $ \Gcal_{AB}$, then the induced metric on $M_{d+1}$ is defined to be
\begin{align}
g_{\mu\nu}\equiv e_{\mu}{}^{A}e_{\nu}{}^{B} \Gcal_{AB}\ .
\end{align}
These ingredients can be taken to obey
\begin{align}
n^{A}{}_{I}e_{\mu A}=0 \, \quad n^{A}{}_{i}n_{Aj}=\delta_{ij}\ ,
\end{align}
where all $A,B,\ldots$ indices are raised and lowered with $ \Gcal_{AB}$ and we assume throughout that the directions normal to the brane are spacelike.

  All tensors on $\Mcal _{D+1}$ can then have their components projected onto the tangent and normal directions to $M_{d+1}$.  The projector onto the tangent directions is
\begin{align}
\Pcal_{\parallel}^{A}{}_{B}\equiv e^{\mu A}e_{\mu B}=\delta^{A}_{B}-n^{A}{}_{i }n_{B}{}^{i}\ ,\label{app:eq:PParallel}
\end{align}
where $g_{\mu\nu}$, $ \Gcal_{AB}$, and $\delta_{ij}$ are used to raise and lower the appropriate indices. The projector onto the normal directions is the complement to the above:
\begin{align}
\Pcal_{\perp}^{A}{}_{B}\equiv\delta^{A}_{B}-\Pcal_{\parallel}^{A}{}_{B}=\delta^{A}_{B}-e^{\mu A}e_{\mu B}=n^{A}{}_{i}n_{B}{}^{i}\ . \label{app:eq:PPerp}
\end{align}

Associated to each normal vector $n^{A}{}_{i}$ is an extrinsic curvature $K_{\mu\nu}^{i}$ symmetric under $\mu\longleftrightarrow \nu$ whose form and properties are:
\begin{align}
K_{\mu\nu}^{i}=e_{\mu}{}^{B}e_{\nu}{}^{A}\nabla_{B}n_{A}{}^{i}=-n_{A}{}^{i}e_{\mu}{}^{B}\nabla_{B}e_{\nu}{}^{A}=\frac{1}{2}e_{\mu}{}^{A}e_{\nu}{}^{B}\pounds_{n_{i}} \Gcal_{AB} \ , \quad  K^{\mu i}_{\mu}=\nabla_{A}n^{Ai}\ .\label{app:eq:ExtrinsicCurvatureDefs}
\end{align}
It is useful to define the following combination in order to avoid explicit appearances of the $i$ labels:
\begin{align}
 K_{\mu\nu}^{A}\equiv K_{\mu\nu}^{i}n^{A}{}_{i} \ ,
 \end{align}
 from which it follows that $e_{\alpha A}K_{\mu\nu}{}^{A}=0$.
 
 \subsection{Covariant Derivatives}
 
The tangent vectors $e_{\mu}{}^{A}$ transform covariantly under both bulk and brane diffeomorphisms:
\begin{align}
e_{\mu}{}^{A}(X(x))\longrightarrow \frac{\partial x^{\nu}}{\partial y^{\mu}}\frac{\partial Y^{A}}{\partial X^{B}}e_{\nu}{}^{B}(Y(y))\ ,
\end{align}
under $X^{A}\longrightarrow Y^{A}$ and $x^{\mu}\longrightarrow y^{\mu}$.  Given a generic tensor $T^{A\alpha\ldots}$ which transforms covariantly under such diffeomorphisms, the covariant derivative of such tensors along the brane is given by
\begin{align}
 \Dcal_{\mu}T^{A\alpha \ldots}=\partial_{\mu}T^{A\alpha}+ \Gamma^{A}_{BC}T^{C\alpha}\partial_{\mu}X^{B}+ \Gamma^{\alpha}_{\mu\nu}T^{A\nu}+\ldots\label{app:eq:BraneCovariantDerivative}
\end{align}
where $\Gamma^{A}_{BC}$ and $\Gamma^{\alpha}_{\mu\nu}$ are computed from $\Gcal_{AB}$ and $g_{\mu\nu}$, respectively, and the commutator gives
\begin{align}
\left [ \Dcal_{\mu}, \Dcal_{\nu}\right ]T^{A\alpha \ldots}&=R^{\alpha}{}_{\sigma\mu\nu}T^{A\sigma \ldots}+\Rcal^{A}{}_{BCD}\partial_{\mu}X^{C}\partial_{\nu}X^{D}T^{B \alpha \ldots}+\ldots\label{app:eq:DCommutator}
\end{align}
where the curvatures corresponding to $g_{\mu\nu}$ and $\Gcal_{AB}$ are denoted by $R_{\mu\nu\rho\sigma}$ and $\Rcal_{ABCD}$, respectively.  

The covariant derivative $\Dcal_{\mu}$ of the tangent vectors $e_{\mu}{}^{A}$ determine the extrinsic curvature tensors \eqref{app:eq:ExtrinsicCurvatureDefs}.  This follows from the Gauss-Weingarten relations, in which one projects $e_{\mu}{}^{B}\nabla_{B}e_{\nu}{}^{A}$ onto its tangent and orthogonal pieces, with the result:
\begin{align}
e_{\mu}{}^{B}\nabla_{B}e_{\nu}{}^{A}&=\Pcal_{\parallel}^{A}{}_{C}e_{\mu}{}^{B}\nabla_{B}e_{\nu}{}^{C}+\Pcal_{\perp}^{A}{}_{C}e_{\mu}{}^{B}\nabla_{B}e_{\nu}{}^{C}\nn
&=  \Gamma^{\rho}_{\mu\nu}e_{\rho}{}^{A}-K_{\mu\nu}^{A} \label{app:eq:GaussWeingarten1}
\end{align}
where $ \Gamma^{\rho}_{\mu\nu}$ is the standard Christoffel symbol associated to $g_{\mu\nu}$.  Rearranging the above and comparing to \eqref{app:eq:BraneCovariantDerivative} implies
\begin{align}
 \Dcal_{\mu}e_{\nu}{}^{A}&=-K_{\mu\nu}{}^{A}\nn
 \implies \Dcal_{\mu}\Pcal_{\parallel}^{AB}&=-2K_{\mu}{}^{\nu(A}e_{\nu}{}^{ B)}\nn
 \implies \Dcal_{\mu}\Pcal_{\perp}^{AB}&=2K_{\mu}{}^{\nu(A}e_{\nu}{}^{ B)}\ .\label{app:eq:DOnTangentVector}
\end{align}

  It is additionally useful to define the covariant derivatives $\Dcal^{\perp}_{\mu}$ and $\Dcal^{\parallel}_{\mu}$ via
\begin{align}
\Dcal^{\perp}_{\mu}\equiv \Pcal_{\perp}\cdot \Dcal_{\mu}\cdot \Pcal_{\perp} \ , \quad \Dcal^{\parallel}_{\mu}\equiv \Pcal_{\parallel}\cdot \Dcal_{\mu}\cdot \Pcal_{\parallel} \label{app:eq:DPerpParallel} \ ,
\end{align}
schematically, which naturally act on the spaces of normal and tangent tensors, respectively.  The former derivative is of particular importance in this paper and we will require the commutator of $\Dcal^{\perp}_{\mu}$ on a vector $\phi^{A}$ normal to $M_{d+1}$, such that $\Pcal_{\perp}^{AB}\phi_{B}=\phi^{A}$:
\begin{align}
\left [\Dcal_{\mu}^{\perp},\Dcal_{\nu}^{\perp}\right ]\phi^{A}&=\Pcal^{A B}_{\perp }\Rcal_{BCDE}\phi^{C}e_{\mu}{}^{D}e_{\nu}{}^{E}+2K_{[\mu}{}^{\alpha A}K_{\nu]\alpha B}\phi^{B}\ .\label{app:eq:DPerpCommutator}
\end{align}

\subsection{Gauss-Codazzi Relations}

The Gauss-Codazzi relations express various projections of the bulk Riemann tensor $\Rcal_{ABCD}$ in terms quantities defined on the brane $M_{d+1}$, namely the brane curvature $R_{\mu\nu\rho\sigma}$ and the extrinsic curvatures $K_{\mu\nu}{}^{A}$ \eqref{app:eq:ExtrinsicCurvatureDefs}. One such relation follows from combining \eqref{app:eq:DCommutator} and \eqref{app:eq:DOnTangentVector} to find
\begin{align}
2 \Dcal_{[\rho} \Dcal_{\sigma]}e_{\mu}{}^{A}&=-2 \Dcal_{[\rho}K_{\sigma]\mu}{}^{A}\nn
&=R_{\mu}{}^{\kappa}{}_{\rho\sigma}e_{\kappa}{}^{A}+\Rcal^{A}{}_{BCD}e_{\mu}{}^{B}e_{\rho}{}^{C}e_{\sigma}{}^{D} \label{app:eq:GaussCodazzi} \ .
\end{align} Contracting with $e_{\nu A}$ gives and expression for the induced Riemann tensor on $M_{d+1}$
\begin{align}
R_{\mu\nu\rho\sigma}&=\Rcal_{ABCD}e_{\mu}{}^{A}e_{\nu}{}^{B}e_{\rho}{}^{C}e_{\sigma}{}^{D}-2K_{\nu[\rho}{}^{A}K_{\sigma]\mu A}\ ,
\end{align}
where $e_{\alpha A}K_{\mu\nu}{}^{A}=0$ was used.  Projecting onto the normal directions instead gives
 \begin{align}
 -2\Dcal_{\perp[\rho}K_{\sigma]\mu}{}^{A}&=\Pcal_{\perp}^{AB}\Rcal_{BCDE}e_{\mu }{}^{C}e_{\rho}{}^{D}e_{\sigma}{}^{E}\ .
 \end{align}
 
 \section{Normal Coordinates \label{app:NormalCoordinates}}
 
For the background-field method calculation considered in this paper, we are interested in taking the induced metric $\partial_{\mu}X^{A}\partial_{\nu}X^{B}\Gcal_{AB}(X)$ which appears in, e.g., the Polyakov action \eqref{PolyakovAction} and creating a perturbed version, denoted here by $\gamma_{\mu\nu}$, by introducing a set of fluctuations $\delta X^{A}$ about the $X^{A}$.  The fluctuations are to be path-integrated over to generate the quantum effective action $\Gamma[X]$,
\begin{align}
  e^{i\Gamma[X]}=\int\Dcal\delta X\, e^{iS[X,\delta X]}\ ,
  \end{align}
  schematically.
  
    If we were to introduce the $\delta X^{A}$ in the naive way by replacing $X^{A}\longrightarrow X^{A}+\delta X^{A}$ and working with
\begin{align}
  \gamma^{\rm naive}_{\mu\nu}(X,\delta X)=\partial_{\mu }\left (X^{A}+\delta X^{A}\right )\partial_{\nu}\left (X^{B}+\delta X^{B}\right )\Gcal_{AB}(X+\delta X)\label{app:eq:NaiveShiftedInducedMetric}\ ,
  \end{align}
  the resulting calculations would be cumbersome because the $\delta X^{A}$'s are not proper bulk tensors and the expansion would not be manifestly covariant.  A more clever expansion involves the use of normal coordinates around the point $X^{A}$ in which the directions and magnitudes of geodesics\footnote{The following is equivalent to the procedure used in \cite{AlvarezGaume:1981hn,Mukhi:1985vy}, for instance, to address the same problem, though the derivations are different.} emanating from $X^{A}$ are used in place of the $\delta X^{A}$'s.    Specifically, we can switch from $\delta X^{A}\longrightarrow \chi^{A}$ with $\chi^{A}$ a true tensor by letting
\begin{align}
\chi^{A}\equiv -\sigma^{A}\left (X,X+\delta X\right )\ .\label{app:eq:NormalCoordinates}
\end{align}
Above, $\sigma(X_{1},X_{2})$ is the geodesic interval for the spacetime (see App.~\ref{app:CovariantHeatKernels}) and $\sigma^{A\ldots A'\ldots}(X_{1},X_{2})=\nabla^{A}_{(X_{1})}\ldots\nabla^{A'}_{(X_{2})}\ldots\sigma(X_{1},X_{2})$.  Given the geodesic connecting $X$ and $X+\delta X$, the direction of $\chi^{A}$ corresponds to the tangent vector of this geodesic at the point $X$ and the magnitude of $\chi^{A}$ is a measure of the separation of the two points, i.e., the size of $\delta X$.   We follow \cite{Poisson:2011nh} when quoting various properties of $\sigma$ below.

Letting $X_{1}=X$ and $X_{2}\equiv X+\delta X$, we start the analysis of the induced metric by first introducing $x$-dependence into \eqref{app:eq:NormalCoordinates} as it appears when considering hypersurface embeddings into the bulk spacetime:
\begin{align}
\chi^{A}(x)=-\sigma^{A}(X_{1}(x),X_{2}(x))\ .
\end{align}
Taking a derivative and using \eqref{app:eq:NormalCoordinates} yields
\begin{align}
\partial_{\mu}\chi^{A}(x)&=-\partial_{\mu}X_{2}^{A'}(x)\sigma^{A}{}_{A'}\left (X_{1},X_{2}\right )-\partial_{\mu}(X_{1}^{B})\left (\sigma_{B}{}^{A}(X_{1},X_{2})+ \Gamma^{A}_{BC}\chi^{C}\right )\ .
\end{align}
In the near coincident limit where $X_{1}\approx X_{2}$, $\sigma_{B}{}^{A}(X_{1},X_{2})$ can be expanded as
\begin{align}
\sigma_{B}{}^{A}(X_{1},X_{2})&\approx \delta^{A}_{B}-\frac{1}{3}\Rcal^{A}{}_{CBD}\sigma^{C}\sigma^{D}+\mathcal{O}\left (\sigma^{3}\right )\ ,
\end{align}
which allows us to rearrange the preceding result as
\begin{align}
\partial_{\mu}X_{2}^{A'}(x)\sigma^{A}{}_{A'}(X_{1},X_{2})
&= \partial_{\mu}X_{1}^{A}+ \Dcal_{\mu}\chi^{A}-\frac{1}{3}\Rcal^{A}{}_{CBD}\chi^{C}\chi^{D}\partial_{\mu}X_{1}^{B}+\mathcal{O}\left (\chi^{3}\right ) \ ,\label{app:eq:NormalCoordinatesDerivativeRelation}
\end{align}
where $\Dcal_{\mu}$ is the brane derivative which is covariant under both brane and bulk diffeomorphisms, as defined in App.~\ref{app:Hypersurfaces}.

Next, we also use the fact that at coincidence
\begin{align}
\sigma^{A}{}_{A'}(X_{1},X_{2})&\approx-\Gcal^{B}{}_{A'}(X_{1},X_{2})\left (\delta^{A}_{B}+\frac{1}{6}\Rcal^{A}{}_{CBD}(X_{1})\sigma^{C}\sigma^{D}+\mathcal{O}\left (\sigma^{3}\right )\right )\ ,
\end{align}
where $\Gcal_{BA'}(X_{1},X_{2})$ is the parallel propagator, as defined in \cite{Poisson:2011nh}, to note that
\begin{align}
&\quad \partial_{\mu}X_{1}^{A'}(x)\sigma^{A}{}_{A'}\partial_{\nu}X_{1}^{B'}(x)\sigma^{B}{}_{B'}\times\left (\Gcal_{AB}(X_{1})-\frac{1}{3}\Rcal_{ACBD}(X_{1})\sigma^{C}\sigma^{D}\right )\nn
&=\partial_{\mu}X_{2}^{A'}(x)\partial_{\nu}X_{2}^{B'}(x)\Gcal_{A'B'}(X_{2})+\mathcal{O}\left (\sigma^{3}\right )\ ,\label{app:eq:NormalCoordinatesRewriting}
\end{align}
where $\Gcal^{C}{}_{A'}(X_{1},X_{2})\Gcal^{D}{}_{B'}(X_{1},X_{2})\Gcal_{CD}(X_{1})=\Gcal_{A'B'}(X_{2})$ was used.
After replacing $X_{2}\longrightarrow X+\delta X$ and removing the primes, the final line above is found to be precisely $\gamma^{\rm naive}_{\mu\nu}(X,\delta X)$ \eqref{app:eq:NaiveShiftedInducedMetric}.  Finally, using \eqref{app:eq:NormalCoordinatesDerivativeRelation} in \eqref{app:eq:NormalCoordinatesRewriting} we find the equivalent, covariant expression of interest:
\begin{align}
\gamma^{\rm naive}_{\mu\nu}(X,\delta X)&= \partial_{\mu}X^{A}\partial_{\nu}X^{B}\Gcal_{AB}(X)+2 \Dcal_{(\mu}\chi^{A}\partial_{\nu)}X^{B}\Gcal_{AB}(X)\nn
&\quad + \Dcal_{\mu}\chi^{A} \Dcal_{\nu}\chi^{B}\Gcal_{AB}(X)-\Rcal_{ABCD}(X)\partial_{\mu}X^{A}\chi^{B}\partial_{\nu}X^{C}\chi^{D}+\mathcal{O}(\chi^{3})\nn
&\equiv \gamma_{\mu\nu}(X,\chi)\ .\label{app:eq:NormalCoordinateMetricExpansion}
\end{align}
For instance, the $\Ocal(\chi^{2})$ terms in the perturbed Polyakov action in \eqref{PolyakovActionQuadratic} come from expanding out
\begin{align}
S_{\rm Poly}=\int\rd^{d+1}x\, \sqrt{-g}\left (-\frac{1}{2}g^{\mu\nu}\gamma_{\mu\nu}(X,\chi)+\frac{(d-1)}{2}\right )\ .
\end{align}

\section{Covariant Heat Kernel Methods\label{app:CovariantHeatKernels}}

The functional determinants which arise in one-loop computations can be efficiently computed through the use of covariant heat kernel methods.  We review the construction here, following \cite{Barvinsky:1985an}. See, e.g., \cite{Gilkey:1975iq,Proceedings:1985fja,Avramidi:1986mj} for alternative presentations.

\subsection{General Scenario}

Functional determinants arise from elementary gaussian integrals:
\begin{align}
\int\Dcal\phi\, \exp\left [i \phi\cdot \Ocal\cdot\phi\right ]= \exp\left [-\frac{c}{2}\ln \Det\Ocal\right ]=\exp\left [-\frac{c}{2}\Tr\ln\Ocal\right ]\ ,
\end{align}
where $\Ocal$ is some differential operator of interest, the constant is given by $c=1$ ($c=-1$) for bosons (fermions), and overall normalizations and relevant indices were omitted.  Heat kernel methods start by representing the functional trace as an integral\footnote{Only the $s\longrightarrow 0$ end of the integral contributes after a proper $i\epsilon$ prescription and \eqref{app:eq:GenericTrLnIntegralExpression} holds up to divergent terms independent of $\Ocal$. Similarly, one has $\frac{1}{\Ocal^{n}}=\frac{-1}{i^{n}\Gamma[n]}\int_{0}^{\infty}\rd s\, s^{n-1}e^{is\Ocal}$.\label{app:foot:HKPropagators}}
\begin{align}
\Tr\ln\Ocal=-\int_{0}^{\infty}\frac{\rd s}{s}\, \int\rd^{d+1}x\, \tr \langle x|e^{is\mathcal{O}}|x\rangle\ ,\label{app:eq:GenericTrLnIntegralExpression}
\end{align}
where $\Tr$ indicates both a functional trace and a trace over whatever indices are associated with $\mathcal{O}$, while $\tr$ is only a trace in the latter sense.  The states $|x\rangle$ carry any indices associated with $\mathcal{O}$, suppressed above.  

The utility of this construction is that we can compute $\langle x|e^{is\mathcal{O}}|x\rangle$ by first considering the off-diagonal matrix element $\langle x|e^{is\mathcal{O}}|x'\rangle$ which can be interpreted as the quantum-mechanical amplitude to go from $x'\longrightarrow x$ in ``time\footnote{$s$ does not typically have units of time.}" s under the influence of Hamiltonian $H=-\mathcal{O}$.  Writing $\langle x|e^{is\mathcal{O}}|x'\rangle\equiv \langle x|x';s\rangle$, this matrix element obeys an effective Schr\"odinger equation\footnote{An explicit, simple example: when $\mathcal{O}=\partial^{2}$, we have $\langle x|x'\rangle=\delta^{d+1}(x-x')$ and $\langle x|\partial^{2}|x'\rangle\equiv \partial^{2}_{x}\delta^{d+1}(x-x')$.  The Schr\"odinger equation then comes from $i\partial_{s}\langle x|e^{i\partial^{2}}|x'\rangle=-\langle x|\partial^{2}e^{i\partial^{2}}|x'\rangle=-\int\rd^{d+1}y\, \langle x|\partial^{2}|y\rangle\langle y|e^{i\partial^{2}}|x'\rangle =-\int\rd^{d+1}y\,\partial^{2}_{x}\delta^{d+1}(x-y)\langle y|e^{i\partial^{2}}|x'\rangle =-\partial^{2}_{x}\left (\int\rd^{d+1}y\,\delta^{d+1}(x-y)\langle y|e^{i\partial^{2}}|x'\rangle\right ) =-\partial_{x}^{2}\langle x|e^{i\partial^{2}}|x'\rangle$.}
\begin{align}
i\partial_{s}\langle x|x';s\rangle=-\mathcal{O}_{x}\langle x|x';s\rangle\ .
\end{align}
In typical cases, one can then use the above to solve for the coincident limit result $\lim _{x'\to x}\langle x|x';s\rangle$ in a power-series expansion in $s$, and the $\Ocal(s^{0})$ term in the series determines the logarithmically divergent contribution to \eqref{app:eq:GenericTrLnIntegralExpression}, due to the $s\longrightarrow 0$ end of the integral, which is often the quantity of interest.

\subsection{Canonical Scenario}

We now restrict our attention to the canonical scenario in which the operator $\mathcal{O}$ takes on the form
\begin{align}
\mathcal{O}\longrightarrow \Dcal^{2}+\Ucal^{AB}\ ,
\end{align}
where $A,B$ are some set of indices,  $\Ucal^{AB}$ is a symmetric matrix constructed from local fields, and $\Dcal$ is a covariant derivative whose internal indices are suppressed.  For simplicity, we also restrict the following discussion to the case where $\mathcal{O}$ acts on fields with a single vector index, but generalizations are straightforward.  We denote the commutator of $\Dcal_{\mu}$ on a generic tensor field $T^{A\alpha\ldots}$ by
\begin{align}
\left [\Dcal_{\mu},\Dcal_{\nu}\right ]T^{A\alpha\ldots}\equiv R^{\alpha}{}_{\beta\mu\nu}T^{A\alpha\ldots}+\Fcal_{\mu\nu}{}^{A}{}_{B}T^{B\alpha\ldots}+\ldots
\end{align}
where $\Fcal_{\mu\nu (AB)}=\Fcal_{(\mu\nu) AB}=0$.
The effective Schr\"odinger equation is then
\begin{align}
i\partial_{s}\langle x,A|x',Z';s\rangle=-\left (\Dcal^{2}+\Ucal^{AB}\right )\langle x,B|x',Z';s\rangle\ ,\label{app:eq:HeatKernelSchrodingerEq}
\end{align}
where $A,Z'$ are internal indices and the operator only acts on unprimed indices and coordinates.

The Schr\"odinger equation can then be solved by employing the ansatz \cite{Barvinsky:1985an}
\begin{align}
\langle x,A|x',Z';s\rangle&\equiv i\frac{ \sqrt{-\det\left (-\sigma_{\mu\nu'}(x,x')\right )}}{(4\pi is)^{(d+1)/2}}\exp\left (\frac{i\sigma(x,x')}{2s}\right )\sum_{n=0}^{\infty}(is)^{n}\mathsf{a}_{n}^{AZ'}(x,x')\ ,\label{app:eq:HeatKernelPointParticleAmplitudeAnsatz} 
\end{align}
which reduces \eqref{app:eq:HeatKernelSchrodingerEq} to a set of simple recursion relations for the $a_{n}$'s. Above, $\sigma(x,x')$ is the world function which characterizes the geodesic distance between two points $x$ and $x'$ on a given manifold and $\sigma_{\mu_1\ldots\mu_m\nu'_1\ldots\nu'_n}(x,x')\equiv \nabla_{\mu _{1}}\ldots\nabla_{\mu _{m}}\nabla_{\nu'_{1}}\ldots\nabla_{\nu'_{n}}\sigma(x,x')$.  One method for computing $\sigma(x,x')$ is to consider the action
\begin{align}
S[y^{\mu}]&=\int\rd t\, \frac{1}{4}g_{\mu\nu}(y)\frac{\rd y^{\mu }}{\rd t}\frac{\rd y^{\nu}}{\rd t}\ , \label{app:eq:HeatKernelPointParticleAction}
\end{align} 
whose equation of motion is simply the geodesic equation. The on-shell value of the action then determines $\sigma(x,x')$ via
\begin{align}
S[x^{\mu}]_{\rm on-shell}&\equiv \frac{\sigma(x,x')}{2s}\ , \label{app:eq:WorldFunction}
\end{align}
where the action is evaluated the geodesic $y^{\mu}(t)$ satisfying $y^{\mu}(0)=x^{\mu}$ and $y^{\mu}(s)=x'^{\mu}$.  The $\sim e^{\frac{i\sigma}{2s}}$ factor in the ansatz \eqref{app:eq:HeatKernelPointParticleAmplitudeAnsatz} can be roughly understood as arising from the point-particle's action's \eqref{app:eq:HeatKernelPointParticleAction} contribution to $\langle x|x';s\rangle\sim \int\Dcal x\,  e^{iS}$, morally speaking.  A detailed review of the world functions and related geometric quantities can be found in \cite{Poisson:2011nh}. For a review focused on heat kernel applications, see \cite{Proceedings:1985fja}.

   The recursion relations stemming from using \eqref{app:eq:HeatKernelPointParticleAmplitudeAnsatz} in \eqref{app:eq:HeatKernelSchrodingerEq} are to be solved subject to
   \begin{align}
    \lim _{x'\to x}\mathsf{a}_{0}^{AZ'}(x,x')=\mathsf{g}^{AZ}\ ,
    \end{align} where $\mathsf{g}^{AZ}$ is the field-space metric which is compatible with the covariant derivative $\Dcal$.  This condition is necessary to reproduce the known short-distance behavior of the propagator.  The construction is well-reviewed in the references listed at the beginning of this appendix and we focus only on the ultimate results in the below.

We focus on the logarithmically divergent terms in the trace, which only occur in even dimensions  where $d+1=2n$, $n\in\mathbb{Z}$.  In the dimensional regularization scheme used in \cite{Brown:1976wc,Brown:1977pq} and reviewed in \cite{Barvinsky:1985an}, only the term $\propto s^{0}$ in \eqref{app:eq:HeatKernelPointParticleAmplitudeAnsatz} contributes and working in $d+1-2\varepsilon$ dimensions, the logarithmically divergent piece is captured by a pole in $\varepsilon$, as usual	:
\begin{align}
\Tr\ln\left (\Dcal^{2}+\Ucal\right )&\supset -\frac{1}{\varepsilon}\frac{i}{(4\pi)^{n}}\int\rd^{d+1}x\, \sqrt{-g}\, \tr [\mathsf{a}_{n}(x)]\ ,\label{app:eq:LogDivergencesFromHeatKernel}
\end{align}
where we used the notation $[A(x)]=\lim_{x'\to x}A(x,x')$ to denote coincidence limits and $\mathsf{g}_{AZ}$ to perform the trace over indices.   Repeating standard calculations in our conventions, we find the following results for various low dimensional cases:
\begin{align}
\tr[\mathsf{a}_{1}(x)]&= \frac{\Ncal R}{6}+\Ucal\nn
\tr[\mathsf{a}_{2}(x)]&=\frac{1}{2} \Ucal_{AB} \Ucal^{AB} -  \frac{1}{180} \Ncal  R_{\alpha \beta} R^{\alpha \beta} + \frac{1}{6} \Ucal R + \frac{1}{72} \Ncal  R^2 + \frac{1}{180} \Ncal  R_{\alpha \beta \gamma \delta} R^{\alpha \beta \gamma \delta} \nn
&\quad-  \frac{1}{12} \mathcal{F}_{\beta \alpha BA} \mathcal{F}^{\beta \alpha BA} + \frac{1}{6} \Dcal^{2}\Ucal + \frac{1}{30} \Ncal  \Dcal^{2}R\nn
\tr[\mathsf{a}_{3}(x)]&=\frac{1}{6} \Ucal_{A}{}^{C} \Ucal^{AB} \Ucal_{BC} -  \frac{1}{180} \Ucal R_{\alpha \beta} R^{\alpha \beta} + \frac{1}{5670} \Ncal  R_{\alpha}{}^{\gamma} R^{\alpha \beta} R_{\beta \gamma} + \frac{1}{12} \Ucal_{AB} \Ucal^{AB} R \nn
&\quad-  \frac{1}{1080} \Ncal  R_{\alpha \beta} R^{\alpha \beta} R + \frac{1}{72} \Ucal R^2 + \frac{1}{1296} \Ncal  R^3 -  \frac{1}{1890} \Ncal  R^{\alpha \beta} R^{\gamma \delta} R_{\alpha \gamma \beta \delta} \nn
&\quad+ \frac{1}{180} \Ucal R_{\alpha \beta \gamma \delta} R^{\alpha \beta \gamma \delta} + \frac{1}{1080} \Ncal  R R_{\alpha \beta \gamma \delta} R^{\alpha \beta \gamma \delta} -  \frac{1}{5670} \Ncal  R^{\alpha \beta} R_{\alpha}{}^{\gamma \delta \varepsilon} R_{\beta \gamma \delta \varepsilon}\nn
&\quad + \frac{1}{1890} \Ncal  R_{\alpha \beta}{}^{\varepsilon \zeta} R^{\alpha \beta \gamma \delta} R_{\gamma \delta \varepsilon \zeta} -  \frac{1}{12} \Ucal^{ac} \mathcal{F}_{\beta \alpha cb} \mathcal{F}^{\beta \alpha}{}_{A}{}^{B} -  \frac{1}{72} R \mathcal{F}_{\beta \alpha BA} \mathcal{F}^{\beta \alpha BA} \nn
&\quad+ \frac{1}{90} R^{\alpha \gamma} \mathcal{F}_{\alpha}{}^{\beta ab} \mathcal{F}_{\gamma \beta ab} -  \frac{1}{30} \mathcal{F}_{\alpha}{}^{\beta}{}_{B}{}^{C} \mathcal{F}^{\alpha \gamma BA} \mathcal{F}_{\gamma \beta ac} -  \frac{1}{60} R_{\beta \alpha \gamma \delta} \mathcal{F}^{\beta \alpha BA} \mathcal{F}^{\gamma \delta}{}_{BA} \nn
&\quad+ \frac{1}{6} \Ucal^{AB} \Dcal^{2}\Ucal_{AB} + \frac{1}{36} R \Dcal^{2}\Ucal + \frac{1}{30} \Ucal \Dcal^{2}R + \frac{1}{180} \Ncal  R \Dcal^{2}R \nn
&\quad+ \frac{1}{12} \Dcal _{\alpha}\Ucal_{AB} \Dcal ^{\alpha}\Ucal^{AB} + \frac{1}{30} \Dcal _{\alpha}\Ucal \Dcal ^{\alpha}R + \frac{17}{5040} \Ncal  \Dcal _{\alpha}R \Dcal ^{\alpha}R + \frac{1}{90} R^{\alpha \beta} \Dcal _{\beta}\Dcal _{\alpha}\Ucal\nn
&\quad + \frac{1}{60} \Dcal^{4}\Ucal + \frac{1}{280} \Ncal  \Dcal^{4}R + \frac{1}{420} \Ncal  R_{\alpha \beta} \Dcal ^{\beta}\Dcal ^{\alpha}R\nn
&\quad -  \frac{1}{180} \Dcal _{\alpha}\mathcal{F}^{\alpha}{}_{\beta}{}^{BA} \Dcal _{\gamma}\mathcal{F}^{\gamma \beta}{}_{BA} -  \frac{1}{630} \Ncal  R^{\alpha \beta} \Dcal _{\gamma}\Dcal ^{\gamma}R_{\alpha \beta} -  \frac{1}{30} \mathcal{F}^{\beta \alpha BA} \Dcal _{\gamma}\Dcal ^{\gamma}\mathcal{F}_{\beta \alpha BA} \nn
&\quad-  \frac{1}{1260} \Ncal  \Dcal _{\beta}R_{\alpha \gamma} \Dcal ^{\gamma}R^{\alpha \beta} -  \frac{1}{2520} \Ncal  \Dcal _{\gamma}R_{\alpha \beta} \Dcal ^{\gamma}R^{\alpha \beta} -  \frac{1}{45} \Dcal _{\gamma}\mathcal{F}_{\alpha}{}^{\beta}{}_{BA} \Dcal ^{\gamma}\mathcal{F}^{\alpha}{}_{\beta}{}^{BA}\nn
&\quad + \frac{1}{567} \Ncal  R_{\alpha \gamma \beta \delta} \Dcal ^{\delta}\Dcal ^{\gamma}R^{\alpha \beta} + \frac{11}{5670} \Ncal  R^{\alpha \beta \gamma \delta} \Dcal _{\varepsilon}\Dcal ^{\varepsilon}R_{\alpha \beta \gamma \delta} + \frac{1}{560} \Ncal  \Dcal _{\varepsilon}R_{\alpha \beta \gamma \delta} \Dcal ^{\varepsilon}R^{\alpha \beta \gamma \delta}\label{app:eq:anHeatKernelResults}
\end{align}
where $\Ncal=\mathsf{g}^{A}{}_{A}$ is the dimensionality of the vector space and $\Ucal\equiv \Ucal^{A}{}_{A}$.  The Bianchi identities $R_{[\mu\nu\rho]\sigma}=\Dcal_{[\mu}R_{\nu\rho]\sigma\alpha}=\Dcal_{[\mu}\Fcal_{\nu\rho]AB}=0$ were used to simplify, but no integrations by parts were performed.

 \section{Naive DBI Calculation in $d+1=6$ \label{app:Naive6DDBICalculation}}
 
The computation of \eqref{DBIFunctionalDet} when $d+1=6$ proceeds similarly to the $d+1=4$ case.  The steps are simply longer and more burdensome, so we have relegated them to this appendix.

The logarithmically divergent terms arising from \eqref{DBIFunctionalDet} are given by \eqref{app:eq:LogDivergencesFromHeatKernel} and \eqref{app:eq:anHeatKernelResults}.  After removing various total derivatives and using the dimension $d+1=6$ topological Gauss-Bonnet term $\mathcal{L}_{\rm GB}$, explicitly given by
\begin{align}
\mathcal{L}^{(6)}_{\rm GB}[g]&\equiv -\frac{1}{6}\epsilon^{\mu\nu\rho\sigma\tau\theta}\epsilon^{\alpha\beta\kappa\lambda \delta\zeta}R_{\mu\nu\alpha\beta}R_{\rho\sigma\kappa\lambda}R_{\tau\theta\delta\zeta}\nn
&=\frac{64}{3} R_{\alpha}{}^{\gamma} R^{\alpha \beta} R_{\beta \gamma} - 16 R_{\alpha \beta} R^{\alpha \beta} R + \frac{4}{3} R^3 + 32 R^{\alpha \beta} R^{\gamma \delta} R_{\alpha \gamma \beta \delta} + 4 R R_{\alpha \beta \gamma \delta} R^{\alpha \beta \gamma \delta}\nn
&\quad - 32 R^{\alpha \beta} R_{\alpha}{}^{\gamma \delta \varepsilon} R_{\beta \gamma \delta \varepsilon} -  \frac{32}{3} R_{\alpha}{}^{\varepsilon}{}_{\gamma}{}^{\zeta} R^{\alpha \beta \gamma \delta} R_{\beta \varepsilon \delta \zeta} + \frac{16}{3} R_{\alpha \beta}{}^{\varepsilon \zeta} R^{\alpha \beta \gamma \delta} R_{\gamma \delta \varepsilon \zeta}\label{app:eq:GaussBonnet6D}\ ,
\end{align}
 we find
\begin{align}
&\quad\int\rd^{6}x\sqrt{-\tilde{g}}\, [\mathsf{a}_{3}(x)]\nn
&= \int\rd^{6}x\sqrt{-g}\,\Big(- \frac{1}{378} R_{\alpha}{}^{\zeta} R^{\alpha \beta} R_{\beta \zeta} + \frac{1}{1440} R^3 -  \frac{1}{1260} R^{\alpha \beta} R^{\zeta \delta} R_{\alpha \zeta \beta \delta}
  + \frac{1}{1440} R R_{\alpha \beta \zeta \delta} R^{\alpha \beta \zeta \delta}\nn
  &\quad + \frac{1}{504} R^{\alpha \beta} R_{\alpha}{}^{\zeta \delta \varepsilon} R_{\beta \zeta \delta \varepsilon} + \frac{1}{15120} R_{\alpha \beta}{}^{\varepsilon \zeta} R^{\alpha \beta \zeta \delta} R_{\zeta \delta \varepsilon \zeta} 
 + \frac{R_{\beta \zeta} R^{\beta \zeta} \nabla_{\alpha}\nabla^{\alpha}\gamma}{210 \gamma} -  \frac{R^2 \nabla_{\alpha}\nabla^{\alpha}\gamma}{96 \gamma} \nn
  &\quad-  \frac{R_{\beta \zeta \delta \varepsilon} R^{\beta \zeta \delta \varepsilon} \nabla_{\alpha}\nabla^{\alpha}\gamma}{224 \gamma} -  \frac{1}{336} \nabla_{\alpha}R \nabla^{\alpha}R
  + \frac{R \nabla_{\alpha}\gamma \nabla^{\alpha}R}{168 \gamma} + \frac{\nabla_{\alpha}\nabla_{\beta}\nabla^{\beta}\gamma \nabla^{\alpha}R}{840 \gamma} + \frac{R^{\beta \zeta} \nabla_{\alpha}R_{\beta \zeta} \nabla^{\alpha}\gamma}{420 \gamma}\nn
  &\quad + \frac{R_{\beta \zeta} R^{\beta \zeta} \nabla_{\alpha}\gamma \nabla^{\alpha}\gamma}{1260 \gamma^2} 
 -  \frac{R^2 \nabla_{\alpha}\gamma \nabla^{\alpha}\gamma}{288 \gamma^2} -  \frac{11 R_{\beta \zeta \delta \varepsilon} R^{\beta \zeta \delta \varepsilon} \nabla_{\alpha}\gamma \nabla^{\alpha}\gamma}{10080 \gamma^2} -  \frac{R \nabla_{\alpha}\nabla_{\beta}\nabla^{\beta}\gamma \nabla^{\alpha}\gamma}{840 \gamma^2}\nn
  &\quad
  -  \frac{R_{\beta \zeta} \nabla_{\alpha}\nabla^{\zeta}\nabla^{\beta}\gamma \nabla^{\alpha}\gamma}{210 \gamma^2} + \frac{253 R \nabla_{\alpha}\nabla^{\alpha}\gamma \nabla_{\beta}\nabla^{\beta}\gamma}{5040 \gamma^2} -  \frac{13 \nabla_{\alpha}\gamma \nabla^{\alpha}R \nabla_{\beta}\nabla^{\beta}\gamma}{210 \gamma^2}
  + \frac{211 R \nabla_{\alpha}\gamma \nabla^{\alpha}\gamma \nabla_{\beta}\nabla^{\beta}\gamma}{3360 \gamma^3}\nn
  &\quad + \frac{181 \nabla_{\alpha}\nabla_{\zeta}\nabla^{\zeta}\gamma \nabla^{\alpha}\gamma \nabla_{\beta}\nabla^{\beta}\gamma}{3360 \gamma^3}
  + \frac{5 \nabla^{\alpha}R \nabla_{\beta}\nabla^{\beta}\nabla_{\alpha}\gamma}{168 \gamma} -  \frac{5 R \nabla^{\alpha}\gamma \nabla_{\beta}\nabla^{\beta}\nabla_{\alpha}\gamma}{168 \gamma^2} \nn
  &\quad+ \frac{25 \nabla_{\alpha}\nabla^{\beta}\gamma \nabla^{\alpha}\gamma \nabla_{\beta}\nabla_{\zeta}\nabla^{\zeta}\gamma}{336 \gamma^3}
  + \frac{R_{\alpha \zeta} \nabla^{\alpha}\gamma \nabla_{\beta}\nabla^{\zeta}\nabla^{\beta}\gamma}{420 \gamma^2} -  \frac{5 R_{\alpha \beta} \nabla^{\alpha}R \nabla^{\beta}\gamma}{168 \gamma} + \frac{\nabla_{\alpha}\nabla_{\beta}\gamma \nabla^{\alpha}R \nabla^{\beta}\gamma}{420 \gamma^2}
\nn
  &\quad  -  \frac{17 R_{\alpha}{}^{\zeta} R_{\beta \zeta} \nabla^{\alpha}\gamma \nabla^{\beta}\gamma}{840 \gamma^2} + \frac{19 R_{\alpha \beta} R \nabla^{\alpha}\gamma \nabla^{\beta}\gamma}{560 \gamma^2} + \frac{R^{\zeta \delta} R_{\alpha \zeta \beta \delta} \nabla^{\alpha}\gamma \nabla^{\beta}\gamma}{840 \gamma^2}
  + \frac{11 R_{\alpha}{}^{\zeta \delta \varepsilon} R_{\beta \zeta \delta \varepsilon} \nabla^{\alpha}\gamma \nabla^{\beta}\gamma}{1680 \gamma^2}\nn
  &\quad + \frac{115 R_{\beta \zeta} \nabla_{\alpha}\nabla^{\zeta}\gamma \nabla^{\alpha}\gamma \nabla^{\beta}\gamma}{6048 \gamma^3} -  \frac{\nabla_{\alpha}\gamma \nabla^{\alpha}R \nabla_{\beta}\gamma \nabla^{\beta}\gamma}{280 \gamma^3}
  + \frac{R \nabla_{\alpha}\gamma \nabla^{\alpha}\gamma \nabla_{\beta}\gamma \nabla^{\beta}\gamma}{96 \gamma^4} -  \frac{19 R \nabla^{\alpha}\gamma \nabla_{\beta}\nabla_{\alpha}\gamma \nabla^{\beta}\gamma}{1680 \gamma^3} \nn
  &\quad-  \frac{349 \nabla_{\alpha}\nabla^{\zeta}\gamma \nabla^{\alpha}\gamma \nabla_{\beta}\nabla_{\zeta}\gamma \nabla^{\beta}\gamma}{10080 \gamma^4} 
 -  \frac{109 \nabla_{\alpha}\gamma \nabla^{\alpha}\gamma \nabla_{\beta}\nabla_{\zeta}\nabla^{\zeta}\gamma \nabla^{\beta}\gamma}{2240 \gamma^4} -  \frac{\nabla^{\alpha}\gamma \nabla_{\beta}\nabla_{\zeta}\nabla^{\zeta}\gamma \nabla^{\beta}\nabla_{\alpha}\gamma}{210 \gamma^3}\nn
  &\quad -  \frac{\nabla_{\beta}\nabla_{\zeta}\nabla^{\zeta}\gamma \nabla^{\beta}\nabla_{\alpha}\nabla^{\alpha}\gamma}{240 \gamma^2}
  + \frac{R_{\alpha}{}^{\zeta} R_{\beta \zeta} \nabla^{\beta}\nabla^{\alpha}\gamma}{90 \gamma} -  \frac{R_{\alpha \beta} R \nabla^{\beta}\nabla^{\alpha}\gamma}{504 \gamma} -  \frac{R^{\zeta \delta} R_{\alpha \zeta \beta \delta} \nabla^{\beta}\nabla^{\alpha}\gamma}{1260 \gamma}\nn
  &\quad -  \frac{11 R_{\alpha}{}^{\zeta \delta \varepsilon} R_{\beta \zeta \delta \varepsilon} \nabla^{\beta}\nabla^{\alpha}\gamma}{2520 \gamma} 
 + \frac{11 R \nabla_{\beta}\nabla_{\alpha}\gamma \nabla^{\beta}\nabla^{\alpha}\gamma}{5040 \gamma^2} + \frac{R^{\beta \zeta} \nabla^{\alpha}\gamma \nabla_{\zeta}R_{\alpha \beta}}{420 \gamma} -  \frac{R_{\alpha}{}^{\beta} \nabla^{\alpha}\gamma \nabla_{\zeta}R_{\beta}{}^{\zeta}}{420 \gamma} 
\nn
  &\quad + \frac{\nabla_{\alpha}\nabla^{\beta}\gamma \nabla^{\alpha}\gamma \nabla_{\zeta}R_{\beta}{}^{\zeta}}{420 \gamma^2} -  \frac{\nabla_{\alpha}\gamma \nabla^{\alpha}\gamma \nabla^{\beta}\gamma \nabla_{\zeta}R_{\beta}{}^{\zeta}}{140 \gamma^3} + \frac{\nabla^{\alpha}\gamma \nabla^{\beta}\nabla_{\alpha}\gamma \nabla_{\zeta}R_{\beta}{}^{\zeta}}{420 \gamma^2} 
 + \frac{\nabla_{\alpha}\nabla^{\beta}\gamma \nabla^{\alpha}\gamma \nabla_{\zeta}\nabla_{\beta}\nabla^{\zeta}\gamma}{336 \gamma^3}\nn
  &\quad + \frac{\nabla_{\alpha}\gamma \nabla^{\alpha}\gamma \nabla^{\beta}\gamma \nabla_{\zeta}\nabla_{\beta}\nabla^{\zeta}\gamma}{140 \gamma^4} -  \frac{13 \nabla^{\alpha}\gamma \nabla^{\beta}\nabla_{\alpha}\gamma \nabla_{\zeta}\nabla_{\beta}\nabla^{\zeta}\gamma}{1680 \gamma^3}
  -  \frac{421 \nabla_{\alpha}\nabla^{\alpha}\gamma \nabla_{\beta}\nabla^{\beta}\gamma \nabla_{\zeta}\nabla^{\zeta}\gamma}{5040 \gamma^3} \nn
  &\quad-  \frac{353 \nabla_{\alpha}\gamma \nabla^{\alpha}\gamma \nabla_{\beta}\nabla^{\beta}\gamma \nabla_{\zeta}\nabla^{\zeta}\gamma}{1120 \gamma^4}
  -  \frac{43 R_{\alpha \beta} \nabla^{\alpha}\gamma \nabla^{\beta}\gamma \nabla_{\zeta}\nabla^{\zeta}\gamma}{160 \gamma^3} -  \frac{293 \nabla_{\alpha}\gamma \nabla^{\alpha}\gamma \nabla_{\beta}\gamma \nabla^{\beta}\gamma \nabla_{\zeta}\nabla^{\zeta}\gamma}{6720 \gamma^5} 
\nn
  &\quad + \frac{19 \nabla^{\alpha}\gamma \nabla_{\beta}\nabla_{\alpha}\gamma \nabla^{\beta}\gamma \nabla_{\zeta}\nabla^{\zeta}\gamma}{560 \gamma^4} + \frac{25 \nabla^{\alpha}\gamma \nabla_{\beta}\nabla^{\beta}\gamma \nabla_{\zeta}\nabla^{\zeta}\nabla_{\alpha}\gamma}{96 \gamma^3} 
 -  \frac{13 \nabla_{\alpha}\nabla^{\beta}\gamma \nabla^{\alpha}\gamma \nabla_{\zeta}\nabla^{\zeta}\nabla_{\beta}\gamma}{1680 \gamma^3}\nn
  &\quad + \frac{141 \nabla_{\alpha}\gamma \nabla^{\alpha}\gamma \nabla^{\beta}\gamma \nabla_{\zeta}\nabla^{\zeta}\nabla_{\beta}\gamma}{2240 \gamma^4} 
 -  \frac{\nabla^{\alpha}\gamma \nabla^{\beta}\nabla_{\alpha}\gamma \nabla_{\zeta}\nabla^{\zeta}\nabla_{\beta}\gamma}{14 \gamma^3} -  \frac{25 \nabla^{\beta}\nabla_{\alpha}\nabla^{\alpha}\gamma \nabla_{\zeta}\nabla^{\zeta}\nabla_{\beta}\gamma}{336 \gamma^2}\nn
  &\quad + \frac{R_{\alpha \beta} \nabla^{\alpha}\gamma \nabla_{\zeta}\nabla^{\zeta}\nabla^{\beta}\gamma}{420 \gamma^2}
  -  \frac{1}{840} \nabla_{\zeta}R_{\alpha \beta} \nabla^{\zeta}R^{\alpha \beta} -  \frac{291 R_{\beta \zeta} \nabla_{\alpha}\gamma \nabla^{\alpha}\gamma \nabla^{\beta}\gamma \nabla^{\zeta}\gamma}{2240 \gamma^4} + \frac{\nabla^{\alpha}\gamma \nabla^{\beta}\gamma \nabla_{\zeta}R_{\alpha \beta} \nabla^{\zeta}\gamma}{35 \gamma^3}\nn
  &\quad
  -  \frac{179 \nabla_{\alpha}\gamma \nabla^{\alpha}\gamma \nabla_{\beta}\gamma \nabla^{\beta}\gamma \nabla_{\zeta}\gamma \nabla^{\zeta}\gamma}{1344 \gamma^6} + \frac{179 \nabla_{\alpha}\gamma \nabla^{\alpha}\gamma \nabla^{\beta}\gamma \nabla_{\zeta}\nabla_{\beta}\gamma \nabla^{\zeta}\gamma}{560 \gamma^5} 
 -  \frac{2 \nabla^{\alpha}\gamma \nabla^{\beta}\gamma \nabla_{\zeta}\nabla_{\beta}\nabla_{\alpha}\gamma \nabla^{\zeta}\gamma}{35 \gamma^4}\nn
  &\quad + \frac{839 R_{\beta \zeta} \nabla^{\alpha}\gamma \nabla^{\beta}\gamma \nabla^{\zeta}\nabla_{\alpha}\gamma}{6048 \gamma^3}
  + \frac{187 \nabla^{\alpha}\gamma \nabla_{\beta}\nabla_{\zeta}\gamma \nabla^{\beta}\gamma \nabla^{\zeta}\nabla_{\alpha}\gamma}{5040 \gamma^4} -  \frac{19 R_{\beta \zeta} \nabla^{\beta}\nabla^{\alpha}\gamma \nabla^{\zeta}\nabla_{\alpha}\gamma}{2160 \gamma^2}
\nn
  &\quad  -  \frac{217 \nabla^{\alpha}\gamma \nabla^{\beta}\gamma \nabla_{\zeta}\nabla_{\beta}\gamma \nabla^{\zeta}\nabla_{\alpha}\gamma}{1440 \gamma^4} + \frac{\nabla^{\beta}\nabla^{\alpha}\gamma \nabla_{\zeta}\nabla_{\beta}\gamma \nabla^{\zeta}\nabla_{\alpha}\gamma}{72 \gamma^3}
  -  \frac{R_{\beta \zeta} \nabla^{\alpha}\gamma \nabla^{\zeta}\nabla_{\alpha}\nabla^{\beta}\gamma}{420 \gamma^2} \nn
  &\quad-  \frac{143 R_{\alpha \zeta} \nabla^{\beta}\nabla^{\alpha}\gamma \nabla^{\zeta}\nabla_{\beta}\gamma}{15120 \gamma^2} + \frac{25 R_{\alpha \zeta} \nabla^{\alpha}\gamma \nabla^{\zeta}\nabla_{\beta}\nabla^{\beta}\gamma}{336 \gamma^2} -  \frac{R_{\beta \zeta} \nabla_{\alpha}\nabla^{\alpha}\gamma \nabla^{\zeta}\nabla^{\beta}\gamma}{1260 \gamma^2} -  \frac{\nabla_{\alpha}R_{\beta \zeta} \nabla^{\alpha}\gamma \nabla^{\zeta}\nabla^{\beta}\gamma}{105 \gamma^2}\nn
  &\quad + \frac{R_{\beta \zeta} \nabla_{\alpha}\gamma \nabla^{\alpha}\gamma \nabla^{\zeta}\nabla^{\beta}\gamma}{80 \gamma^3} -  \frac{\nabla_{\alpha}\nabla_{\beta}\nabla_{\zeta}\gamma \nabla^{\alpha}\gamma \nabla^{\zeta}\nabla^{\beta}\gamma}{84 \gamma^3} + \frac{13 \nabla_{\alpha}\nabla_{\zeta}\nabla_{\beta}\gamma \nabla^{\alpha}\gamma \nabla^{\zeta}\nabla^{\beta}\gamma}{420 \gamma^3} \nn
  &\quad-  \frac{\nabla^{\alpha}\gamma \nabla_{\beta}R_{\alpha \zeta} \nabla^{\zeta}\nabla^{\beta}\gamma}{420 \gamma^2} -  \frac{\nabla^{\alpha}\gamma \nabla_{\beta}\nabla_{\alpha}\nabla_{\zeta}\gamma \nabla^{\zeta}\nabla^{\beta}\gamma}{336 \gamma^3} + \frac{\nabla_{\alpha}\gamma \nabla^{\alpha}\gamma \nabla_{\beta}\nabla_{\zeta}\gamma \nabla^{\zeta}\nabla^{\beta}\gamma}{160 \gamma^4} \nn
  &\quad+ \frac{13 \nabla^{\alpha}\gamma \nabla_{\beta}\nabla_{\zeta}\nabla_{\alpha}\gamma \nabla^{\zeta}\nabla^{\beta}\gamma}{1680 \gamma^3} -  \frac{\nabla^{\alpha}\gamma \nabla_{\zeta}R_{\alpha \beta} \nabla^{\zeta}\nabla^{\beta}\gamma}{60 \gamma^2} + \frac{37 \nabla^{\alpha}\gamma \nabla_{\zeta}\nabla_{\alpha}\nabla_{\beta}\gamma \nabla^{\zeta}\nabla^{\beta}\gamma}{1680 \gamma^3}\nn
  &\quad -  \frac{\nabla_{\alpha}\nabla^{\alpha}\gamma \nabla_{\zeta}\nabla_{\beta}\gamma \nabla^{\zeta}\nabla^{\beta}\gamma}{560 \gamma^3} -  \frac{89 \nabla_{\alpha}\gamma \nabla^{\alpha}\gamma \nabla_{\zeta}\nabla_{\beta}\gamma \nabla^{\zeta}\nabla^{\beta}\gamma}{1680 \gamma^4} + \frac{19 \nabla^{\alpha}\gamma \nabla_{\zeta}\nabla_{\beta}\nabla_{\alpha}\gamma \nabla^{\zeta}\nabla^{\beta}\gamma}{1680 \gamma^3}\nn
  &\quad -  \frac{R_{\beta \zeta} \nabla^{\alpha}\gamma \nabla^{\zeta}\nabla^{\beta}\nabla_{\alpha}\gamma}{420 \gamma^2} + \frac{\nabla_{\zeta}R_{\alpha \beta} \nabla^{\zeta}\nabla^{\beta}\nabla^{\alpha}\gamma}{210 \gamma} + \frac{\nabla_{\zeta}\nabla_{\alpha}\nabla_{\beta}\gamma \nabla^{\zeta}\nabla^{\beta}\nabla^{\alpha}\gamma}{336 \gamma^2} -  \frac{13 \nabla_{\zeta}\nabla_{\beta}\nabla_{\alpha}\gamma \nabla^{\zeta}\nabla^{\beta}\nabla^{\alpha}\gamma}{1680 \gamma^2} \nn
  &\quad-  \frac{13 R_{\alpha \zeta \beta \delta} \nabla^{\alpha}\gamma \nabla^{\beta}\gamma \nabla^{\delta}\nabla^{\zeta}\gamma}{840 \gamma^3} + \frac{R_{\alpha \beta \zeta \delta} \nabla^{\beta}\nabla^{\alpha}\gamma \nabla^{\delta}\nabla^{\zeta}\gamma}{2520 \gamma^2} -  \frac{R_{\alpha \zeta \beta \delta} \nabla^{\beta}\nabla^{\alpha}\gamma \nabla^{\delta}\nabla^{\zeta}\gamma}{945 \gamma^2} \nn
  &\quad+ \frac{47 R_{\alpha \delta \beta \zeta} \nabla^{\beta}\nabla^{\alpha}\gamma \nabla^{\delta}\nabla^{\zeta}\gamma}{7560 \gamma^2}\Big) \label{app:eq:DBINaiveLogDivergences6D}\ ,
\end{align}
which is not DBI-invariant. However,  after adding total derivatives and using the same on-shell conditions employed in simplifying \eqref{DBINaiveLogDivergences4D},  \eqref{app:eq:DBINaiveLogDivergences6D} can be dramatically simplified to the form
\begin{align}
\,\sqrt{-\tilde{g}}\,[\mathsf{a}_{3}(x)]&= \sqrt{-g}\,\Big(- \frac{187}{9450} \langle K^{6}\rangle + \frac{113}{14175} \langle K^{3}\rangle^{2} -  \frac{5}{1512}\langle K^{2}\rangle^{3}\nn
&\quad + \frac{223}{2100} \langle K^{2}\rangle \nabla_{\varepsilon}K_{\gamma \delta} \nabla^{\varepsilon}K^{\gamma \delta}+{\rm total \ derivatives}\Big)\ ,\label{app:eq:DBICovariantLogDivergences6D}
\end{align}
in the condensed notation of App.~\ref{app:Conventions}. The total derivatives added to the action in order to simplify were $\mathcal{L}_{\rm TD}=\nabla_{\alpha}J^{\alpha}+\frac{10729}{108864000}\mathcal{L}^{(6)}_{\rm GB}[g]$ where
\begin{align}
J^{\alpha}&=\frac{23}{2100} K_{\beta}{}^{\delta} K^{\beta \gamma} K_{\gamma}{}^{\varepsilon} \nabla^{\alpha}K_{\delta \varepsilon} + \frac{223}{2100} K_{\beta \gamma} K^{\beta \gamma} K^{\delta \varepsilon} \nabla^{\alpha}K_{\delta \varepsilon} + \frac{619}{1814400} K^{\alpha \beta} K_{\gamma}{}^{\varepsilon} K^{\gamma \delta} \nabla_{\beta}K_{\delta \varepsilon}\nn
&\quad -  \frac{44 K^{\alpha \gamma} K_{\beta}{}^{\delta} K_{\gamma}{}^{\varepsilon} K_{\delta}{}^{\zeta} K_{\varepsilon \zeta} \nabla^{\beta}\phi}{2835 \gamma} + \frac{115 K^{\alpha}{}_{\beta} K_{\gamma}{}^{\varepsilon} K^{\gamma \delta} K_{\delta}{}^{\zeta} K_{\varepsilon \zeta} \nabla^{\beta}\phi}{12096 \gamma} -  \frac{283 K^{\alpha \gamma} K_{\beta \gamma} K_{\delta}{}^{\zeta} K^{\delta \varepsilon} K_{\varepsilon \zeta} \nabla^{\beta}\phi}{45360 \gamma}\nn
&\quad + \frac{K^{\alpha \gamma} K_{\beta}{}^{\delta} K_{\gamma \delta} K_{\varepsilon \zeta} K^{\varepsilon \zeta} \nabla^{\beta}\phi}{1890 \gamma} -  \frac{19853 K^{\alpha}{}_{\beta} K_{\gamma \delta} K^{\gamma \delta} K_{\varepsilon \zeta} K^{\varepsilon \zeta} \nabla^{\beta}\phi}{362880 \gamma} -  \frac{701 K_{\gamma}{}^{\varepsilon} K^{\gamma \delta} \nabla^{\alpha}\nabla_{\beta}K_{\delta \varepsilon} \nabla^{\beta}\phi}{72576 \gamma}\nn
&\quad -  \frac{1}{100} K^{\alpha \beta} K_{\beta}{}^{\gamma} K^{\delta \varepsilon} \nabla_{\gamma}K_{\delta \varepsilon} + \frac{7 K_{\beta}{}^{\gamma} \nabla^{\alpha}K^{\delta \varepsilon} \nabla^{\beta}\phi \nabla_{\gamma}K_{\delta \varepsilon}}{25920 \gamma} + \frac{913 K^{\alpha \gamma} \nabla_{\beta}K^{\delta \varepsilon} \nabla^{\beta}\phi \nabla_{\gamma}K_{\delta \varepsilon}}{181440 \gamma} \nn
&\quad+ \frac{143 K_{\beta}{}^{\delta} K_{\varepsilon \zeta} K^{\varepsilon \zeta} \nabla^{\alpha}K_{\gamma \delta} \nabla^{\beta}\phi \nabla^{\gamma}\phi}{18144 \gamma^2} + \frac{K_{\beta}{}^{\delta} K_{\delta}{}^{\varepsilon} K_{\varepsilon}{}^{\zeta} \nabla^{\alpha}K_{\gamma \zeta} \nabla^{\beta}\phi \nabla^{\gamma}\phi}{42 \gamma^2}\nn
&\quad + \frac{4369 K_{\beta}{}^{\delta} K_{\gamma}{}^{\varepsilon} K_{\delta}{}^{\zeta} \nabla^{\alpha}K_{\varepsilon \zeta} \nabla^{\beta}\phi \nabla^{\gamma}\phi}{181440 \gamma^2} -  \frac{7 K_{\beta}{}^{\delta} K_{\gamma \delta} K^{\varepsilon \zeta} \nabla^{\alpha}K_{\varepsilon \zeta} \nabla^{\beta}\phi \nabla^{\gamma}\phi}{51840 \gamma^2}\nn
&\quad + \frac{\nabla^{\alpha}\nabla_{\gamma}K_{\delta \varepsilon} \nabla_{\beta}K^{\delta \varepsilon} \nabla^{\beta}\phi \nabla^{\gamma}\phi}{210 \gamma^2} -  \frac{149 K^{\alpha \delta} K_{\beta}{}^{\varepsilon} K_{\varepsilon}{}^{\zeta} \nabla^{\beta}\phi \nabla_{\gamma}K_{\delta \zeta} \nabla^{\gamma}\phi}{45360 \gamma^2} \nn
&\quad-  \frac{K^{\alpha \delta} K_{\beta}{}^{\varepsilon} K_{\delta}{}^{\zeta} \nabla^{\beta}\phi \nabla_{\gamma}K_{\varepsilon \zeta} \nabla^{\gamma}\phi}{210 \gamma^2} -  \frac{1777 K^{\alpha}{}_{\beta} K_{\delta}{}^{\zeta} K^{\delta \varepsilon} \nabla^{\beta}\phi \nabla_{\gamma}K_{\varepsilon \zeta} \nabla^{\gamma}\phi}{362880 \gamma^2} \nn
&\quad+ \frac{149 K^{\alpha \delta} K_{\beta \delta} K^{\varepsilon \zeta} \nabla^{\beta}\phi \nabla_{\gamma}K_{\varepsilon \zeta} \nabla^{\gamma}\phi}{45360 \gamma^2} -  \frac{7 K^{\gamma \delta} \nabla^{\alpha}K_{\beta}{}^{\varepsilon} \nabla^{\beta}\phi \nabla_{\delta}K_{\gamma \varepsilon}}{25920 \gamma} + \frac{K^{\alpha \delta} K_{\beta}{}^{\varepsilon} K_{\gamma}{}^{\zeta} \nabla^{\beta}\phi \nabla^{\gamma}\phi \nabla_{\delta}K_{\varepsilon \zeta}}{210 \gamma^2} \nn
&\quad-  \frac{K^{\alpha}{}_{\beta} K_{\gamma}{}^{\delta} K^{\varepsilon \zeta} \nabla^{\beta}\phi \nabla^{\gamma}\phi \nabla_{\delta}K_{\varepsilon \zeta}}{42 \gamma^2} -  \frac{323 K^{\alpha \varepsilon} K_{\beta}{}^{\zeta} K_{\gamma \zeta} K_{\delta}{}^{\eta} K_{\varepsilon \eta} \nabla^{\beta}\phi \nabla^{\gamma}\phi \nabla^{\delta}\phi}{10080 \gamma^3}\nn
&\quad + \frac{7 K^{\alpha}{}_{\beta} K_{\gamma}{}^{\varepsilon} K_{\delta}{}^{\zeta} K_{\varepsilon}{}^{\eta} K_{\zeta \eta} \nabla^{\beta}\phi \nabla^{\gamma}\phi \nabla^{\delta}\phi}{25920 \gamma^3} -  \frac{499 K^{\alpha}{}_{\beta} K_{\gamma}{}^{\varepsilon} K_{\delta \varepsilon} K_{\zeta \eta} K^{\zeta \eta} \nabla^{\beta}\phi \nabla^{\gamma}\phi \nabla^{\delta}\phi}{90720 \gamma^3} \nn
&\quad-  \frac{2 K_{\beta}{}^{\varepsilon} K_{\gamma}{}^{\zeta} \nabla^{\alpha}\nabla_{\delta}K_{\varepsilon \zeta} \nabla^{\beta}\phi \nabla^{\gamma}\phi \nabla^{\delta}\phi}{105 \gamma^3} -  \frac{83 K_{\beta}{}^{\varepsilon} \nabla^{\alpha}K_{\gamma}{}^{\zeta} \nabla^{\beta}\phi \nabla^{\gamma}\phi \nabla_{\delta}K_{\varepsilon \zeta} \nabla^{\delta}\phi}{6480 \gamma^3} \nn
&\quad+ \frac{149 K^{\alpha}{}_{\beta} \nabla^{\beta}\phi \nabla_{\gamma}K^{\varepsilon \zeta} \nabla^{\gamma}\phi \nabla_{\delta}K_{\varepsilon \zeta} \nabla^{\delta}\phi}{90720 \gamma^3} -  \frac{7 K^{\alpha}{}_{\beta} \nabla^{\beta}\phi \nabla_{\varepsilon}K_{\gamma \delta} \nabla^{\varepsilon}K^{\gamma \delta}}{51840 \gamma} \nn
&\quad+ \frac{3173 K_{\beta}{}^{\zeta} K_{\gamma \zeta} K_{\delta}{}^{\eta} \nabla^{\alpha}K_{\varepsilon \eta} \nabla^{\beta}\phi \nabla^{\gamma}\phi \nabla^{\delta}\phi \nabla^{\varepsilon}\phi}{60480 \gamma^4} -  \frac{3173 K^{\alpha}{}_{\beta} K_{\gamma}{}^{\eta} K_{\delta \eta} K_{\varepsilon}{}^{\theta} K_{\zeta \theta} \nabla^{\beta}\phi \nabla^{\gamma}\phi \nabla^{\delta}\phi \nabla^{\varepsilon}\phi \nabla^{\zeta}\phi}{120960 \gamma^5}
\end{align}
The total derivative current above can also be phrased in terms of covariant derivative of $\ln\gamma$, as in \eqref{DBI4DNaiveTotalDerivativeTerm}.

\section{Divergences for $\Mcal_{D+1}=\mathsf{M}_{4}\times S_{N}$\label{app:MTimesSNDivergences}}

In this appendix, we collect the lengthy expression which contains the one-loop, logarithmic divergences for branes embedded in the product manifold $\Mcal_{D+1}=\mathsf{M}_{4}\times S_{N}$ where $\mathsf{M}_{4}$ is $d+1=4$ dimensional Minkowski space.  This computation corresponds to the scenario in Sec.~\ref{sec:ProductManifolds} and the relevant geometry for $S_{N}$ is given in \eqref{NSphereGeometry}.

The result is that the trace of the second Seeley-DeWitt coefficient which determines the $d+1=4$ dimensional logarithmic divergence via \eqref{app:eq:LogDivergencesFromHeatKernel} is:
\begin{align}
\tr[\mathsf{a}_{2}]&=
\frac{2 \phi_{\alpha}{}^{f} \phi^{\alpha b} \phi_{\beta f} \phi^{\beta}{}_{b}}{3 L^4} + \frac{(-7 + 3 N) \phi_{\alpha b} \phi^{\alpha b} \phi_{\beta f} \phi^{\beta f}}{6 L^4} -  \frac{4 \phi_{\alpha}{}^{f} \phi^{\alpha b} \phi_{\beta}{}^{g} \phi^{\beta}{}_{b} \phi_{\gamma g} \phi^{\gamma}{}_{f}}{3 L^4} \nn
&\quad-  \frac{(-15 + N) \phi_{\alpha b} \phi^{\alpha b} \phi_{\beta}{}^{g} \phi^{\beta f} \phi_{\gamma g} \phi^{\gamma}{}_{f}}{6 L^4} + \frac{(-7 + N) \phi_{\alpha b} \phi^{\alpha b} \phi_{\beta f} \phi^{\beta f} \phi_{\gamma g} \phi^{\gamma g}}{6 L^4} \nn
&\quad-  \frac{(-40 + N) \phi_{\alpha}{}^{f} \phi^{\alpha b} \phi_{\beta}{}^{g} \phi^{\beta}{}_{b} \phi_{\gamma}{}^{h} \phi^{\gamma}{}_{f} \phi_{\delta h} \phi^{\delta}{}_{g}}{60 L^4} + \frac{(-90 + N) \phi_{\alpha b} \phi^{\alpha b} \phi_{\beta}{}^{g} \phi^{\beta f} \phi_{\gamma}{}^{h} \phi^{\gamma}{}_{f} \phi_{\delta h} \phi^{\delta}{}_{g}}{90 L^4} \nn
&\quad+ \frac{(-40 + 3 N) \phi_{\alpha}{}^{f} \phi^{\alpha b} \phi_{\beta f} \phi^{\beta}{}_{b} \phi_{\gamma}{}^{h} \phi^{\gamma g} \phi_{\delta h} \phi^{\delta}{}_{g}}{120 L^4} -  \frac{(-25 + N) \phi_{\alpha b} \phi^{\alpha b} \phi_{\beta f} \phi^{\beta f} \phi_{\gamma}{}^{h} \phi^{\gamma g} \phi_{\delta h} \phi^{\delta}{}_{g}}{30 L^4}\nn
&\quad + \frac{(-12 + N) \phi_{\alpha b} \phi^{\alpha b} \phi_{\beta f} \phi^{\beta f} \phi_{\gamma g} \phi^{\gamma g} \phi_{\delta h} \phi^{\delta h}}{72 L^4} + \frac{\phi^{\alpha b} \phi^{\beta f} \phi_{\alpha}{}^{\gamma}{}_{f} \phi_{\beta \gamma b}}{3 L^2} -  \frac{\phi^{\alpha b} \phi^{\beta f} \phi_{\alpha}{}^{\gamma}{}_{b} \phi_{\beta \gamma f}}{3 L^2}\nn
&\quad -  \frac{N \phi^{\alpha b} \phi^{\beta}{}_{b} \phi^{\gamma f} \phi^{\delta}{}_{f} \phi_{\alpha \gamma}{}^{g} \phi_{\beta \delta g}}{45 L^2} -  \frac{\phi_{\alpha}{}^{f} \phi^{\alpha b} \phi_{\beta \gamma f} \phi^{\beta \gamma}{}_{b}}{L^2} -  \frac{(-7 + N) \phi_{\alpha b} \phi^{\alpha b} \phi_{\beta \gamma f} \phi^{\beta \gamma f}}{6 L^2}\nn
&\quad + (\tfrac{1}{6} -  \tfrac{1}{90} N) \phi_{\alpha}{}^{\gamma f} \phi^{\alpha \beta b} \phi_{\beta}{}^{\delta}{}_{f} \phi_{\gamma \delta b} + (- \tfrac{1}{6} -  \tfrac{1}{180} N) \phi_{\alpha}{}^{\gamma}{}_{b} \phi^{\alpha \beta b} \phi_{\beta}{}^{\delta f} \phi_{\gamma \delta f} -  \frac{2 \phi_{\alpha}{}^{f} \phi^{\alpha b} \phi^{\beta}{}_{b} \phi^{\gamma g} \phi_{\beta}{}^{\delta}{}_{g} \phi_{\gamma \delta f}}{3 L^2} \nn
&\quad+ \frac{N \phi^{\alpha b} \phi^{\beta}{}_{b} \phi^{\gamma f} \phi^{\delta}{}_{f} \phi_{\alpha \beta}{}^{g} \phi_{\gamma \delta g}}{45 L^2} + \frac{2 \phi_{\alpha}{}^{f} \phi^{\alpha b} \phi^{\beta}{}_{b} \phi^{\gamma g} \phi_{\beta}{}^{\delta}{}_{f} \phi_{\gamma \delta g}}{3 L^2} -  \frac{N \phi_{\alpha}{}^{f} \phi^{\alpha b} \phi^{\beta}{}_{b} \phi^{\gamma}{}_{f} \phi_{\beta}{}^{\delta g} \phi_{\gamma \delta g}}{90 L^2} \nn
&\quad+ \frac{N \phi_{\alpha b} \phi^{\alpha b} \phi^{\beta f} \phi^{\gamma}{}_{f} \phi_{\beta}{}^{\delta g} \phi_{\gamma \delta g}}{90 L^2} + \frac{N \phi_{\alpha}{}^{f} \phi^{\alpha b} \phi^{\beta g} \phi^{\gamma}{}_{g} \phi^{\delta h} \phi^{\varepsilon}{}_{h} \phi_{\beta \delta b} \phi_{\gamma \varepsilon f}}{45 L^2}\nn
&\quad + \frac{N \phi_{\alpha}{}^{f} \phi^{\alpha b} \phi^{\beta}{}_{b} \phi^{\gamma}{}_{f} \phi_{\delta}{}^{h} \phi^{\delta g} \phi_{\beta}{}^{\varepsilon}{}_{g} \phi_{\gamma \varepsilon h}}{90 L^2} + \tfrac{1}{90} (45 + N) \phi_{\alpha \beta}{}^{f} \phi^{\alpha \beta b} \phi_{\gamma \delta f} \phi^{\gamma \delta}{}_{b}\nn
&\quad + \frac{2 \phi_{\alpha}{}^{f} \phi^{\alpha b} \phi_{\beta}{}^{g} \phi^{\beta}{}_{b} \phi_{\gamma \delta g} \phi^{\gamma \delta}{}_{f}}{L^2} + \frac{(-13 + N) \phi_{\alpha b} \phi^{\alpha b} \phi_{\beta}{}^{g} \phi^{\beta f} \phi_{\gamma \delta g} \phi^{\gamma \delta}{}_{f}}{6 L^2} \nn
&\quad+ \tfrac{1}{72} (-12 + N) \phi_{\alpha \beta b} \phi^{\alpha \beta b} \phi_{\gamma \delta f} \phi^{\gamma \delta f} + \frac{(-12 + N) \phi_{\alpha}{}^{f} \phi^{\alpha b} \phi_{\beta f} \phi^{\beta}{}_{b} \phi_{\gamma \delta g} \phi^{\gamma \delta g}}{36 L^2} \nn
&\quad-  \frac{(-12 + N) \phi_{\alpha b} \phi^{\alpha b} \phi_{\beta f} \phi^{\beta f} \phi_{\gamma \delta g} \phi^{\gamma \delta g}}{36 L^2} -  \frac{N \phi_{\alpha}{}^{f} \phi^{\alpha b} \phi^{\beta g} \phi^{\gamma}{}_{g} \phi^{\delta h} \phi^{\varepsilon}{}_{h} \phi_{\beta \gamma b} \phi_{\delta \varepsilon f}}{45 L^2} \nn
&\quad+ \frac{\phi_{\alpha}{}^{f} \phi^{\alpha b} \phi^{\beta}{}_{b} \phi_{\gamma}{}^{h} \phi^{\gamma g} \phi^{\delta}{}_{g} \phi_{\beta}{}^{\varepsilon}{}_{h} \phi_{\delta \varepsilon f}}{3 L^2} + \tfrac{1}{45} (-15 + N) \phi_{\alpha}{}^{f} \phi^{\alpha b} \phi_{\beta}{}^{\delta g} \phi^{\beta \gamma}{}_{b} \phi_{\gamma}{}^{\varepsilon}{}_{g} \phi_{\delta \varepsilon f}\nn
&\quad -  \frac{N \phi_{\alpha b} \phi^{\alpha b} \phi_{\beta}{}^{g} \phi^{\beta f} \phi^{\gamma h} \phi^{\delta}{}_{h} \phi_{\gamma}{}^{\varepsilon}{}_{f} \phi_{\delta \varepsilon g}}{90 L^2} + \tfrac{1}{90} (30 + N) \phi_{\alpha}{}^{f} \phi^{\alpha b} \phi_{\beta}{}^{\delta}{}_{f} \phi^{\beta \gamma}{}_{b} \phi_{\gamma}{}^{\varepsilon g} \phi_{\delta \varepsilon g}\nn
&\quad -  \frac{\phi_{\alpha}{}^{f} \phi^{\alpha b} \phi^{\beta}{}_{b} \phi_{\gamma}{}^{h} \phi^{\gamma g} \phi^{\delta}{}_{g} \phi_{\beta}{}^{\varepsilon}{}_{f} \phi_{\delta \varepsilon h}}{3 L^2} + (-1 -  \tfrac{1}{45} N) \phi_{\alpha}{}^{f} \phi^{\alpha b} \phi_{\beta \gamma}{}^{g} \phi^{\beta \gamma}{}_{b} \phi_{\delta \varepsilon g} \phi^{\delta \varepsilon}{}_{f}\nn
&\quad -  \frac{\phi_{\alpha}{}^{f} \phi^{\alpha b} \phi_{\beta}{}^{g} \phi^{\beta}{}_{b} \phi_{\gamma}{}^{h} \phi^{\gamma}{}_{f} \phi_{\delta \varepsilon h} \phi^{\delta \varepsilon}{}_{g}}{L^2} + \frac{\phi_{\alpha b} \phi^{\alpha b} \phi_{\beta}{}^{g} \phi^{\beta f} \phi_{\gamma}{}^{h} \phi^{\gamma}{}_{f} \phi_{\delta \varepsilon h} \phi^{\delta \varepsilon}{}_{g}}{L^2}\nn
&\quad -  \frac{(-12 + N) \phi_{\alpha}{}^{f} \phi^{\alpha b} \phi_{\beta f} \phi^{\beta}{}_{b} \phi_{\gamma}{}^{h} \phi^{\gamma g} \phi_{\delta \varepsilon h} \phi^{\delta \varepsilon}{}_{g}}{36 L^2} + \frac{(-12 + N) \phi_{\alpha b} \phi^{\alpha b} \phi_{\beta f} \phi^{\beta f} \phi_{\gamma}{}^{h} \phi^{\gamma g} \phi_{\delta \varepsilon h} \phi^{\delta \varepsilon}{}_{g}}{36 L^2}\nn
&\quad + (\tfrac{1}{3} -  \tfrac{1}{36} N) \phi_{\alpha}{}^{f} \phi^{\alpha b} \phi_{\beta \gamma f} \phi^{\beta \gamma}{}_{b} \phi_{\delta \varepsilon g} \phi^{\delta \varepsilon g} + (\tfrac{1}{6} -  \tfrac{1}{90} N) \phi_{\alpha}{}^{f} \phi^{\alpha b} \phi_{\beta}{}^{h} \phi^{\beta g} \phi_{\gamma}{}^{\varepsilon}{}_{g} \phi^{\gamma \delta}{}_{b} \phi_{\delta}{}^{\zeta}{}_{h} \phi_{\varepsilon \zeta f}\nn
&\quad + (- \tfrac{1}{6} -  \tfrac{1}{180} N) \phi_{\alpha}{}^{f} \phi^{\alpha b} \phi_{\beta}{}^{h} \phi^{\beta g} \phi_{\gamma}{}^{\varepsilon}{}_{f} \phi^{\gamma \delta}{}_{b} \phi_{\delta}{}^{\zeta}{}_{g} \phi_{\varepsilon \zeta h} + \tfrac{1}{90} (45 + N) \phi_{\alpha}{}^{f} \phi^{\alpha b} \phi_{\beta}{}^{h} \phi^{\beta g} \phi_{\gamma \delta g} \phi^{\gamma \delta}{}_{b} \phi_{\varepsilon \zeta h} \phi^{\varepsilon \zeta}{}_{f} \nn
&\quad+ \tfrac{1}{72} (-12 + N) \phi_{\alpha}{}^{f} \phi^{\alpha b} \phi_{\beta}{}^{h} \phi^{\beta g} \phi_{\gamma \delta f} \phi^{\gamma \delta}{}_{b} \phi_{\varepsilon \zeta h} \phi^{\varepsilon \zeta}{}_{g}\nn
&\quad +{\rm total-derivatives}\label{app:eq:a2NSphereProductSpace}
\end{align}
where for brevity we defined
\begin{align}
\phi_{\mu}{}^{a}\equiv\partial_{\mu}\phi^{a}\ , \quad \phi_{\mu\nu}{}^{a}\equiv \dcal_{\mu}\partial_{\nu}\phi^{a}\ ,
\end{align}
with $\dcal$ defined in \eqref{dcalDef} and all Greek and Latin indices were raised and lowered\footnote{ $\dcal_{\mu}$ is not compatible with $g_{\mu\nu}$ and $\partial_{\mu}$ is not compatible with $\g_{ab}$ and our raising and lowering conventions correspond to putting all metric factors outside of all derivatives.  Index placement is simply a form of shorthand here.  For instance, $\phi_{\mu}{}^{\nu }{}_{a}$ stands for $g^{\nu\alpha}\g_{ab}\dcal_{\mu}\partial_{\alpha}\phi^{b}$, which is not equal to, say, $\dcal_{\mu}\left (g^{\nu\alpha}\partial_{\alpha}\left (\g_{ab}\phi^{b}\right )\right )$. } with $g_{\mu\nu}$ and  $\g_{ab}$, respectively.

\section{Quantum corrections for the Special Galileon \label{app:SpecGal}}

In the main text we have focused on computing (one-loop) quantum corrections for generic brane setups which allowed us to gain new insights into quantum corrections of lower-dimensional scalar field theories that non-linearly realise symmetries of the higher-dimensional setup.  While theories such as DBI and NLSM can be obtained as limiting cases of this general scenario, this is not true for all exceptional scalar theories.  In particular, while there does exist a geometric interpretation of the special galileon \cite{Novotny:2016jkh}, it requires ingredients beyond those considered in the present work.

 In this appendix, we consider loop corrections to the special galileon and analyze the results along similar lines to those considered in the main text. An extensive study of the quantum corrections to the special galileon can be found in \cite{Preucil:2019nxt} and the following results can also be found in that work.  Below we are simply interested in understanding aspects of the special galileon in the context of the present paper's focus.
\\  

\nin{\bf The Special Galileon}: The special galileon \cite{Hinterbichler:2015pqa} is given by the following action\footnote{Note that we have implicitly applied a Galileon duality transformation \cite{deRham:2013hsa} -- see \cite{deRham:2014lqa,Kampf:2014rka,Noller:2015eda} for further details and extensions -- to eliminate the cubic Galileon interaction. If the Galileon also couples to other fields, then care must be taken to consistently apply the duality field re-definition to these other coupling terms as well.}

\begin{equation}
S=\int \mathrm{d}^{4}x \left[
-\frac{1}{2}\pi\Box\pi 
+ \frac{1}{\Lambda_3^6} \pi\epsilon^{\mu\nu\rho\sigma}{\epsilon}{^{\alpha \beta \gamma}_{\sigma}}\, \pi_{\mu\alpha}\,\pi_{\nu\beta}\pi_{\rho\gamma}
\right],
\label{SGact}
\end{equation}
where $\pi$ is a scalar field, $\pi_{\mu\alpha} \equiv \partial_\mu\partial_\alpha\pi$, we work in Euclidean signature, and $\Lambda_3$ is the strong coupling scale of the theory (any would-be dimensionless coefficient of the quartic interaction has been absorbed into $\Lambda_3$). In addition to the standard galileon symmetries, under which $\pi \longrightarrow \pi+c+b_{\mu}x^{\mu}$ with $c, b^{\mu}$ constant, the theory then obeys the following enhanced symmetry \cite{Hinterbichler:2015pqa}
\begin{align} \label{specGalSym}
\pi &\to \pi + s_{\mu\nu}\Big[x^\mu x^\nu + \frac{24}{\Lambda_3^6} \pi^\mu \pi^\nu \Big],
\end{align}
where $s_{\mu\nu}$ is a constant, traceless, and infinitesimal matrix. The lagrangian \eqref{SGact} changes by a total derivative under \eqref{specGalSym}. 
\\

\nin{\bf One-loop divergences I}: Following the approach of \cite{Heisenberg:2019wjv,Heisenberg:2020cyi} we may then straightforwardly compute (one-loop) quantum corrections for the Special Galileon. In the notation of \cite{Heisenberg:2020cyi}, which is only used in this appendix, the key ingredients for this computation are the inverse effective metric $M$ and the effective potential $U$. These satisfy
\comment{
\begin{align}
M^{\mu\nu} &= \frac{1}{6}\,\varepsilon^{\mu\alpha\rho\sigma}{{\varepsilon}^{\nu}}_{\alpha\rho\sigma} -12\frac{\tilde{c}_4}{\Lambda^6}{\varepsilon}^{\mu\alpha\rho\sigma}{{\varepsilon}^{\nu\beta\gamma}}_{\sigma}\,\partial_{\alpha}\partial_{\beta} \bar{\pi} \partial_{\rho}\partial_{\gamma} \bar{\pi} \\
U &= \nabla_{\nu} L^{\nu}+L_{\nu}L^{\nu}, \text{ where } L^{\rho}\equiv\frac{1}{2}M^{\mu\nu}{\Gamma}{^{\rho}_{\mu\nu}}.
\end{align}
}
\begin{align}
M^{\mu\nu} &= \delta^{\mu\nu} -\frac{12}{\Lambda^6}{\varepsilon}^{\mu\alpha\rho\sigma}{{\varepsilon}^{\nu\beta\gamma}}_{\sigma}\, \pi_{\alpha\beta} \pi_{\rho\gamma},
&U &= \nabla_{\nu} L^{\nu}+L_{\nu}L^{\nu}, \text{ where } L^{\rho}\equiv\frac{1}{2}M^{\mu\nu}{\Gamma}{^{\rho}_{\mu\nu}}.
\label{MandU}
\end{align}
Note that the covariant derivative and connection $\Gamma$ are defined with respect to the effective metric $M$. All indices in \eqref{MandU} and \eqref{OneLoopGour} are raised and lowered with respect to this effective metric and, in \eqref{OneLoopGour}, curvature tensors are evaluated for $M$ as well. 
Using heat-kernel techniques, the quantum corrections to the Special Galileon at one-loop order can then be expressed in a closed form in terms of geometrical curvature invariants of the effective metric $M_{\mu\nu}$ and the potential $U$ \cite{Heisenberg:2019wjv,Heisenberg:2020cyi}
\begin{align}\label{OneLoopGour}
\Gamma_{1}^{\mathrm{div}}=-\frac{1}{4\varepsilon}\frac{1}{(4\pi)^2}\int_{\mathcal{M}}\mathrm{d}^4x\,\sqrt{M}\left\{\frac{R_{\mu\nu}R^{\mu\nu}}{30}
+\frac{R^2}{60}-\frac{RU}{3}+U^2\right\}\ .
\end{align}
Note that we assume we are working in four dimensions, so have dropped an additional contribution of Gauss-Bonnet form (which therefore vanishes in four dimensions, up to total derivatives). As a concrete example, let us focus on the 4-point function evaluated in this way. We find
\begin{align}
\Gamma_{1,4}^{\rm div} &=\frac{6}{5}\frac{1}{(4\pi)^2}\frac{1}{\varepsilon}\frac{1}{\Lambda_3^{12}} \,\int \mathrm{d}^4x \, \Big[8 \pi^{\alpha \gamma{}}_{\alpha{}}\pi^{\sigma{}}_{\beta{}\sigma{}}\pi^{\beta{}\delta{}}_{\gamma{}}\pi^{\rho{}}_{\delta{}\rho{}}-2 \pi^{}_{\alpha{}\gamma{}\beta{}}\pi^{\alpha{}\gamma{}\beta{}}_{}\pi^{}_{\delta{}\sigma{}\rho{}}\pi^{\delta{}\sigma{}\rho{}}_{}+4 \pi^{\alpha{}\gamma{}}_{\alpha{}}\pi^{\beta{}}_{\gamma{}\beta{}}\pi^{}_{\delta{}\sigma{}\rho{}}\pi^{\delta{}\sigma{}\rho{}}_{} \nn 
&-4 \pi^{\alpha{}\gamma{}}_{\alpha{}}\pi^{\sigma{}}_{\beta{}\delta{}}\pi^{\beta{}\delta{}}_{\gamma{}}\pi^{\rho{}}_{\sigma{}\rho{}}-6 \pi^{\alpha{}\gamma{}}_{\alpha{}}\pi^{\beta{}}_{\gamma{}\beta{}}\pi^{\delta{}\sigma{}}_{\delta{}}\pi^{\rho{}}_{\sigma{}\rho{}}+8 \pi^{\alpha{}\gamma{}}_{}\pi^{\beta{}\delta{}}_{\beta{}}\pi^{\sigma{}\rho{}}_{\sigma{}}\pi^{}_{\alpha{}\gamma{}\delta{}\rho{}}-8 \pi^{\alpha{}\gamma{}}_{}\pi^{\beta{}\delta{}}_{\beta{}}\pi^{\sigma{}\rho{}}_{\delta{}}\pi^{}_{\alpha{}\gamma{}\sigma{}\rho{}} \nn 
&-4 \pi^{\alpha{}\gamma{}}_{}\pi^{}_{\beta{}\delta{}\sigma{}}\pi^{\beta{}\delta{}\sigma{}}_{}\pi^{\rho{}}_{\alpha{}\gamma{}\rho{}}+4 \pi^{\alpha{}\gamma{}}_{}\pi^{\beta{}\delta{}}_{\beta{}}\pi^{\sigma{}}_{\delta{}\sigma{}}\pi^{\rho{}}_{\alpha{}\gamma{}\rho{}}-4 \pi^{\alpha{}\gamma{}}_{}\pi^{\beta{}\delta{}}_{}\pi^{\sigma{}\rho{}}_{\alpha{}\gamma{}}\pi^{}_{\beta{}\delta{}\sigma{}\rho{}}-2 \pi^{\alpha{}\gamma{}}_{}\pi^{\beta{}\delta{}}_{}\pi^{\sigma{}}_{\alpha{}\gamma{}\sigma{}}\pi^{\rho{}}_{\beta{}\delta{}\rho{}} \nn 
&-4 \pi^{\alpha{}}_{\alpha{}}\pi^{\gamma{}\beta{}}_{\gamma{}}\pi^{\delta{}\sigma{}}_{\delta{}}\pi^{\rho{}}_{\beta{}\sigma{}\rho{}}+8 \pi^{\alpha{}\gamma{}}_{}\pi^{\beta{}\delta{}}_{\alpha{}}\pi^{\sigma{}}_{\beta{}\sigma{}}\pi^{\rho{}}_{\gamma{}\delta{}\rho{}}-2 \pi^{\alpha{}\gamma{}}_{}\pi^{\beta{}\delta{}}_{}\pi^{\sigma{}}_{\alpha{}\beta{}\sigma{}}\pi^{\rho{}}_{\gamma{}\delta{}\rho{}}-2 \pi^{\beta{}}_{\alpha{}}\pi^{\alpha{}\gamma{}}_{}\pi^{\rho{}}_{\beta{}\sigma{}\rho{}}\pi^{\delta{}\sigma{}}_{\gamma{}\delta{}} \nn 
&-4 \pi^{\alpha{}}_{\alpha{}}\pi^{\gamma{}\beta{}}_{}\pi^{\rho{}}_{\beta{}\sigma{}\rho{}}\pi^{\delta{}\sigma{}}_{\gamma{}\delta{}}-8 \pi^{\alpha{}\gamma{}}_{}\pi^{\beta{}}_{\alpha{}\beta{}}\pi^{\delta{}\sigma{}}_{\delta{}}\pi^{\rho{}}_{\gamma{}\sigma{}\rho{}}+4 \pi^{\alpha{}}_{\alpha{}}\pi^{\delta{}\sigma{}}_{\beta{}}\pi^{\gamma{}\beta{}}_{\gamma{}}\pi^{\rho{}}_{\delta{}\sigma{}\rho{}}+8 \pi^{\alpha{}\gamma{}}_{}\pi^{\beta{}\delta{}}_{}\pi^{\sigma{}}_{\alpha{}\gamma{}\beta{}}\pi^{\rho{}}_{\delta{}\sigma{}\rho{}} \nn 
&- \pi^{\alpha{}}_{\alpha{}}\pi^{\gamma{}}_{\gamma{}}\pi^{\beta{}\delta{}\sigma{}}_{\beta{}}\pi^{\rho{}}_{\delta{}\sigma{}\rho{}}+4 \pi^{\alpha{}}_{\alpha{}}\pi^{\gamma{}\beta{}}_{}\pi^{\delta{}\sigma{}}_{\gamma{}\beta{}}\pi^{\rho{}}_{\delta{}\sigma{}\rho{}}-4 \pi^{\alpha{}\gamma{}}_{}\pi^{\beta{}}_{\alpha{}\gamma{}}\pi^{\delta{}}_{\beta{}\delta{}}\pi^{\sigma{}\rho{}}_{\sigma{}\rho{}}+4 \pi^{\alpha{}\gamma{}}_{}\pi^{\beta{}}_{\alpha{}\beta{}}\pi^{\delta{}}_{\gamma{}\delta{}}\pi^{\sigma{}\rho{}}_{\sigma{}\rho{}} \nn 
&+4 \pi^{\alpha{}}_{\alpha{}}\pi^{}_{\gamma{}\beta{}\delta{}}\pi^{\gamma{}\beta{}\delta{}}_{}\pi^{\sigma{}\rho{}}_{\sigma{}\rho{}}-4 \pi^{\alpha{}}_{\alpha{}}\pi^{\delta{}}_{\beta{}\delta{}}\pi^{\gamma{}\beta{}}_{\gamma{}}\pi^{\sigma{}\rho{}}_{\sigma{}\rho{}}-4 \pi^{\alpha{}\gamma{}}_{}\pi^{\beta{}\delta{}}_{}\pi^{}_{\alpha{}\gamma{}\beta{}\delta{}}\pi^{\sigma{}\rho{}}_{\sigma{}\rho{}}- \pi^{}_{\alpha{}\gamma{}}\pi^{\alpha{}\gamma{}}_{}\pi^{\beta{}\delta{}}_{\beta{}\delta{}}\pi^{\sigma{}\rho{}}_{\sigma{}\rho{}} \nn 
&-2 \pi^{\alpha{}}_{\alpha{}}\pi^{\gamma{}}_{\gamma{}}\pi^{\beta{}\delta{}}_{\beta{}\delta{}}\pi^{\sigma{}\rho{}}_{\sigma{}\rho{}}+4 \pi^{\beta{}}_{\alpha{}}\pi^{\alpha{}\gamma{}}_{}\pi^{\delta{}}_{\gamma{}\beta{}\delta{}}\pi^{\sigma{}\rho{}}_{\sigma{}\rho{}}+6 \pi^{\alpha{}}_{\alpha{}}\pi^{\gamma{}\beta{}}_{}\pi^{\delta{}}_{\gamma{}\beta{}\delta{}}\pi^{\sigma{}\rho{}}_{\sigma{}\rho{}} \Big],
\label{SpecGal-naive4point}
\end{align}
where indices are raised and lowered with a flat Euclidean metric again and the second subscript in $\Gamma$ indicates the order in fields.
While this can be simplified further via integrating-by-parts, the form given makes the galileon $\pi \to \pi + c + b_\mu x^\mu$ symmetry of the four point function manifest (every field enters with at least two derivatives). However, while the 4-point function (and the one-loop corrections computed in this way in general) are manifestly invariant under this linear symmetry, they are {\it not} invariant under the non-linear special galileon symmetry. That this is the result of the above `naive' and off-shell calculation is of course not a surprise in light of the discussion in the main text and mirrors the analogous result for DBI. 

\noindent {\bf On-shell equivalence I}: 
Just as for DBI, we expect to be able to obtain an on-shell version of \eqref{OneLoopGour} that manifestly respects the special galileon symmetry. Again focusing on the 4-point function \eqref{SpecGal-naive4point} as an example, consider the corresponding $2 \to 2$ scattering amplitude, which satisfies
\begin{align}
{\cal A}_{2 \to 2} = -\frac{6}{5} \frac{1}{(4\pi)^2 \epsilon} \frac{1}{\Lambda_3^{12}} ( s^2+st+t^2)^3.
\label{SpecGal-naiveAmp}
\end{align}
where we have used massless on-shell kinematics: $s+t+u=0$. Instead of directly manipulating \eqref{SpecGal-naive4point}, we can use this result to identify an on-shell form of the four-point function, which yields the same amplitude and is therefore (physically) equivalent. Using this we can find the following on-shell equivalent expression for the 4-point function
\begin{align}
\Gamma_{1,4}^{\rm div}[\pi_{\rm cl}] &=\frac{24}{5} \frac{1}{(4\pi)^2\varepsilon}\frac{1}{\Lambda_3^{12}} \,\int \mathrm{d}^4x \, \pi^{\alpha\beta\gamma} (\pi^{\delta\sigma}_{\alpha} \pi_{\gamma\sigma\rho} \pi^{\rho}_{\beta\delta} -  \pi_{\alpha\beta\gamma} \pi_{\delta\sigma\rho} \pi^{\delta\sigma\rho})\ ,
\label{SpecGal-naive4point-equivalent}
\end{align}
Note that this is of the form $(\partial^3\pi)^n$, so trivially invariant under the `lowering' $s_{\mu\nu} x^\mu x^\nu$ part of the Special Galilean transformation \eqref{specGalSym}.
\\

\noindent {\bf One-loop divergences II}: 
We now compare the above `naive' calculation with the result of \cite{Preucil:2019nxt}, which uses the equations of motion to write the one-loop effective action in a symmetry invariant way from the start. A key realization of \cite{Preucil:2019nxt} is that when the on-shell conditions are imposed in the background field method, the effective metric for fluctuations reduces to the form
\begin{align} \label{metricPN}
g_{\mu\nu} = \delta_{\mu\nu} +\frac{1}{\alpha^2}\pi^{\beta}_{\mu}\pi_{\beta\nu} \ , \quad \alpha^2 \equiv \Lambda_3^6/24
\end{align}
and the above is known to be special-galileon-covariant \cite{Novotny:2016jkh,Roest:2020vny}. This is a special phenomenon: no similar on-shell simplifications occur at the level of the fluctuation metric for DBI, for instance.
In four dimensions, the on-shell one-loop effective action can be written as \eqref{app:eq:anHeatKernelResults} \cite{Barvinsky:1984jd,Barvinsky:1990up,deRham:2014wfa,Preucil:2019nxt}\footnote{Note that we are supressing a factor of $\mu^{-2\varepsilon}$ in comparison with the result of \cite{Preucil:2019nxt} here.}
\begin{align}
\Gamma_{1}^{\rm div}[\pi_{\rm cl}] = - \frac{1}{120} \frac{1}{(4\pi)^2\varepsilon}\int \mathrm{d}^4 x \, \sqrt{g} \left[R_{\mu\nu}R^{\mu\nu} + \frac{1}{2}R^2 \right],
\end{align}
where indices are raised and lowered with the special galileon effective metric $g_{\mu\nu}$ \eqref{metricPN} and curvature tensors are evaluated for that effective metric. This result is invariant under the special galileon symmetry\footnote{In the notation of \cite{Preucil:2019nxt} the Special Galilean symmetry is $\pi \to \pi + \tfrac{1}{2} \hat s_{\mu\nu} (\alpha^2 x^\mu x^\nu + \pi^\mu \pi^\nu)$, where $s_{\mu\nu} = \tfrac{1}{2} \alpha^2 \hat s_{\mu\nu}$.} by construction. Evaluating the resulting 4-point function for comparison with the previous `naive' calculation, we find
\begin{align}
\Gamma_{1,4}^{\rm div}[\pi_{\rm cl}] = - \frac{1}{120\alpha^4} \frac{1}{(4\pi)^2\varepsilon}\,\int \mathrm{d}^4x \, &\Big[\pi_{\alpha{}\gamma{}}^{\delta{}} \pi^{\alpha{}\beta{}\gamma{}} \pi_{\beta{}}^{\sigma{}\rho{}} \pi_{\delta{}\sigma{}\rho{}} - 2 \pi^{\alpha{}}_{\alpha{}}{}^{\gamma{}} \pi_{\beta{}}^{\sigma{}\rho{}} \pi_{\gamma{}}^{\beta{}\delta{}} \pi_{\delta{}\sigma{}\rho{}} + \tfrac{1}{2} \pi_{\alpha{}\beta{}\gamma{}} \pi^{\alpha{}\beta{}\gamma{}} \pi_{\delta{}\sigma{}\rho{}} \pi^{\delta{}\sigma{}\rho{}} \nn 
&-  \pi^{\alpha{}\gamma{}}_{\alpha{}} \pi_{\beta{}\gamma{}}^{\beta{}} \pi_{\delta{}\sigma{}\rho{}} \pi^{\delta{}\sigma{}\rho{}} + \pi^{\alpha{}\gamma{}}_{\alpha{}} \pi_{\beta{}\delta{}}^{\sigma{}} \pi_{\gamma{}}^{\beta{}\delta{}} \pi_{\sigma{}\rho{}}^{\rho{}} + \tfrac{1}{2} \pi^{\alpha{}\gamma{}}_{\alpha{}} \pi_{\beta{}\gamma{}}^{\beta{}} \pi^{\delta{}\sigma{}}_{\delta{}} \pi_{\sigma{}\rho{}}^{\rho{}} \Big].
\label{SpecGal-4point-PN}
\end{align}
Just as for \eqref{SpecGal-naive4point-equivalent}, the  $(\partial^3\pi)^n$ form of this expression yields trivial invariance under the `lowering' $s_{\mu\nu} x^\mu x^\nu$ part of the Special Galilean transformation \eqref{specGalSym}.
\\

\noindent {\bf On-shell equivalence II}: Computing the contact contribution of \eqref{SpecGal-4point-PN} to $\Acal_{2\to 2}$, it is readily verified that the result is \eqref{SpecGal-naiveAmp} and hence the preceding action is on-shell equivalent to \eqref{SpecGal-naive4point}.
This establishes the on-shell equivalence between the three formulations of $\Gamma_{1,4}$: \eqref{SpecGal-naive4point}, \eqref{SpecGal-naive4point-equivalent} and \eqref{SpecGal-4point-PN}. 
\\

\noindent {\bf Discussion}:
The special galileon example reiterates and extends several points discussed in the main text. The presence of a non-linear symmetry means that a naive calculation of the off-shell one-loop effective action yields a result that does not respect this symmetry. However, just as for DBI, going on-shell one can recover a manifestly symmetric physical result, both via the brute force method outlined above as well as via the more elegant formulation of \cite{Preucil:2019nxt}. While ultimately results obtained in these different ways are physically equivalent, this outlines several ways the computation can proceed and invariance of the physical result under a non-linear symmetry of the system can be made manifest.   The geometric construction of the special galileon \cite{Novotny:2016jkh} involves complex bulk spacetimes and K\"ahler forms, which is the reason why they fail to be captured by our methods.  It would be worthwhile to explore whether other interesting models of a similar origin exist and, if so, how to extend our present construction to include such additional ingredients.

\bibliographystyle{utphys}
\bibliography{Bibliography}

\providecommand{\href}[2]{#2}\begingroup\raggedright\begin{thebibliography}{100}

\bibitem{Morgan:2007zza}
F.~Morgan, ``{Colloquium: Soap bubble clusters},''
  \href{http://dx.doi.org/10.1103/RevModPhys.79.821}{{\em Rev. Mod. Phys.} {\bf
  79} (2007)  821--827}.

\bibitem{Silverstein:2003hf}
E.~Silverstein and D.~Tong, ``{Scalar speed limits and cosmology: Acceleration
  from D-cceleration},''
  \href{http://dx.doi.org/10.1103/PhysRevD.70.103505}{{\em Phys. Rev. D} {\bf
  70} (2004)  103505}, \href{http://arxiv.org/abs/hep-th/0310221}{{\tt
  arXiv:hep-th/0310221}}.

\bibitem{Alishahiha:2004eh}
M.~Alishahiha, E.~Silverstein, and D.~Tong, ``{DBI in the sky},''
  \href{http://dx.doi.org/10.1103/PhysRevD.70.123505}{{\em Phys. Rev. D} {\bf
  70} (2004)  123505}, \href{http://arxiv.org/abs/hep-th/0404084}{{\tt
  arXiv:hep-th/0404084}}.

\bibitem{Polchinski:1996na}
J.~Polchinski, ``{Tasi lectures on D-branes},'' in {\em {Theoretical Advanced
  Study Institute in Elementary Particle Physics (TASI 96): Fields, Strings,
  and Duality}}, pp.~293--356.
\newblock 11, 1996.
\newblock \href{http://arxiv.org/abs/hep-th/9611050}{{\tt
  arXiv:hep-th/9611050}}.

\bibitem{Maldacena:1997re}
J.~M. Maldacena, ``{The Large N limit of superconformal field theories and
  supergravity},'' \href{http://dx.doi.org/10.1023/A:1026654312961}{{\em Int.
  J. Theor. Phys.} {\bf 38} (1999)  1113--1133},
  \href{http://arxiv.org/abs/hep-th/9711200}{{\tt arXiv:hep-th/9711200}}.

\bibitem{Cheung:2014dqa}
C.~Cheung, K.~Kampf, J.~Novotny, and J.~Trnka, ``{Effective Field Theories from
  Soft Limits of Scattering Amplitudes},''
  \href{http://dx.doi.org/10.1103/PhysRevLett.114.221602}{{\em Phys. Rev.
  Lett.} {\bf 114} (2015) no.~22, 221602},
  \href{http://arxiv.org/abs/1412.4095}{{\tt arXiv:1412.4095 [hep-th]}}.

\bibitem{Cheung:2016drk}
C.~Cheung, K.~Kampf, J.~Novotny, C.-H. Shen, and J.~Trnka, ``{A Periodic Table
  of Effective Field Theories},''
  \href{http://dx.doi.org/10.1007/JHEP02(2017)020}{{\em JHEP} {\bf 02} (2017)
  020}, \href{http://arxiv.org/abs/1611.03137}{{\tt arXiv:1611.03137
  [hep-th]}}.

\bibitem{Elvang:2018dco}
H.~Elvang, M.~Hadjiantonis, C.~R. Jones, and S.~Paranjape, ``{Soft Bootstrap
  and Supersymmetry},'' \href{http://dx.doi.org/10.1007/JHEP01(2019)195}{{\em
  JHEP} {\bf 01} (2019)  195}, \href{http://arxiv.org/abs/1806.06079}{{\tt
  arXiv:1806.06079 [hep-th]}}.

\bibitem{Low:2019ynd}
I.~Low and Z.~Yin, ``{Soft Bootstrap and Effective Field Theories},''
  \href{http://dx.doi.org/10.1007/JHEP11(2019)078}{{\em JHEP} {\bf 11} (2019)
  078}, \href{http://arxiv.org/abs/1904.12859}{{\tt arXiv:1904.12859
  [hep-th]}}.

\bibitem{ArkaniHamed:2008gz}
N.~Arkani-Hamed, F.~Cachazo, and J.~Kaplan, ``{What is the Simplest Quantum
  Field Theory?},'' \href{http://dx.doi.org/10.1007/JHEP09(2010)016}{{\em JHEP}
  {\bf 09} (2010)  016}, \href{http://arxiv.org/abs/0808.1446}{{\tt
  arXiv:0808.1446 [hep-th]}}.

\bibitem{Cachazo:2015ksa}
F.~Cachazo, S.~He, and E.~Y. Yuan, ``{New Double Soft Emission Theorems},''
  \href{http://dx.doi.org/10.1103/PhysRevD.92.065030}{{\em Phys. Rev. D} {\bf
  92} (2015) no.~6, 065030}, \href{http://arxiv.org/abs/1503.04816}{{\tt
  arXiv:1503.04816 [hep-th]}}.

\bibitem{Cachazo:2016njl}
F.~Cachazo, P.~Cha, and S.~Mizera, ``{Extensions of Theories from Soft
  Limits},'' \href{http://dx.doi.org/10.1007/JHEP06(2016)170}{{\em JHEP} {\bf
  06} (2016)  170}, \href{http://arxiv.org/abs/1604.03893}{{\tt
  arXiv:1604.03893 [hep-th]}}.

\bibitem{Padilla:2016mno}
A.~Padilla, D.~Stefanyszyn, and T.~Wilson, ``{Probing Scalar Effective Field
  Theories with the Soft Limits of Scattering Amplitudes},''
  \href{http://dx.doi.org/10.1007/JHEP04(2017)015}{{\em JHEP} {\bf 04} (2017)
  015}, \href{http://arxiv.org/abs/1612.04283}{{\tt arXiv:1612.04283
  [hep-th]}}.

\bibitem{Guerrieri:2017ujb}
A.~L. Guerrieri, Y.-t. Huang, Z.~Li, and C.~Wen, ``{On the exactness of soft
  theorems},'' \href{http://dx.doi.org/10.1007/JHEP12(2017)052}{{\em JHEP} {\bf
  12} (2017)  052}, \href{http://arxiv.org/abs/1705.10078}{{\tt
  arXiv:1705.10078 [hep-th]}}.

\bibitem{Li:2017fsb}
Z.-z. Li, H.-h. Lin, and S.-q. Zhang, ``{On the Symmetry Foundation of Double
  Soft Theorems},'' \href{http://dx.doi.org/10.1007/JHEP12(2017)032}{{\em JHEP}
  {\bf 12} (2017)  032}, \href{http://arxiv.org/abs/1710.00480}{{\tt
  arXiv:1710.00480 [hep-th]}}.

\bibitem{Bogers:2018kuw}
M.~P. Bogers and T.~Brauner, ``{Geometry of Multiflavor Galileon-Like
  Theories},'' \href{http://dx.doi.org/10.1103/PhysRevLett.121.171602}{{\em
  Phys. Rev. Lett.} {\bf 121} (2018) no.~17, 171602},
  \href{http://arxiv.org/abs/1802.08107}{{\tt arXiv:1802.08107 [hep-th]}}.

\bibitem{Bogers:2018zeg}
M.~P. Bogers and T.~Brauner, ``{Lie-algebraic classification of effective
  theories with enhanced soft limits},''
  \href{http://dx.doi.org/10.1007/JHEP05(2018)076}{{\em JHEP} {\bf 05} (2018)
  076}, \href{http://arxiv.org/abs/1803.05359}{{\tt arXiv:1803.05359
  [hep-th]}}.

\bibitem{Rodina:2018pcb}
L.~Rodina, ``{Scattering Amplitudes from Soft Theorems and Infrared
  Behavior},'' \href{http://dx.doi.org/10.1103/PhysRevLett.122.071601}{{\em
  Phys. Rev. Lett.} {\bf 122} (2019) no.~7, 071601},
  \href{http://arxiv.org/abs/1807.09738}{{\tt arXiv:1807.09738 [hep-th]}}.

\bibitem{Yin:2018hht}
Z.~Yin, ``{The Infrared Structure of Exceptional Scalar Theories},''
  \href{http://dx.doi.org/10.1007/JHEP03(2019)158}{{\em JHEP} {\bf 03} (2019)
  158}, \href{http://arxiv.org/abs/1810.07186}{{\tt arXiv:1810.07186
  [hep-th]}}.

\bibitem{Roest:2019oiw}
D.~Roest, D.~Stefanyszyn, and P.~Werkman, ``{An Algebraic Classification of
  Exceptional EFTs},'' \href{http://dx.doi.org/10.1007/JHEP08(2019)081}{{\em
  JHEP} {\bf 08} (2019)  081}, \href{http://arxiv.org/abs/1903.08222}{{\tt
  arXiv:1903.08222 [hep-th]}}.

\bibitem{Bonifacio:2019rpv}
J.~Bonifacio, K.~Hinterbichler, L.~A. Johnson, A.~Joyce, and R.~A. Rosen,
  ``{Matter Couplings and Equivalence Principles for Soft Scalars},''
  \href{http://dx.doi.org/10.1007/JHEP07(2020)056}{{\em JHEP} {\bf 07} (2020)
  056}, \href{http://arxiv.org/abs/1911.04490}{{\tt arXiv:1911.04490
  [hep-th]}}.

\bibitem{Bern:2019prr}
Z.~Bern, J.~J. Carrasco, M.~Chiodaroli, H.~Johansson, and R.~Roiban, ``{The
  Duality Between Color and Kinematics and its Applications},''
  \href{http://arxiv.org/abs/1909.01358}{{\tt arXiv:1909.01358 [hep-th]}}.

\bibitem{Cachazo:2014xea}
F.~Cachazo, S.~He, and E.~Y. Yuan, ``{Scattering Equations and Matrices: From
  Einstein To Yang-Mills, DBI and NLSM},''
  \href{http://dx.doi.org/10.1007/JHEP07(2015)149}{{\em JHEP} {\bf 07} (2015)
  149}, \href{http://arxiv.org/abs/1412.3479}{{\tt arXiv:1412.3479 [hep-th]}}.

\bibitem{Cheung:2017ems}
C.~Cheung, C.-H. Shen, and C.~Wen, ``{Unifying Relations for Scattering
  Amplitudes},'' \href{http://dx.doi.org/10.1007/JHEP02(2018)095}{{\em JHEP}
  {\bf 02} (2018)  095}, \href{http://arxiv.org/abs/1705.03025}{{\tt
  arXiv:1705.03025 [hep-th]}}.

\bibitem{Gerstein:1971fm}
I.~S. Gerstein, R.~Jackiw, S.~Weinberg, and B.~W. Lee, ``{Chiral loops},''
\href{http://dx.doi.org/10.1103/PhysRevD.3.2486}{{\em Phys. Rev.} {\bf D3}
  (1971)  2486--2492}.

\bibitem{Weinberg1}
S.~Weinberg, {\em {The Quantum theory of fields. Vol. 1: Foundations}}.
\newblock Cambridge University Press,
2005.
\newblock

\bibitem{Weinberg2}
S.~Weinberg, {\em {The quantum theory of fields. Vol. 2: Modern applications}}.
\newblock Cambridge University Press,
2013.
\newblock

\bibitem{Abbott:1981ke}
L.~F. Abbott, ``{Introduction to the Background Field Method},''
{\em Acta Phys. Polon.} {\bf B13} (1982)  33.

\bibitem{Honerkamp:1971sh}
J.~Honerkamp, ``{Chiral multiloops},''
  \href{http://dx.doi.org/10.1016/0550-3213(72)90299-4}{{\em Nucl. Phys. B}
  {\bf 36} (1972)  130--140}.

\bibitem{AlvarezGaume:1981hn}
L.~Alvarez-Gaume, D.~Z. Freedman, and S.~Mukhi, ``{The Background Field Method
  and the Ultraviolet Structure of the Supersymmetric Nonlinear Sigma Model},''
\href{http://dx.doi.org/10.1016/0003-4916(81)90006-3}{{\em Annals Phys.} {\bf
  134} (1981)  85}.

\bibitem{Barvinsky:1985an}
A.~O. Barvinsky and G.~A. Vilkovisky, ``{The Generalized Schwinger-Dewitt
  Technique in Gauge Theories and Quantum Gravity},''
\href{http://dx.doi.org/10.1016/0370-1573(85)90148-6}{{\em Phys. Rept.} {\bf
  119} (1985)  1--74}.

\bibitem{Creminelli:2000gh}
P.~Creminelli and A.~Strumia, ``{Collider signals of brane fluctuations},''
  \href{http://dx.doi.org/10.1016/S0550-3213(00)00711-2}{{\em Nucl. Phys. B}
  {\bf 596} (2001)  125--135}, \href{http://arxiv.org/abs/hep-ph/0007267}{{\tt
  arXiv:hep-ph/0007267}}.

\bibitem{Contino:2001nj}
R.~Contino, L.~Pilo, R.~Rattazzi, and A.~Strumia, ``{Graviton loops and brane
  observables},'' \href{http://dx.doi.org/10.1088/1126-6708/2001/06/005}{{\em
  JHEP} {\bf 06} (2001)  005}, \href{http://arxiv.org/abs/hep-ph/0103104}{{\tt
  arXiv:hep-ph/0103104}}.

\bibitem{Cembranos:2003mr}
J.~Cembranos, A.~Dobado, and A.~L. Maroto, ``{Brane world dark matter},''
  \href{http://dx.doi.org/10.1103/PhysRevLett.90.241301}{{\em Phys. Rev. Lett.}
  {\bf 90} (2003)  241301}, \href{http://arxiv.org/abs/hep-ph/0302041}{{\tt
  arXiv:hep-ph/0302041}}.

\bibitem{Nielsen:1975fs}
N.~K. Nielsen, ``{On the Gauge Dependence of Spontaneous Symmetry Breaking in
  Gauge Theories},''
\href{http://dx.doi.org/10.1016/0550-3213(75)90301-6}{{\em Nucl. Phys.} {\bf
  B101} (1975)  173--188}.

\bibitem{Fukuda:1975di}
R.~Fukuda and T.~Kugo, ``{Gauge Invariance in the Effective Action and
  Potential},''
\href{http://dx.doi.org/10.1103/PhysRevD.13.3469}{{\em Phys. Rev.} {\bf D13}
  (1976)  3469}.

\bibitem{Aitchison:1983ns}
I.~Aitchison and C.~Fraser, ``{Gauge Invariance and the Effective Potential},''
  \href{http://dx.doi.org/10.1016/0003-4916(84)90209-4}{{\em Annals Phys.} {\bf
  156} (1984)  1}.

\bibitem{Hart:1984jy}
C.~Hart, ``{Theory and renormalization of the gauge invariant effective
  action},'' \href{http://dx.doi.org/10.1103/PhysRevD.28.1993}{{\em Phys. Rev.
  D} {\bf 28} (1983)  1993--2006}.

\bibitem{Georgi:1991ch}
H.~Georgi, ``{On-shell effective field theory},''
  \href{http://dx.doi.org/10.1016/0550-3213(91)90244-R}{{\em Nucl. Phys. B}
  {\bf 361} (1991)  339--350}.

\bibitem{Isidori:2001bm}
G.~Isidori, G.~Ridolfi, and A.~Strumia, ``{On the metastability of the standard
  model vacuum},'' \href{http://dx.doi.org/10.1016/S0550-3213(01)00302-9}{{\em
  Nucl. Phys. B} {\bf 609} (2001)  387--409},
  \href{http://arxiv.org/abs/hep-ph/0104016}{{\tt arXiv:hep-ph/0104016}}.

\bibitem{Andreassen:2014eha}
A.~Andreassen, W.~Frost, and M.~D. Schwartz, ``{Consistent Use of Effective
  Potentials},'' \href{http://dx.doi.org/10.1103/PhysRevD.91.016009}{{\em Phys.
  Rev.} {\bf D91} (2015) no.~1, 016009},
\href{http://arxiv.org/abs/1408.0287}{{\tt arXiv:1408.0287 [hep-ph]}}.

\bibitem{DiLuzio:2015iua}
L.~Di~Luzio, G.~Isidori, and G.~Ridolfi, ``{Stability of the electroweak ground
  state in the Standard Model and its extensions},''
  \href{http://dx.doi.org/10.1016/j.physletb.2015.12.009}{{\em Phys. Lett. B}
  {\bf 753} (2016)  150--160}, \href{http://arxiv.org/abs/1509.05028}{{\tt
  arXiv:1509.05028 [hep-ph]}}.

\bibitem{Andreassen:2016cvx}
A.~Andreassen, D.~Farhi, W.~Frost, and M.~D. Schwartz, ``{Precision decay rate
  calculations in quantum field theory},''
  \href{http://dx.doi.org/10.1103/PhysRevD.95.085011}{{\em Phys. Rev. D} {\bf
  95} (2017) no.~8, 085011}, \href{http://arxiv.org/abs/1604.06090}{{\tt
  arXiv:1604.06090 [hep-th]}}.

\bibitem{Howe:1986vm}
P.~S. Howe, G.~Papadopoulos, and K.~Stelle, ``{The Background Field Method and
  the Nonlinear $\sigma$ Model},''
  \href{http://dx.doi.org/10.1016/0550-3213(88)90379-3}{{\em Nucl. Phys. B}
  {\bf 296} (1988)  26--48}.

\bibitem{Mukhi:1985vy}
S.~Mukhi, ``{The Geometric Background Field Method, Renormalization and the
  Wess-Zumino Term in Nonlinear Sigma Models},''
  \href{http://dx.doi.org/10.1016/0550-3213(86)90502-X}{{\em Nucl. Phys. B}
  {\bf 264} (1986)  640--652}.

\bibitem{Gilkey:1975iq}
P.~B. Gilkey, ``{The Spectral geometry of a Riemannian manifold},''
  \href{http://dx.doi.org/10.4310/jdg/1214433164}{{\em J. Diff. Geom.} {\bf 10}
  (1975) no.~4, 601--618}.

\bibitem{Avramidi:1986mj}
I.~G. Avramidi, {\em {Covariant methods for the calculation of the effective
  action in quantum field theory and investigation of higher derivative quantum
  gravity}}.
\newblock PhD thesis, Moscow State U., 1986.
\newblock
\href{http://arxiv.org/abs/hep-th/9510140}{{\tt arXiv:hep-th/9510140
  [hep-th]}}.
\newblock

\bibitem{Hinterbichler:2010xn}
K.~Hinterbichler, M.~Trodden, and D.~Wesley, ``{Multi-field galileons and
  higher co-dimension branes},''
  \href{http://dx.doi.org/10.1103/PhysRevD.82.124018}{{\em Phys. Rev.} {\bf
  D82} (2010)  124018},
\href{http://arxiv.org/abs/1008.1305}{{\tt arXiv:1008.1305 [hep-th]}}.

\bibitem{Pajer:2018egx}
E.~Pajer and D.~Stefanyszyn, ``{Symmetric Superfluids},''
  \href{http://dx.doi.org/10.1007/JHEP06(2019)008}{{\em JHEP} {\bf 06} (2019)
  008}, \href{http://arxiv.org/abs/1812.05133}{{\tt arXiv:1812.05133
  [hep-th]}}.

\bibitem{Grall:2019qof}
T.~Grall, S.~Jazayeri, and E.~Pajer, ``{Symmetric Scalars},''
  \href{http://dx.doi.org/10.1088/1475-7516/2020/05/031}{{\em JCAP} {\bf 05}
  (2020)  031}, \href{http://arxiv.org/abs/1909.04622}{{\tt arXiv:1909.04622
  [hep-th]}}.

\bibitem{Cheung:2020qxc}
C.~Cheung, J.~Mangan, and C.-H. Shen, ``{Hidden Conformal Invariance of Scalar
  Effective Field Theories},'' \href{http://arxiv.org/abs/2005.13027}{{\tt
  arXiv:2005.13027 [hep-th]}}.

\bibitem{Goon:2010xh}
G.~L. Goon, K.~Hinterbichler, and M.~Trodden, ``{Stability and superluminality
  of spherical DBI galileon solutions},''
  \href{http://dx.doi.org/10.1103/PhysRevD.83.085015}{{\em Phys. Rev.} {\bf
  D83} (2011)  085015},
\href{http://arxiv.org/abs/1008.4580}{{\tt arXiv:1008.4580 [hep-th]}}.

\bibitem{deRham:2010eu}
C.~de~Rham and A.~J. Tolley, ``{DBI and the Galileon reunited},''
  \href{http://dx.doi.org/10.1088/1475-7516/2010/05/015}{{\em JCAP} {\bf 1005}
  (2010)  015},
\href{http://arxiv.org/abs/1003.5917}{{\tt arXiv:1003.5917 [hep-th]}}.

\bibitem{Goon:2011uw}
G.~Goon, K.~Hinterbichler, and M.~Trodden, ``{A New Class of Effective Field
  Theories from Embedded Branes},''
  \href{http://dx.doi.org/10.1103/PhysRevLett.106.231102}{{\em Phys. Rev.
  Lett.} {\bf 106} (2011)  231102}, \href{http://arxiv.org/abs/1103.6029}{{\tt
  arXiv:1103.6029 [hep-th]}}.

\bibitem{Goon:2012dy}
G.~Goon, K.~Hinterbichler, A.~Joyce, and M.~Trodden, ``{Galileons as
  Wess-Zumino Terms},'' \href{http://dx.doi.org/10.1007/JHEP06(2012)004}{{\em
  JHEP} {\bf 1206} (2012)  004},
\href{http://arxiv.org/abs/1203.3191}{{\tt arXiv:1203.3191 [hep-th]}}.

\bibitem{Creminelli:2014zxa}
P.~Creminelli, M.~Serone, G.~Trevisan, and E.~Trincherini, ``{Inequivalence of
  Coset Constructions for Spacetime Symmetries},''
  \href{http://dx.doi.org/10.1007/JHEP02(2015)037}{{\em JHEP} {\bf 02} (2015)
  037},
\href{http://arxiv.org/abs/1403.3095}{{\tt arXiv:1403.3095 [hep-th]}}.

\bibitem{deRham:2014wfa}
C.~de~Rham and R.~H. Ribeiro, ``{Riding on irrelevant operators},''
  \href{http://dx.doi.org/10.1088/1475-7516/2014/11/016}{{\em JCAP} {\bf 1411}
  (2014) no.~11, 016},
\href{http://arxiv.org/abs/1405.5213}{{\tt arXiv:1405.5213 [hep-th]}}.

\bibitem{Appelquist:1980ae}
T.~Appelquist and C.~W. Bernard, ``{The Nonlinear $\sigma$ Model in the Loop
  Expansion},'' \href{http://dx.doi.org/10.1103/PhysRevD.23.425}{{\em Phys.
  Rev. D} {\bf 23} (1981)  425}.

\bibitem{Boulware:1981ns}
D.~G. Boulware and L.~S. Brown, ``{SYMMETRIC SPACE SCALAR FIELD THEORY},''
  \href{http://dx.doi.org/10.1016/0003-4916(82)90192-0}{{\em Annals Phys.} {\bf
  138} (1982)  392}.

\bibitem{Akhoury:1982hv}
R.~Akhoury and Y.-P. Yao, ``{The Nonlinear $\sigma$ Model as an Effective
  Lagrangian},'' \href{http://dx.doi.org/10.1103/PhysRevD.25.3361}{{\em Phys.
  Rev. D} {\bf 25} (1982)  3361}.

\bibitem{Gasser:1983yg}
J.~Gasser and H.~Leutwyler, ``{Chiral Perturbation Theory to One Loop},''
  \href{http://dx.doi.org/10.1016/0003-4916(84)90242-2}{{\em Annals Phys.} {\bf
  158} (1984)  142}.

\bibitem{Gaillard:1985uh}
M.~Gaillard, ``{The Effective One Loop Lagrangian With Derivative Couplings},''
  \href{http://dx.doi.org/10.1016/0550-3213(86)90264-6}{{\em Nucl. Phys. B}
  {\bf 268} (1986)  669--692}.

\bibitem{Costa:1988ef}
K.~Costa and F.~Liebrand, ``{Normal Coordinate Methods and Heavy Higgs
  Effects},'' \href{http://dx.doi.org/10.1103/PhysRevD.40.2014}{{\em Phys. Rev.
  D} {\bf 40} (1989)  2014}.

\bibitem{Alonso:2016oah}
R.~Alonso, E.~E. Jenkins, and A.~V. Manohar, ``{Geometry of the Scalar
  Sector},'' \href{http://dx.doi.org/10.1007/JHEP08(2016)101}{{\em JHEP} {\bf
  08} (2016)  101}, \href{http://arxiv.org/abs/1605.03602}{{\tt
  arXiv:1605.03602 [hep-ph]}}.

\bibitem{xAct}
J.~M. Mart\'in-Garc\'ia, ``{xAct, {Efficient} tensor computer algebra for the
  {Wolfram} {Language}},''. \url{http://www.xact.es/}.

\bibitem{Nutma:2013zea}
T.~Nutma, ``{xTras : A field-theory inspired xAct package for mathematica},''
  \href{http://dx.doi.org/10.1016/j.cpc.2014.02.006}{{\em Comput. Phys.
  Commun.} {\bf 185} (2014)  1719--1738},
  \href{http://arxiv.org/abs/1308.3493}{{\tt arXiv:1308.3493 [cs.SC]}}.

\bibitem{Coleman:1969sm}
S.~R. Coleman, J.~Wess, and B.~Zumino, ``{Structure of phenomenological
  Lagrangians. 1.},''
\href{http://dx.doi.org/10.1103/PhysRev.177.2239}{{\em Phys.Rev.} {\bf 177}
  (1969)  2239--2247}.

\bibitem{Callan:1969sn}
J.~Callan, Curtis~G., S.~R. Coleman, J.~Wess, and B.~Zumino, ``{Structure of
  phenomenological Lagrangians. 2.},''
\href{http://dx.doi.org/10.1103/PhysRev.177.2247}{{\em Phys.Rev.} {\bf 177}
  (1969)  2247--2250}.

\bibitem{Low:2020ubn}
I.~Low, L.~Rodina, and Z.~Yin, ``{Double Copy in Higher Derivative Operators of
  Nambu-Goldstone Bosons},'' \href{http://arxiv.org/abs/2009.00008}{{\tt
  arXiv:2009.00008 [hep-th]}}.

\bibitem{Bellucci:2002ji}
S.~Bellucci, E.~Ivanov, and S.~Krivonos, ``{AdS / CFT equivalence
  transformation},'' \href{http://dx.doi.org/10.1103/PhysRevD.66.086001,
  10.1103/PhysRevD.67.049901, 10.1103/PhysRevD.66.086001
  10.1103/PhysRevD.67.049901}{{\em Phys.Rev.} {\bf D66} (2002)  086001},
\href{http://arxiv.org/abs/hep-th/0206126}{{\tt arXiv:hep-th/0206126
  [hep-th]}}.

\bibitem{Elvang:2012st}
H.~Elvang, D.~Z. Freedman, L.-Y. Hung, M.~Kiermaier, R.~C. Myers, and
  S.~Theisen, ``{On renormalization group flows and the a-theorem in 6d},''
  \href{http://dx.doi.org/10.1007/JHEP10(2012)011}{{\em JHEP} {\bf 10} (2012)
  011}, \href{http://arxiv.org/abs/1205.3994}{{\tt arXiv:1205.3994 [hep-th]}}.

\bibitem{Hinterbichler:2012fr}
K.~Hinterbichler, A.~Joyce, J.~Khoury, and G.~E. Miller, ``{DBI Realizations of
  the Pseudo-Conformal Universe and Galilean Genesis Scenarios},''
  \href{http://dx.doi.org/10.1088/1475-7516/2012/12/030}{{\em JCAP} {\bf 12}
  (2012)  030}, \href{http://arxiv.org/abs/1209.5742}{{\tt arXiv:1209.5742
  [hep-th]}}.

\bibitem{DiVecchia:2017uqn}
P.~Di~Vecchia, R.~Marotta, and M.~Mojaza, ``{Double-soft behavior of the
  dilaton of spontaneously broken conformal invariance},''
  \href{http://dx.doi.org/10.1007/JHEP09(2017)001}{{\em JHEP} {\bf 09} (2017)
  001}, \href{http://arxiv.org/abs/1705.06175}{{\tt arXiv:1705.06175
  [hep-th]}}.

\bibitem{Aharony:1999ti}
O.~Aharony, S.~S. Gubser, J.~M. Maldacena, H.~Ooguri, and Y.~Oz, ``{Large N
  field theories, string theory and gravity},''
  \href{http://dx.doi.org/10.1016/S0370-1573(99)00083-6}{{\em Phys. Rept.} {\bf
  323} (2000)  183--386}, \href{http://arxiv.org/abs/hep-th/9905111}{{\tt
  arXiv:hep-th/9905111}}.

\bibitem{Johnson:2005mqa}
C.~V. Johnson, \href{http://dx.doi.org/10.1017/CBO9780511606540}{{\em
  {D-branes}}}.
\newblock Cambridge Monographs on Mathematical Physics. Cambridge University
  Press, 2005.

\bibitem{Baumann:2014nda}
D.~Baumann and L.~McAllister, ``{Inflation and String Theory},''
\href{http://arxiv.org/abs/1404.2601}{{\tt arXiv:1404.2601 [hep-th]}}.

\bibitem{Altland:2006si}
A.~Altland and B.~Simons, {\em {Condensed matter field theory}}.
\newblock
2006.
\newblock

\bibitem{Akhoury:1990px}
R.~Akhoury and A.~Alfakih, ``{Invariant background field method for chiral
  Lagrangians including Wess-Zumino terms},''
  \href{http://dx.doi.org/10.1016/0003-4916(91)90276-E}{{\em Annals Phys.} {\bf
  210} (1991)  81--111}.

\bibitem{Curtright:1984dz}
T.~L. Curtright and C.~K. Zachos, ``{Geometry, Topology and Supersymmetry in
  Nonlinear Models},''
  \href{http://dx.doi.org/10.1103/PhysRevLett.53.1799}{{\em Phys. Rev. Lett.}
  {\bf 53} (1984)  1799}.

\bibitem{Gates:1984nk}
J.~Gates, S.J., C.~Hull, and M.~Rocek, ``{Twisted Multiplets and New
  Supersymmetric Nonlinear Sigma Models},''
  \href{http://dx.doi.org/10.1016/0550-3213(84)90592-3}{{\em Nucl. Phys. B}
  {\bf 248} (1984)  157--186}.

\bibitem{Shmakova:1999ai}
M.~Shmakova, ``{One loop corrections to the D3-brane action},''
  \href{http://dx.doi.org/10.1103/PhysRevD.62.104009}{{\em Phys. Rev. D} {\bf
  62} (2000)  104009}, \href{http://arxiv.org/abs/hep-th/9906239}{{\tt
  arXiv:hep-th/9906239}}.

\bibitem{Wen:2020qrj}
C.~Wen and S.-Q. Zhang, ``{D3-Brane Loop Amplitudes from M5-Brane Tree
  Amplitudes},'' \href{http://dx.doi.org/10.1007/JHEP07(2020)098}{{\em JHEP}
  {\bf 07} (2020)  098}, \href{http://arxiv.org/abs/2004.02735}{{\tt
  arXiv:2004.02735 [hep-th]}}.

\bibitem{Elvang:2020kuj}
H.~Elvang, M.~Hadjiantonis, C.~R. Jones, and S.~Paranjape, ``{Electromagnetic
  Duality and D3-Brane Scattering Amplitudes Beyond Leading Order},''
  \href{http://arxiv.org/abs/2006.08928}{{\tt arXiv:2006.08928 [hep-th]}}.

\bibitem{Klein:2018ylk}
R.~Klein, E.~Malek, D.~Roest, and D.~Stefanyszyn, ``{No-go theorem for a gauge
  vector as a spacetime Goldstone mode},''
  \href{http://dx.doi.org/10.1103/PhysRevD.98.065001}{{\em Phys. Rev. D} {\bf
  98} (2018) no.~6, 065001}, \href{http://arxiv.org/abs/1806.06862}{{\tt
  arXiv:1806.06862 [hep-th]}}.

\bibitem{Cheung:2018oki}
C.~Cheung, K.~Kampf, J.~Novotny, C.-H. Shen, J.~Trnka, and C.~Wen, ``{Vector
  Effective Field Theories from Soft Limits},''
  \href{http://dx.doi.org/10.1103/PhysRevLett.120.261602}{{\em Phys. Rev.
  Lett.} {\bf 120} (2018) no.~26, 261602},
  \href{http://arxiv.org/abs/1801.01496}{{\tt arXiv:1801.01496 [hep-th]}}.

\bibitem{Novotny:2016jkh}
J.~Novotny, ``{Geometry of special Galileons},''
  \href{http://dx.doi.org/10.1103/PhysRevD.95.065019}{{\em Phys. Rev.} {\bf
  D95} (2017) no.~6, 065019},
\href{http://arxiv.org/abs/1612.01738}{{\tt arXiv:1612.01738 [hep-th]}}.

\bibitem{Preucil:2019nxt}
F.~P\v{r}eu\v{c}il and J.~Novotn\'y, ``{Special Galileon at one loop},''
  \href{http://dx.doi.org/10.1007/JHEP11(2019)166}{{\em JHEP} {\bf 11} (2019)
  166}, \href{http://arxiv.org/abs/1909.06214}{{\tt arXiv:1909.06214
  [hep-th]}}.

\bibitem{Poisson:2011nh}
E.~Poisson, A.~Pound, and I.~Vega, ``{The Motion of point particles in curved
  spacetime},'' \href{http://dx.doi.org/10.12942/lrr-2011-7}{{\em Living Rev.
  Rel.} {\bf 14} (2011)  7},
\href{http://arxiv.org/abs/1102.0529}{{\tt arXiv:1102.0529 [gr-qc]}}.

\bibitem{Proceedings:1985fja}
B.~S. DeWitt and R.~Stora, eds., {\em {Relativity, groups and topology:
  Proceedings, 40th Summer School of Theoretical Physics - Session 40}: {Les
  Houches, France, June 27 - August 4, 1983, vol. 2}}.
\newblock North-holland, Amsterdam, 1984.

\bibitem{Brown:1976wc}
L.~S. Brown, ``{Stress Tensor Trace Anomaly in a Gravitational Metric: Scalar
  Fields},'' \href{http://dx.doi.org/10.1103/PhysRevD.15.1469}{{\em Phys. Rev.
  D} {\bf 15} (1977)  1469}.

\bibitem{Brown:1977pq}
L.~S. Brown and J.~P. Cassidy, ``{Stress Tensor Trace Anomaly in a
  Gravitational Metric: General Theory, Maxwell Field},''
  \href{http://dx.doi.org/10.1103/PhysRevD.15.2810}{{\em Phys. Rev. D} {\bf 15}
  (1977)  2810}.

\bibitem{Hinterbichler:2015pqa}
K.~Hinterbichler and A.~Joyce, ``{Hidden symmetry of the Galileon},''
  \href{http://dx.doi.org/10.1103/PhysRevD.92.023503}{{\em Phys. Rev.} {\bf
  D92} (2015) no.~2, 023503},
\href{http://arxiv.org/abs/1501.07600}{{\tt arXiv:1501.07600 [hep-th]}}.

\bibitem{deRham:2013hsa}
C.~de~Rham, M.~Fasiello, and A.~J. Tolley, ``{Galileon Duality},''
  \href{http://dx.doi.org/10.1016/j.physletb.2014.03.061}{{\em Phys. Lett. B}
  {\bf 733} (2014)  46--51}, \href{http://arxiv.org/abs/1308.2702}{{\tt
  arXiv:1308.2702 [hep-th]}}.

\bibitem{deRham:2014lqa}
C.~De~Rham, L.~Keltner, and A.~J. Tolley, ``{Generalized galileon duality},''
  \href{http://dx.doi.org/10.1103/PhysRevD.90.024050}{{\em Phys. Rev. D} {\bf
  90} (2014) no.~2, 024050}, \href{http://arxiv.org/abs/1403.3690}{{\tt
  arXiv:1403.3690 [hep-th]}}.

\bibitem{Kampf:2014rka}
K.~Kampf and J.~Novotny, ``{Unification of Galileon Dualities},''
  \href{http://dx.doi.org/10.1007/JHEP10(2014)006}{{\em JHEP} {\bf 10} (2014)
  006}, \href{http://arxiv.org/abs/1403.6813}{{\tt arXiv:1403.6813 [hep-th]}}.

\bibitem{Noller:2015eda}
J.~Noller and J.~H.~C. Scargill, ``{The decoupling limit of Multi-Gravity:
  Multi-Galileons, Dualities and More},''
  \href{http://dx.doi.org/10.1007/JHEP05(2015)034}{{\em JHEP} {\bf 05} (2015)
  034}, \href{http://arxiv.org/abs/1503.02700}{{\tt arXiv:1503.02700
  [hep-th]}}.

\bibitem{Heisenberg:2019wjv}
L.~Heisenberg and C.~F. Steinwachs, ``{Geometrized quantum Galileons},''
  \href{http://dx.doi.org/10.1088/1475-7516/2020/02/031}{{\em JCAP} {\bf 02}
  (2020)  031}, \href{http://arxiv.org/abs/1909.07111}{{\tt arXiv:1909.07111
  [hep-th]}}.

\bibitem{Heisenberg:2020cyi}
L.~Heisenberg, J.~Noller, and J.~Zosso, ``{Horndeski under the quantum
  loupe},'' \href{http://arxiv.org/abs/2004.11655}{{\tt arXiv:2004.11655
  [hep-th]}}.

\bibitem{Roest:2020vny}
D.~Roest, ``{The Special Galileon as Goldstone of Diffeomorphisms},''
  \href{http://arxiv.org/abs/2004.09559}{{\tt arXiv:2004.09559 [hep-th]}}.

\bibitem{Barvinsky:1984jd}
A.~Barvinsky and G.~Vilkovisky, ``{The generalized Schwinger-De Witt technique
  and the unique effective action in quantum gravity},''
  \href{http://dx.doi.org/10.1016/0370-2693(83)90506-3}{{\em Phys. Lett. B}
  {\bf 131} (1983)  313--318}.

\bibitem{Barvinsky:1990up}
A.~Barvinsky and G.~Vilkovisky, ``{Covariant perturbation theory. 2: Second
  order in the curvature. General algorithms},''
  \href{http://dx.doi.org/10.1016/0550-3213(90)90047-H}{{\em Nucl. Phys. B}
  {\bf 333} (1990)  471--511}.

\end{thebibliography}\endgroup

\end{document}